\documentclass[12pt,english]{article}
\usepackage{lmodern}
\usepackage[T1]{fontenc}
\usepackage[latin9]{inputenc}
\usepackage{geometry}
\usepackage{mathrsfs}
\usepackage{amsmath}
\usepackage{amssymb}
\usepackage{esint} 
\usepackage{lipsum}
\makeatletter
\usepackage{hyperref}
\usepackage{cite}
\usepackage{babel}
\usepackage{mathdots}
\usepackage{amsfonts}
\usepackage{graphicx}
\usepackage{youngtab}
\usepackage{multicol}
\usepackage{slashed}
\usepackage{cleveref}
\usepackage{epsfig} 

\crefname{equation}{Eq.}{Eqs.}
\crefname{figure}{Fig.}{Figs.}
\crefname{table}{Table}{Tables}
\crefname{Section}{Section}{Sections}
\newcommand{\crefrangeconjunction}{-}
\allowdisplaybreaks

\makeatother
\numberwithin{equation}{section}

\newcommand{\el}[1]{\label{#1}}
\newcommand{\er}[1]{\eqref{#1}}

\newcommand{\ci}[1]{}
\newcommand{\ba}{\begin{eqnarray}}
\newcommand{\ea}{\end{eqnarray}}
\newcommand{\be}{\begin{equation}}
\newcommand{\ee}{\end{equation}}
\newcommand{\ke}{\rangle}
\newcommand{\br}{\langle}
\newcommand{\lb}{\left(}
\newcommand{\rb}{\right)}
\newcommand{\lc}{\left.}
\newcommand{\rc}{\right.}
\newcommand{\lsb}{\left[}
\newcommand{\rsb}{\right]}

\newcommand{\nn}{\nonumber \\}
\newcommand{\p}{\partial}

\newcommand{\hpl}{{+}{\textstyle\frac{1}{2}}}
\newcommand{\hmi}{{-}{\textstyle\frac{1}{2}}}

\newcommand{\al}{\alpha}

\def\thalf{\tfrac{1}{2}}

\newcommand{\postscript}[2]{\setlength{\epsfxsize}{#2\hsize}
   \centerline{\epsfbox{#1}}}
\newcommand{\comment}[1]{}

\def\xa{{\alpha}}

\def\xe{{\epsilon}}

\def\xk{{\kappa}}

\def\xl{{\lambda}}

\def\xt{{\theta}}

\def \Tr {{\rm Tr}}

\def\CD{{\cal D}}

\def\CM{{\cal M}}

\usepackage[usenames,dvipsnames]{xcolor}
\definecolor{rossoCP3}{cmyk}{0,.88,.77,.40}
\begin{document}

\title{\begin{flushright}
\vspace*{.1in}
{\small MPP--2014--314 \\}
{\small LMU-ASC 48/14 \\}
{\small CERN-PH-TH/2014-143 \\}
\end{flushright}
\vspace{0.2in}
\color{rossoCP3} {\bf String Resonances at Hadron Colliders} \color{black}}

\author{{\small \bf Luis A.~Anchordoqui${\bf ^1}$, Ignatios
    Antoniadis${\bf ^2}$\thanks{On leave of absence
from CPHT Ecole Polytechnique, F-91128, Palaiseau Cedex.}, De-Chang
Dai$^{\bf 3,4}$, Wan-Zhe Feng${\bf ^5}$,}\\
  {\small \bf Haim  Goldberg${\bf ^6}$,  Xing Huang${\bf ^7}$, Dieter 
  L\"ust$^{\bf 5,8}$, Dejan  Stojkovic$^{\bf 9,10}$, and Tomasz
   R. Taylor${\bf ^6}$}\\ \\
{\small \em  $^1$Department of Physics \& Astronomy,}\\
{\small \em Lehman College, City University of New York, Bronx NY 10468, USA}\\
{\small \em $^2$Department of Physics,}\\ {\small \em CERN Theory Division,
CH-1211 Geneva 23, Switzerland }\\
{\small \em $^3$Institute of Natural Sciences, Shanghai Key Lab for
  Particle Physics and Cosmology, }\\
{\small \em Shanghai Jiao Tong
  University, Shanghai 200240, China}\\
{\small \em $^4$ Department of Physics \& Astronomy, Center for Astrophysics and Cosmology,} \\
{\small \em Shanghai Jiao Tong
  University, Shanghai 200240, China}\\
{\small \em $^5$Max-Planck-Institut f\"ur Physik}\\{\small \em  Werner-Heisenberg-Institut, 80805 M\"unchen, Germany}\\
{\small \em $^6$Department of Physics,}\\{\small \em Northeastern University, Boston, MA 02115, USA}\\
{\small \em $^7$Department of Physics,} \\
{\small \em National Taiwan Normal University, Taipei, 116, Taiwan}\\
{\small \em $^8$Arnold Sommerfeld Center for Theoretical Physics,}\\
{\small \em Ludwig-Maximilians-Universit\"at M\"unchen, 80333 M\"unchen, Germany}\\
{\small \em $^{9}$HEPCOS, Department of Physics,}\\ {\small \em State University of New York at Buffalo, Buffalo,
  NY 14260-1500, USA}\\
{\small \em $^{10}$Perimeter Institute for Theoretical Physics,}\\
{\small \em 31
  Caroline Street North, Waterloo, Ontario N2J 2Y5, Canada}}
\date{}
\maketitle
\newpage
\begin{abstract}
  \noindent We consider extensions of the standard model based on open
  strings ending on D-branes, with gauge bosons due to strings
  attached to stacks of D-branes and chiral matter due to strings
  stretching between intersecting D-branes. Assuming that the
  fundamental string mass scale $M_s$ is in the TeV range and that the
  theory is weakly coupled, we discuss possible signals of string
  physics at the upcoming HL-LHC run (integrated luminosity $=
  3000~{\rm fb}^{-1}$) with a center-of-mass energy of \mbox{$\sqrt{s}
    = 14~{\rm TeV}$} and at potential future $pp$ colliders, HE-LHC
  and VLHC, operating at $\sqrt{s} = 33$ and $100$~TeV, respectively
  (with the same integrated luminosity).  In such D-brane
  constructions, the dominant contributions to full-fledged string
  amplitudes for all the common QCD parton subprocesses leading to
  dijets and $\gamma$ + jet are completely independent of the details
  of compactification and can be evaluated in a parameter-free
  manner. We make use of these amplitudes evaluated near the first
  $(n=1)$ and second $(n=2)$ resonant poles to determine the discovery
  potential for Regge excitations of the quark, the gluon, and the
  color singlet living on the QCD stack.  We show that for string
  scales as large as 7.1~TeV (6.1~TeV), lowest massive Regge
  excitations are open to discovery at the $\geq 5\sigma$ in dijet
  ($\gamma$ + jet) HL-LHC data. We also show that for $n=1$ the dijet
  discovery potential at HE-LHC and VLHC exceedingly improves: up to
  15~TeV and 41~TeV, respectively. To compute the signal-to-noise
  ratio for $n=2$ resonances, we first carry out a complete
  calculation of all relevant decay widths of the second massive-level
  string states (including decays into massless particles and a
  massive $n=1$ and a massless particle), where we rely on
  factorization and CFT techniques. Helicity wave functions of
  arbitrary higher spin massive bosons are also constructed. We
  demonstrate that for string scales $M_s \lesssim 10.5~{\rm TeV}$
  ($M_s \lesssim 28~{\rm TeV}$) detection of $n=2$ Regge recurrences
  at HE-LHC (VLHC) would become the smoking gun for D-brane string
  compactifications. Our calculations have been performed using a
  semianalytic parton model approach which is cross checked against
  an original software package.  The string event generator interfaces
  with HERWIG and Pythia through BlackMax. The source code is
  publically available in the hepforge repository.
\end{abstract}

\newpage
\tableofcontents

\newpage
\section{Introduction}

One of the most challenging problems in high-energy physics today is
to find out what is the underlying theory that completes the standard
model (SM). Despite its remarkable success, the SM is incomplete with
many unsolved puzzles -- the most striking one being the huge
disparity between the strength of gravity and of the other three known
fundamental interactions corresponding to the electromagnetic, weak,
and strong nuclear forces. Indeed, gravitational interactions are
suppressed by a very high-energy scale, the Planck mass $M_{\rm Pl} =
G_{\rm N}^{-1/2} \sim 10^{19}~{\rm GeV}$, associated to a length
$l_{\rm Pl} \sim 10^{-35}~{\rm m}$, where they are expected to become
important. This hierarchy problem suggests that new physics could be
at play above about the electroweak scale $M_{\rm EW} \sim G_{\rm
  F}^{-1/2} \sim 300~{\rm GeV}$ and has been arguably {\em the} driving
force behind high-energy physics for several decades.

In a quantum theory, the hierarchy implies a severe fine-tuning of the
fundamental parameters in more than 30 decimal places in order to keep
the masses of elementary particles at their observed values. The
reason is that quantum radiative corrections to all masses generated
by the Higgs vacuum expectation value (VEV) are proportional to the
ultraviolet cutoff which in the presence of gravity is fixed by the
Planck mass. As a result, all masses are ``attracted'' to about
$10^{16}$ times heavier than their observed values. A
fine-tuned cancellation of the radiative corrections seems unnatural, even though it is in
principle self-consistent. {\it Naturalness} implies that either the fundamental scale of
gravity must be much smaller than the Planck mass, or else there should exist a
mechanism which ensures this cancellation, perhaps arising from a new
symmetry principle beyond the SM. Low-energy supersymmetry (SUSY) with
all superparticle masses in the TeV region is a textbook
example. Indeed, in the limit of exact SUSY, quadratically
divergent corrections to the Higgs self-energy are exactly cancelled,
while in the softly broken case, they are cutoff by the SUSY 
breaking mass splittings. On the other hand, for low-mass-scale
strings, quadratic divergences are cutoff by the string scale $M_s$,
and low-energy SUSY is not needed~\cite{Antoniadis:1998ig}. These two
diametrically opposite viewpoints are experimentally testable at
high-energy particle colliders,  in particular at the CERN LHC.

The recent discovery of a particle with a mass around
126~GeV~\cite{Chatrchyan:2012ufa,Aad:2012tfa}, which seems to be the
SM Higgs, has possibly plugged the final remaining experimental hole
in the SM, cementing the theory further. The LHC data are so far
compatible with the SM within 2$\sigma$ and its precision tests. It is
also compatible with low-energy SUSY, although with some degree of
fine-tuning in its minimal version. Indeed, in the minimal
supersymmetric standard model (MSSM), the lightest Higgs scalar mass
$m_h$ satisfies the  inequality
\begin{equation}
m_h^2 \lesssim m_Z^2 \cos^2 2 \beta + \frac{3}{(4\pi)^2} \frac{m_t^4}{v^2} \left[ \ln \frac{m_{\tilde t}^2}{m_t^2} + \frac{A^2_t}{m_{\tilde t}^2} \left( 1 - \frac{A_t^2}{12 m_{\tilde t}^2} \right) \right] \lesssim (130~{\rm GeV})^2 \,,
\end{equation}
where the first term in the rhs corresponds to the tree-level
prediction and the second term includes the one loop-corrections due
to the top and stop loops. Here, $m_Z$, $m_t$, $m_{\tilde t}$ are the
$Z$-boson and the top and stop quark masses, respectively; $v =
\sqrt{v_i^2 + v_2^2}$ with $v_i$ is the VEVs of the two Higgses; $\tan
\beta = v_2 /v_1$; and $A_t$ is the trilinear stop scalar coupling. Thus,
a Higgs mass around 126~GeV requires a heavy stop $m_t \simeq 3~{\rm
  TeV}$ for vanishing $A_t$, or $A_t \simeq 3 m_{\tilde t} \simeq
1.5~{\rm TeV}$ in the ``best''-case scenario. These values are obviously
consistent with the present LHC bounds on SUSY searches, but
they are expected to be probed in the next run at double
energy. Theoretically, they imply a fine-tuning of the electroweak
 scale at the percent to per mille level. This fine-tuning can be
alleviated in supersymmetric models beyond the MSSM.

\begin{figure}[tbp]
\postscript{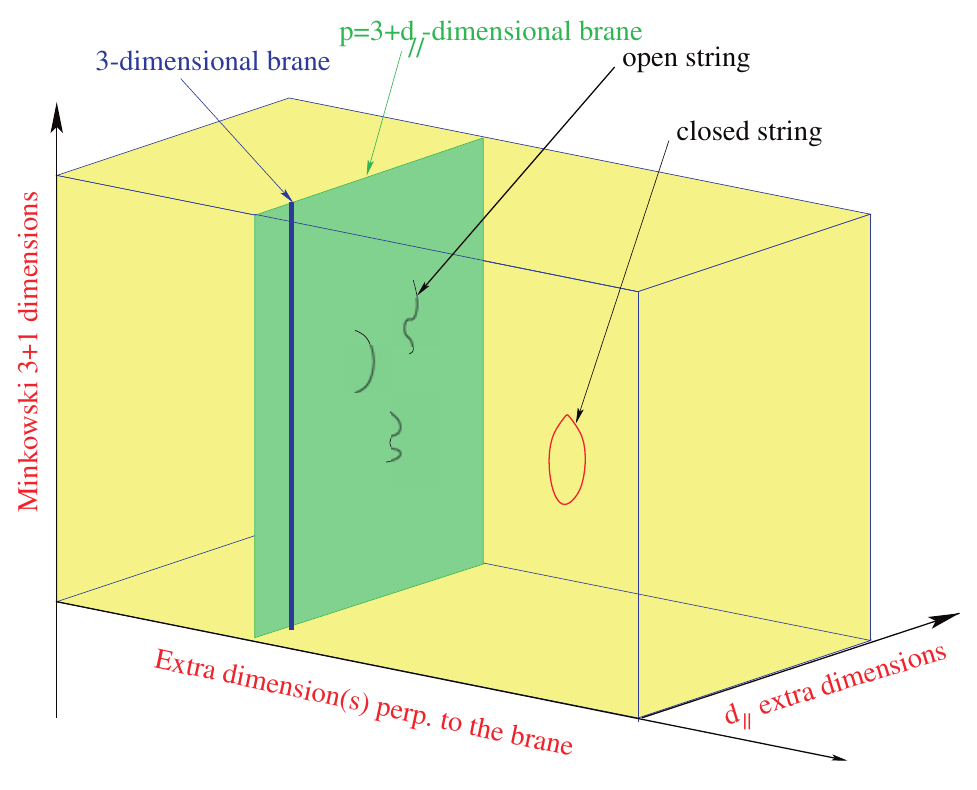}{0.8}
\caption{D-brane
  setup with $d_\parallel$ parallel and $d_\perp$ transverse internal
  directions. }
\label{stringyuniverse}
\end{figure}

Low-mass-scale superstring theory provides a braneworld description
of the SM, which is localized on membranes extending
in $p+3$ spatial dimensions, the so-called D-branes. Gauge
interactions emerge as excitations of open strings with endpoints
attached on the D-branes, whereas gravitational interactions are
described by closed strings that can propagate in all nine spatial
dimensions of string theory [these comprise parallel dimensions
extended along the $(p+3)$-branes and transverse dimensions].  For an
illustration, consider type II string theory compactified on a
six-dimensional torus $T^6$, which includes a D$p$-brane wrapped
around $p-3$ dimensions of $T^6$ with the remaining dimensions along
our familiar (uncompactified) three spatial dimensions. We denote the
radii of the {\it internal} longitudinal directions (of the
D$p$-brane) by $R_i^\parallel$, $i = 1, \dots p-3$ and the radii of
the transverse directions by $R^\perp_j$, $j = 1, \dots 9-p$; see
Fig.~\ref{stringyuniverse}.

The  Planck mass, which is related to the string mass scale  by
\begin{equation}
M_{\rm Pl}^2=\frac{8}{g_s^2} \ M_s^8\
\frac{V_6}{(2\pi)^{6}}\ ,
\label{extradim}
\end{equation}
determines the strength of the gravitational interactions. Here,
\begin{equation}
V_6=(2\pi)^6\ \prod_{i=1}^{p-3} R^\parallel_i\ \prod_{j=1}^{9-p}
R^\perp_j
\end{equation}
is the volume of $T^6$ and $g_s$ is the string coupling. It follows
that the string scale can be chosen hierarchically smaller than the
Planck mass at the expense of introducing $9-p$ large transverse
dimensions felt only by gravity, while keeping the string coupling
small. For example, for a string mass scale $M_s \approx {\cal O}
(1~{\rm TeV})$, the volume of the internal space needs to be as large
as $V_6 M_s^6 \approx {\cal O} (10^{32}).$ On the other hand, the strength of coupling of the gauge theory
living on the D-brane world volume is not enhanced as long as
$R_i^\parallel \sim M_s^{-1}$ remain small, 
\begin{equation}
\frac{1}{g^2} = \frac{1}{2\pi \, g_s}\ M_s{}^{p-3}\ \
\prod_{i=1}^{p-3}R_i^\parallel\ .
\label{catorce}
\end{equation}
The weakness of the effective four-dimensional gravity compared to gauge
interactions (ratio of $v/M_{\rm Pl}$) is then attributed to the
largeness of the transverse space radii $R_i^\perp \sim 10^{32} l_s$
compared to the string length $l_s = M_s^{-1}$.  Should nature be so
cooperative, a whole tower of infinite string excitations will open up
at this low-mass threshold, and new particles of spin $J$ follow the
well-known Regge trajectories of vibrating strings: $J = J_0 + \alpha'
M^2$, where $\alpha'$ is the Regge slope parameter that determines the
fundamental string mass scale
\begin{equation}
M_s={1\over \sqrt{\alpha'}}\, .
\label{Ms}
\end{equation}
Only one assumption will be necessary in order to set up a solid
framework: the string coupling must be small for the validity of the
above D-brane framework and of 
perturbation theory in the computation of scattering amplitudes. In
this case, black hole production and other strong gravity effects
occur at energies above the string scale; therefore, at least the 
lowest few Regge recurrences are available for examination, free from
interference with some complex quantum gravitational phenomena.

In a series of publications, we have computed open string scattering
amplitudes in D-brane models and have discussed the associated
phenomenological aspects of low-mass string Regge recurrences related
to experimental searches for physics beyond the
SM~\cite{Accomando:1999sj,Anchordoqui:2007da,Anchordoqui:2008ac,Anchordoqui:2008hi,Anchordoqui:2008di,Lust:2008qc,Lust:2009pz,Feng:2010yx,Feng:2011qc,Feng:2012bb,Anchordoqui:2009mm,Anchordoqui:2009ja,Anchordoqui:2010zs}.\footnote{String
  Regge resonances in models with low-mass string scale are also
  discussed in
  Refs.~\cite{Cullen:2000ef,Burikham:2003ha,Burikham:2004su,Chemtob:2008cb,Kitazawa:2010gh,Dong:2010jt,Hashi:2011cz,Hashi:2012ka},
  while Kaluza--Klein (KK) graviton exchange into the bulk, which
  appears at the next order in perturbation theory, is discussed
  in Refs.~\cite{Dudas:1999gz,Chialva:2005gt}.}  We have shown that certain
amplitudes to leading order in string coupling (but including all
string $\alpha'$ corrections) are
universal~\cite{Lust:2008qc,Lust:2009pz}. These amplitudes, which
include $2 \to 2$ scattering processes involving four gluons or two
gluons and two quarks, are independent of the details of the
compactification, such as the configuration of branes, the geometry of
the extra dimensions, and whether SUSY is broken or not.\footnote{The
  only remnant of the compactification is the relation between the
  Yang--Mills coupling and the string coupling. We take this relation
  to reduce to field theoretical results in the case where they exist,
  {\it e.g.}, $gg \to gg$. Then, because of the required
  correspondence with field theory, the phenomenological results are
  independent of the compactification of the transverse
  space. However, a different phenomenology would result as a
  consequence of warping one or more parallel
  dimensions~\cite{Hassanain:2009at,Perelstein:2009qi,Anchordoqui:2010vn}.}
This model independence makes it possible to compute the string
corrections to $\gamma$ + jet and dijet signals at the LHC, which, if
traced to low-mass-scale string theory, could with $100~{\rm fb}^{-1}$
of integrated luminosity (at $\sqrt{s} = 14~{\rm TeV}$) probe
deviations from SM physics at a $5\sigma$ significance for $M_s$ as
large as 6.8~TeV~\cite{Anchordoqui:2007da,Anchordoqui:2008di}.
Indeed, the signal for string excitations is spectacularly dazzling:
after operating for only a few months, with merely 2.9 inverse
picobarns of integrated luminosity, the LHC7 CMS experiment ruled out
$M_s < 2.5~{\rm TeV}$ by searching for narrow resonances in the dijet
mass spectrum~\cite{Khachatryan:2010jd}. In fact, the LHC has the
capacity to discover strongly interacting narrow resonances in
practically all ranges up to $\sqrt{s}_{\rm LHC}/2$, and therefore,
since no significance excess above background has been observed thus
far, the ATLAS~\cite{ATLAS:2012pu} and
CMS~\cite{Chatrchyan:2011ns,Chatrchyan:2013qha} experiments have
already excluded $M_s \lesssim 4.5~{\rm TeV}$.

In this work we extend our previous studies in various directions. In
all our previous analyses, the discovery reach was laid out processing
the string amplitudes using a semianalytic parton model approach. To
confront technical detector challenges, however, the standard approach
to data analysis is typically reliant on the existence of Monte Carlo
event simulation tools that allow complete simulation of the
signal. In this paper we are filling this gap by bringing the
excitations of open strings into the ATLAS/CMS analysis software
environment. A complete simulation with full Pythia treatment is quite
a difficult task, because this event generator is set up in the same
way perturbation theory works and consequently handles color flow
lines of ordinary Feynman diagrams. Note that in string theory, there
are processes (like $gg \to g\gamma$) that in ordinary field theory
work only at loop level and their color lines do not follow the normal
lines of tree-level Feynman diagrams. The proposed strategy here is to
incorporate the string amplitudes into
BlackMax~\cite{Dai:2007ki,Dai:2009by}, a comprehensive black hole
event generator for LHC analysis that interfaces (via the Les Houches
accord~\cite{Boos:2001cv}) to HERWIG and Pythia. The parton evolution
and hadronization will then be performed with the correct format for
direct implementation in the official Monte Carlo packages for
simulating an actual experiment at the LHC. The two-step approach
advanced herein can circumvent the color line technicalities and, at
the same time, facilitate the comparison with high-multiplicity events
from gravitational collapse.

Recently the idea of building a 33 TeV and/or 100 TeV circular
proton-proton collider has gained momentum, starting with an
endorsement in the Snowmass Energy Frontier report~\cite{Gershtein:2013iqa}, and
importantly followed by the creation of two parallel initiatives: one
at CERN~\cite{CERN} and one in China~\cite{China}. In this paper we
study the discovery reach and exclusion limits of lowest massive Regge
excitations for the  collider specifications,
\vspace{-10pt}
\begin{center}
\renewcommand{\arraystretch}{1.5}
\setlength{\tabcolsep}{12pt}
\begin{tabular}{c|c|c}
\hline 
\hline
Machine & $\sqrt{s}$ (TeV) & Final integrated luminosity\\
\hline
LHC phase I & $14$ & $300$ fb$^{-1}$ \\
HL-LHC or LHC phase II &  $14$ & $3000$ fb$^{-1}$ \\
HE-LHC & $33$ & $3000$ fb$^{-1}$ \\
VLHC & $100$ & $3000$ fb$^{-1}$ \\
\hline
\hline
\end{tabular}
\end{center}
that are extensively discussed in the Snowmass Energy Frontier
report~\cite{Gershtein:2013iqa}.  For the HE-LHC and VLHC, the second
excited string states may also be within reach. The decay widths of $n
= 2$ resonances into massless particles have been previously obtained
in Refs.~\cite{Dong:2010jt,Hashi:2011cz}. For a full treatment, however, one
still needs to compute the decay widths into one massive $n = 1$
particle and a massless particle. Herein, we obatin all these widths by
factorizing four-point amplitudes with one massive ($n = 1$) and three
massless particles.

The layout of the paper is as follows. We begin in Sec.~\ref{s2} with
an outline of the basic setting of intersecting D-brane models and we
discuss general aspects of the effective low-energy theory inherited
from properties of the overarching string theory. After that, we
particularize the discussion to three- and four-stack intersecting
D-brane configurations that realize the SM by open strings. For
completness, in Sec.~\ref{s3} we provide a summary of previous
results. In particular, we give an overview of all formulae relevant
for the $s$-channel string amplitudes of lowest massive Regge
excitations leading to $\gamma$ + jet and dijets. Readers already
familiar with these topics may skip this section. In Secs.~\ref{s4}
and \ref{s5} we present a complete calculation of all relevant decay
widths of the second massive-level string states. The computation is
performed in a model-independent and universal way, and so our results
hold for all compactifications. Armed with the full-fledged string
amplitudes of all partonic subprocesses, in Sec.~\ref{discoveryreach}
we quantify signal and background rates of $n=1$ and $n=2$ Regge
recurrences in the early LHC phase I, HL-LHC, HE-LHC, and VLHC. In
Sec.~\ref{s7} we describe the input and output of the string event
generator interface (SEGI) with HERWIG and Pythia through BlackMax and
present some illustrative results. Finally in Sec.~\ref{s8} we make a
few observations on the consequences of the overall picture discussed
herein.

A point worth noting at this juncture is that the tensor-to-scalar ratio ($r
= 0.20^{+0.07}_{-0.05}$) inferred from the excess B-mode power
observed by the Background Imaging of Cosmic Extragalactic
Polarization (BICEP2) experiment suggests in simple slow-roll models
an era of inflation with energy densities of order $(10^{16}~{\rm
  GeV})^4$, not far below the Planck density~\cite{Ade:2014xna}. This
presumably suggests that low-mass-scale string compactifications in
connection with large extra dimension are quite hard to
realize. However, one should keep in mind
that there is an ongoing controversy concerning the effect of
background on the BICEP2 result~\cite{Liu:2014mpa,Flauger:2014qra}.

\section{Intersecting D-brane string compactifications}
\label{s2}

D-brane low-mass-scale string compactifications provide a collection
of building block rules that can be used to build up the SM or
something very close to
it~\cite{Blumenhagen:2001te,Cvetic:2001tj,Cvetic:2001nr,Antoniadis:2001np,Ibanez:2001nd,Kiritsis:2002aj,Cremades:2003qj,Honecker:2004kb,Gmeiner:2005vz,Gmeiner:2008xq,Kiritsis:2003mc,Blumenhagen:2005mu,Blumenhagen:2006ci,Honecker:2013kda,Berenstein:2014wva}. The
details of the D-brane construct depend a lot on whether we use
oriented string or unoriented string models. The basic unit of gauge
invariance for oriented string models is a $U(1)$ field, so that a
stack of $N$ identical D-branes eventually generates a $U(N)$ theory
with the associated $U(N)$ gauge group. In the presence of many
D-brane types, the gauge group becomes a product form $\prod U(N_i)$,
where $N_i$ reflects the number of D-branes in each stack. Gauge
bosons (and associated gauginos in a SUSY model) arise from strings
terminating on {\em one} stack of D-branes, whereas chiral matter
fields are obtained from strings stretching between {\em two}
stacks. Each of the two strings end points carries a fundamental
charge with respect to the stack of branes on which it
terminates. Matter fields thus posses quantum numbers associated with
a bifundamental representation.  In orientifold brane configurations,
which are necessary for tadpole cancellation, and thus consistency of
the theory, open strings become in general nonoriented. For
unoriented strings the above rules still apply, but we are allowed
many more choices because the branes come in two different
types. There are branes for which the images under the orientifold are
different from themselves, and also branes that 
are their own images under the orientifold procedure. Stacks of the
first type combine with their mirrors and give rise to $U(N)$ gauge
groups, while stacks of the second type give rise to only $SO(N)$ or
$Sp(N)$ gauge groups.

\subsection{Mass mixing effect}
\label{massmixing}

In three-stack intersecting brane models, one could have one or two 
massive $U(1)$'s, depending on using $Sp(1)$ or $U(2)$ to realize
$SU(2)$; while in four-stack models, one could have two or three massive
$U(1)$'s.  In general, one can have many $U(1)$'s in the intersecting
brane model constructions including hidden sectors, and in these cases
there will be many massive $U(1)$'s, which have been studied
in Refs.~\cite{Cvetic:2011iq,Feng:2014cla,Feng:2014eja}.  Assuming no kinetic mixing,
effectively the Lagrangian for all the $U(1)$'s from an $n$-stack
model can be written as
\begin{align}
\mathscr{L} &
=-\frac{1}{4}\sum_{a}F_{a}^{2}-\frac{1}{2}A_{a}M_{ab}^{2}A_{b}+\sum_{a}\bar{\psi}_{a}(i\slashed\partial+g'_{a}
Q_{a}\slashed A_{a})\psi_{a}\,,
\label{Almiron}
\end{align}
where $\psi_{a}$ denotes the matter fields charged under $U(1)_{a}$
($a,b,\cdots$ label the stack of D-branes), $g'_a$ are the gauge
couplings, and $Q_a$ are the charges. Note that the relation for
$U(N)$ unification, $g'_a = g_a / \sqrt{2N}$, holds only at $M_s$
because the $U (1)$ couplings ($g'_1,\, g'_2,\, g'_3,\, \cdots$) run
differently from the non-Abelian $SU(3)$ ($g_3$) and $SU(2)$
($g_2$)~\cite{Anchordoqui:2011eg}.  The $U(1)$ mass-squared matrix is
of the form~\cite{Ghilencea:2002da,Feng:2014cla}
\begin{equation}
M_{ab}^{2}=g'_{a}g'_{b}K_{ai}\mathcal{G}_{ij}K_{jb}^{T}\,,
\end{equation}
where the integer-entry matrix $K$ contains all the information of
local model constructions -- wrapping numbers which give rise
to correct family multiplicity and the (MS)SM spectrum -- and $\mathcal{G}_{ij}$ is the metric of the complex structure
moduli space.\footnote{
For toroidal models, the explicit form of $\mathcal{G}_{ij}$ can be
derived, see for example Ref.~\cite{Feng:2014cla}. }
In general, the entries of the $U(1)$ mass-squared matrix are all
of order of $M_{s}^{2}$. This $U(1)$ mass-squared matrix is positive semidefinite
which has one zero eigenvalue that corresponds to the hypercharge.
One could diagonalize $M_{ab}^{2}$ using an orthogonal matrix $O$
such that
\begin{equation}
O^{T}M^{2}O=\left(\begin{array}{ccccc}
\lambda_{1}^{2}\\
 & \lambda_{2}^{2} &  & 0\\
 &  & \ddots\\
 & 0 &  & \ddots\\
 &  &  &  & \lambda_{n}^{2}
\end{array}\right)\equiv D^{2}\,,
\end{equation}
where the eigenvalues are sorted from small to large, {\it i.e.},
$\lambda_{i}<\lambda_{j}$ for $i<j$. $\lambda_{1}=0$ corresponds to
the mass of the hypercharge gauge boson $Y_{\mu}\equiv
A_{1,\mu}^{(m)}$. We can define the gauge boson corresponding to the
lightest massive $U(1)$ to be $Z'$. Here we only discuss the case that
there is only one massless $U(1)$, and thus $D^{2}$ contains only one
zero eigenvalue (hypercharge) and all other $U(1)$'s are
massive.\footnote{ The hidden sector could have massless $U(1)$, which
  leads to the hidden photon scenario. Some models ({\it e.g.},
  SM$^{++}$~\cite{Anchordoqui:2012wt,Anchordoqui:2012fq}) may have a
  massless $U(1)_{B-L}$, but it must develop a mass to avoid long-range
  force. We omit this discussion here.}  This transformation also
takes the gauge fields from their original basis into the physical
mass eigenbasis as (with an upper index $^{(m)}$)
\begin{equation}
A_{i}^{(m)}=\sum_{a}O_{ia}^{T}A_{a}\,.
\end{equation}
The column vectors of the orthogonal matrix $O$ are the eigenvectors
of $M^{2}$. Since the eigenvalues are already sorted, the first column
vector gives rise to the hypercharge combination
\begin{equation}
Y_{\mu}=A_{1,\mu}^{(m)}=\sum_{a}O_{1a}^{T}A_{a}\,,
\end{equation}
and the second column vector gives rise to
\begin{equation}
Z_{\mu}'=A_{2,\mu}^{(m)}=\sum_{a}O_{2a}^{T}A_{a}\,,
\end{equation}
and so on. Conversely, one could also write the gauge bosons in the
original basis in terms of the mass eigenstates
\begin{equation}
A_{a}=\sum_{i}O_{ai}A_{i}^{(m)}\,.\label{AinMassB}
\end{equation}

After the mass mixing, the Lagrangian in the $U(1)$ gauge boson mass
eigenbasis reads
\begin{equation}
\mathscr{L}=-\frac{1}{4}\sum_{i}F_{i}^{(m)2}-\frac{1}{2}D_{ii}^{2}(A_{i}^{(m)})^{2}+\sum_{a}\bar{\psi}_{a}(i\slashed\partial+\bar
g_{i}^{(m)}Q_{i}^{(m)}\slashed A_{i}^{(m)})\psi_{a}\,.
\end{equation}
Since the elements in the orthogonal matrix $O$ are in general irrational
numbers (except for the first column, for which the entrees are all fractional
numbers which give rise to to the hypercharge), the gauge charges
in the $U(1)$ mass eigenbasis are not quantized. A matter field
carrying $Q_{a}$ under $U(1)_{a}$, with the gauge coupling $g'_{a}$,
after the mass mixing couples to the gauge field $A_{i}^{(m)}$
in the mass eigenbasis, with strength $\bar g_{i}^{(m)}Q_{i}^{(m)}\equiv\sum_{a}g'_{a}Q_{a}O_{ai}$.
Thus, all the matter fields raised from the D-brane can couple to all the
anomalous $U(1)$'s. Since the elements of the $U(1)$ mass-squared matrix
are around the same order, the entries of the orthogonal matrix $O$
are in general of order $\mathcal{O}(1)$. Thus the anomalous $U(1)$'s
could couple to all the SM particles with sizable strength~\cite{Feng:2014cla}.

\subsection{Higgs mechanism and $Z-Z'$ mixing}

The Higgs field(s) is (are) also realized as (an) open string(s) stretching
between two stacks of D-branes and hence is (are) charged under the two
$U(1)$'s. After the mass mixing, the Higgs field(s) would be also
charged under all the $U(1)$'s in the mass eigenbasis and couple
to all these massive $U(1)$ gauge bosons. Thus, after the electroweak
symmetry breaking, all the gauge boson masses would be corrected. The
covariant derivative reads
\begin{equation}
{\cal D}_{\mu}=\partial_{\mu}-ig_{2}A_{\mu}^{a}T^{a}-i\frac{1}{2}g_{Y}Y_{\mu}-i\sum_{i=2}^{n}
\bar g_{i}^{(m)}Q_{i}^{(m)}A_{i}^{(m)}\,,
\label{Echeverria}
\end{equation}
where $T^{a}=\sigma^{a}/2$ is the $SU(2)$ generator and $Y_{\mu}$
the hypercharge gauge boson. Effectively, the mass terms of all the $U(1)$'s take
the form
\begin{align}
-\mathscr{L}_{m} & ={\cal D}_{\mu}\phi {\cal D}^{\mu}\phi+\frac{1}{2}D_{ii}^{2}(A_{i}^{(m)})^{2}\nonumber \\
 & =\frac{1}{2}\frac{v^{2}}{4}\Big[g_{2}^{2}(A_{\mu}^{1})^{2}+g_{2}^{2}(A_{\mu}^{1})^{2}\nonumber \\
 & \qquad\quad+\big(-g_{2}A_{\mu}^{3}+g_{Y}Y_{\mu}+2\sum_{i=2}^{n}
 \bar g_{i}^{(m)}Q_{i}^{(m)}A_{i}^{(m)}\big)^{2}\Big]+\frac{1}{2}D_{ii}^{2}(A_{i}^{(m)})^{2}\,,\label{LM}
\end{align}
where $v$ is the VEV of the Higgs. $A_{\mu}^{1}$ and $A_{\mu}^{2}$
give rise to $W^{\pm}$ and the mass mixing only occurs within $A_{\mu}^{3},A_{i}^{(m)}$.
One needs to perform another diagonalization to determine the mass
eigenstates of all the massive $U(1)$ gauge bosons. The special form
of Eq.~\eqref{LM} ensures there is only one massless eigenstates
$A_{\mu}^{\gamma}=\frac{1}{\sqrt{g_{2}^{2}+g_{Y}^{2}}}(g_{Y}A_{\mu}^{3}+g_{2}Y_{\mu})$
which will be identified to be the photon. And the electric charge
remains unchanged, i.e., $e=\frac{g_{2}g_{Y}}{\sqrt{g_{2}^{2}+g_{Y}^{2}}}$.
However, the $Z$ boson would be a mixture of $Z_{{\rm SM}}$ and all
the $A_{i}^{(m)}$. The mass of the $Z$ boson is corrected by
\begin{equation}
M_{Z}=\frac{v}{2}\sqrt{g_{2}^{2}+g_{Y}^{2}}+\mathcal{O}\left(\frac{v^{2}}{M_{Z'}^{2}}\right)\,.\label{Zmass}
\end{equation}
Hence, the mass of the $Z'$ gauge boson cannot be very light;
otherwise, 
it would violate the constraints on $Z-Z'$ mixing from the electroweak
precision test~\cite{Erler:2009jh}.  In addition, as mentioned
earlier, all the anomalous $U(1)$'s could couple to all the SM
particles with sizable strength.  LEP II and the LHC both set stringent
bounds on them.  In particular, the bound from LEP II on $Z'$ reads
$M_{Z'}/g_{Z'l^{+}l^{-}}>6\,{\rm TeV}$~\cite{:2005ema,Carena:2004xs}.
Because of the QCD background, LHC could set bounds on the $Z'$ by either
examining the leptonic Drell--Yan processes $pp\to Z' \to
l^+l^-$~\cite{Accomando:2010fz,Chatrchyan:2012it}, or examining the
dijet resonances from a heavy $Z'$~\cite{Chatrchyan:2013qha}.  These
bounds are quite strong.  Though it is difficult for LHC to
distinguish low energy hadronic final states due to the QCD
background, the LHC bound on a leptophobic $Z'$ [for example, $Z'$ for
$U(1)_{B}$] is not that strong~\cite{Dobrescu:2013cmh}.  However, it
is very likely that the $Z'$ from D-brane models would couple to all
the SM particles with sizable strength.  Thus, in general,
unless there is some fine-tuning, this type of $Z'$ has to be quite massive
($\gtrsim2$~TeV) to pass all the current experimental constraints from
colliders.  We also would like to point out here that although in
general $Z'$ [the lightest anomalous $U(1)$] can be much lighter than
the string scale, this is a model-dependent question.  For many cases,
especially for intersecting brane models with fewer extra $U(1)$'s
[e.g., the minimal D-brane model $U(3) \times Sp(1) \times U(1)$
with only one additional (massive) $U(1)$],
the mass of $Z'$ can also be closed to the string scale.

\subsection{SM from D-brane constructs}
\label{SMDb}

While the existence of Regge excitations is a completely universal
feature of string theory, there are many ways of realizing the SM in
such a framework.  Individual models use various D-brane
configurations and compactification spaces. Consequently, these may
lead to very different SM extensions, but as far as the collider
signatures of Regge excitations are concerned, their differences boil
down to a few parameters. The most relevant characteristics is how the
$U(1)_Y$ hypercharge is embedded in the $U(1)$ associated to
$D$-branes. One $U(1)$ (baryon number) comes from the ``QCD'' stack of
three branes, as a subgroup of the $U(3)$ group that contains $SU(3)$
color, but obviously one needs at least one extra $U(1)$. As noted in
Sec~\ref{massmixing}, in D-brane compactifications the hypercharge
always appears as a linear, nonanomalous combination of the baryon
number with one, two, or more $U(1)$s. The precise form of this
combination bears down on the photon couplings; however, the
differences between individual models amount to numerical values of a
few parameters.

The minimal embedding of the SM particle spectrum requires at least
three brane stacks~\cite{Antoniadis:2000ena} leading to three distinct
models of the type $U(3) \times U(2) \times U(1)$ that were classified
in Refs.~\cite{Antoniadis:2000ena, Antoniadis:2004dt}. In such minimal
models the color stack $a$ of three D-branes is intersected by the
(weak doublet) stack $b$ and by one (weak singlet) D-brane
$c$~\cite{Antoniadis:2000ena}. {}For the two-brane stack $b$, there is
a freedom of choosing physical state projections leading either to
$U(2)$ or to the symplectic $Sp(1)$ representation of Weinberg-Salam
$SU(2)_L$.

In the bosonic sector, the open strings terminating on QCD stack $a$
contain the standard $SU(3)$ octet of gluons $g_\mu^a$ and an
additional $U(1)_a$ gauge boson $C_\mu$, most simply the manifestation
of a gauged baryon number symmetry: $U(3)_a\sim SU(3)\times
U(1)_a$. On the $U(2)_b$ stack the open strings correspond to the
electroweak gauge bosons $A_\mu^a$, and again an additional $U(1)_b$
gauge field $X_\mu$.  So the associated gauge groups for these stacks
are $SU(3) \times U(1)_a,$ $SU(2)_L \times U(1)_b$, and $U(1)_c$,
respectively.  We can further simplify the model by eliminating
$X_\mu$; to this end instead we can choose the projections leading to
$Sp(1)$ instead of $U(2)$ \cite{Berenstein:2006pk}. The $U(1)_Y$ boson
$Y_\mu$, which gauges the usual electroweak hypercharge symmetry, is a
linear combination of $C_\mu$, the $U(1)_c$ boson $B_\mu$, and perhaps
a third additional $U(1)$ gauge field, $X_\mu$.\footnote{In the
  notation of (\ref{Almiron}), $C$, $X$,  and $B$ correspond to $A_a,$
  $A_b$, and $A_c$. We will freely switch between these two notations
  depending on which is more convenient for the discussion.}  The fermionic matter
consists of open strings located at the intersection points of the
three stacks.  Concretely, the left-handed quarks are sitting at the
intersection of the $a$ and the $b$ stacks, whereas the right-handed
$u$ quarks come from the intersection of the $a$ and $c$ stacks and
the right-handed $d$ quarks are situated at the intersection of the
$a$ stack with the $c'$ (orientifold mirror) stack. All the scattering
amplitudes between these SM particles essentially only depend on the
local intersection properties of these D-brane stacks.

\begin{table}
\caption{Chiral fermion spectrum of the $U(3) \times Sp(1) \times U(1)$ D-brane model.}
\begin{center}
\begin{tabular}{c|ccccc}
\hline
\hline
 Name &~~Representation~~& ~$Q_a$~& ~$Q_c$~ & ~$Q_Y$~ \\
\hline
~~$U_i$~~ & $({\bar 3},1)$ &    $-1$ & $\phantom{-}1$ & $-\frac{2}{3}$ \\[1mm]
~~$D_i$~~ &  $({\bar 3},1)$ &    $-1$ & $-1$ & $\phantom{-}\frac{1}{3}$  \\[1mm]
~~$L_i$~~ & $(1,2)$&    $\phantom{-}0$ & $\phantom{-}1$ & $-\frac{1}{2}$  \\[1mm]
~~$E_i$~~ &  $(1,1)$&  $\phantom{-}0$ & $-2$ &  $\phantom{-}1$  \\[1mm]
~~$Q_i$~~ & $(3,2)$& $\phantom{-}1$ & $\phantom{-}0$ & $\phantom{-}\frac{1}{6}$ \\[1mm]
\hline
\hline
\end{tabular}
\end{center}
\label{t1}
\end{table}

The chiral fermion spectrum of the $U(3) \times Sp(1) \times U(1)$
D-brane model is given in Table~\ref{t1}.  In such a minimal D-brane
construction,  the coupling strength of $C_\mu$ is down by root 6 
when compared to the $SU(3)_C$ coupling $g_3$, and the hypercharge
\begin{equation}
Q_Y = \frac{1}{6} \, Q_a -\frac{1}{2} \, Q_c
\label{hypercarga}
\end{equation}
is free of anomalies. However, the $Q_a$ (gauged baryon number) is
anomalous.  This anomaly is canceled by the f-D version of the
Green--Schwarz (GS)
mechanism~\cite{Green:1984sg,Witten:1984dg,Dine:1987xk,Atick:1987gy,Lerche:1987qk,Sagnotti:1992qw}. The
vector boson $Y'_\mu$, orthogonal to the hypercharge, must grow a mass
in order to avoid long-range forces between baryons other than gravity
and Coulomb forces. The anomalous mass growth allows the survival of
global baryon number conservation, preventing fast proton
decay~\cite{Ghilencea:2002da}.

In the $U(3) \times Sp(1) \times U(1)$ D-brane model, the
$U(1)_a$ assignments are fixed (they give the baryon number) and the
hypercharge assignments are fixed by the SM. Therefore, the mixing angle
$\theta_P$ between the hypercharge and the $U(1)_a$ is obtained in a
similar manner to the way the Weinberg angle is fixed by the $SU(2)_L$
and the $U(1)_Y$ couplings ($g_2$ and $g_Y$, respectively) in the
SM. The Lagrangian containing the $U(1)_a$ and $U(1)_c$ gauge fields
is given by
\begin{equation}
\mathscr{ L} = g'_1 \, \hat B_\mu \, J_B^\mu + g'_3 \, \hat  C_\mu  \,  J_C^\mu
\label{apache}
\end{equation}
where $\hat B_\mu = \cos \theta_P \, Y_\mu + \sin \theta_P \, Y'_\mu$
and
$\hat C_\mu = -\sin \theta_P \, Y_\mu + \cos \theta_P\, Y'_\mu$
are canonically normalized.  Substitution of these expressions into (\ref{apache}) leads to
\begin{eqnarray}
\mathscr{L}  =  Y_\mu \left(g'_1 \cos \theta_P J_B^\mu - g'_3 \sin \theta_P J_C^\mu \right)
+  Y'_\mu \left(g'_1 \sin \theta_P J_B^\mu + g'_3 \cos \theta_P J_C^\mu \right),
\end{eqnarray}
with $g'_1 \, \cos \theta_P\,  J_B^\mu -   g'_3 \, \sin \theta_P \,  J_C^\mu = g_Y \, J_Y^\mu$. We have seen
that the hypercharge is anomaly free if
$J_Y =  \frac{1}{6} \,  J_C^\mu - \frac{1}{2} \,  J_B^\mu$, yielding
\begin{equation}
g'_1 \cos \theta_P = \frac{1}{2} g_Y \quad {\rm and}  \quad g'_3 \, \sin \theta_P = \frac{1}{6} g_Y \, .
\label{pipita}
\end{equation} 
{}From (\ref{pipita}) we obtain the following relations
\begin{equation}
\tan \theta_P =  \, \frac{g'_1}{3 g'_3}, \quad \quad
\left(\frac{g_Y}{2g'_1}\right)^2 + \left(\frac{g_Y}{6 g'_3}\right)^2 = 1, \quad {\rm and}
\quad \frac{1}{4 {g'_1}^2} + \frac{1}{36 {g'_3}^2} = \frac{1}{g_Y^2} \, .
\label{pellerano}
\end{equation}
We use the evolution of gauge couplings from the weak scale $M_Z$ as
determined by the one-loop beta functions of the SM with three
families of quarks and leptons and one Higgs doublet,
\begin{equation}
{1\over \alpha_i(M)}={1\over \alpha_i(M_Z)}-
{b_i\over 2\pi}\ln{M \over M_Z}\ ; \quad i=2,3,Y,
\end{equation}
where $\alpha_i=g_i^2/4\pi$ and $b_3=-7$, $b_2=-19/6$, $b_Y=41/6$. We
also use the measured values of the couplings at the $Z$ pole
$\alpha_3(M_Z)=0.118\pm 0.003$, $\alpha_2(M_Z)=0.0338$,
$\alpha_Y(M_Z)=0.01014$ (with the errors in $\alpha_{2,Y}$ less than
1\%)~\cite{Beringer:1900zz}.  Running couplings up to 5~TeV, which is
where the phenomenology will be, we get $\kappa \equiv \sin \theta_P
\sim 0.14$. When the theory undergoes electroweak symmetry breaking,
because $Y'$ couples to the Higgs, one gets additional mixing. Hence
$Y'$ is not exactly a mass eigenstate. The explicit form of the
low-energy eigenstates $A_\mu$, $Z_\mu,$ and $Z'_\mu$ is given
in Ref.~\cite{Berenstein:2008xg}.

We pause to summarize the degree of model dependency stemming from the
multiple $U(1)$ content of the minimal model containing three stacks of
D-branes. First, there is an initial choice to be made for the gauge
group living on the $b$ stack. This can be either $Sp(1)$ or
$U(2)$. In the case of $Sp(1)$, the requirement that the hypercharge
remains anomaly free is sufficient to fix its $U(1)_a$ and $U(1)_c$
content, as explicitly presented in Eqs.~(\ref{pipita}) and
(\ref{pellerano}). Consequently, the fermion couplings, as well as the
mixing angle $\theta_P$ between hypercharge and the baryon number
gauge field are wholly determined by the usual SM couplings.  The
alternative selection -- that of $U(2)$ as the gauge group tied to the
$b$ stack -- branches into some further choices. This is because the
$Q_a,\ Q_b,\ Q_c$ content of the hypercharge operator 
\begin{equation}
  Q_Y = c_a \, Q_a + c_b \, Q_b + c_c \, Q_c
\end{equation}
is not uniquely determined by the anomaly cancelation requirement.  In
fact, as seen in Ref.~\cite{Antoniadis:2000ena}, there are three possible
embeddings with one more possibility for the hypercharge combination
besides (\ref{hypercarga}). This final choice does not depend on
further symmetry considerations.

\begin{figure}[tbp]
\postscript{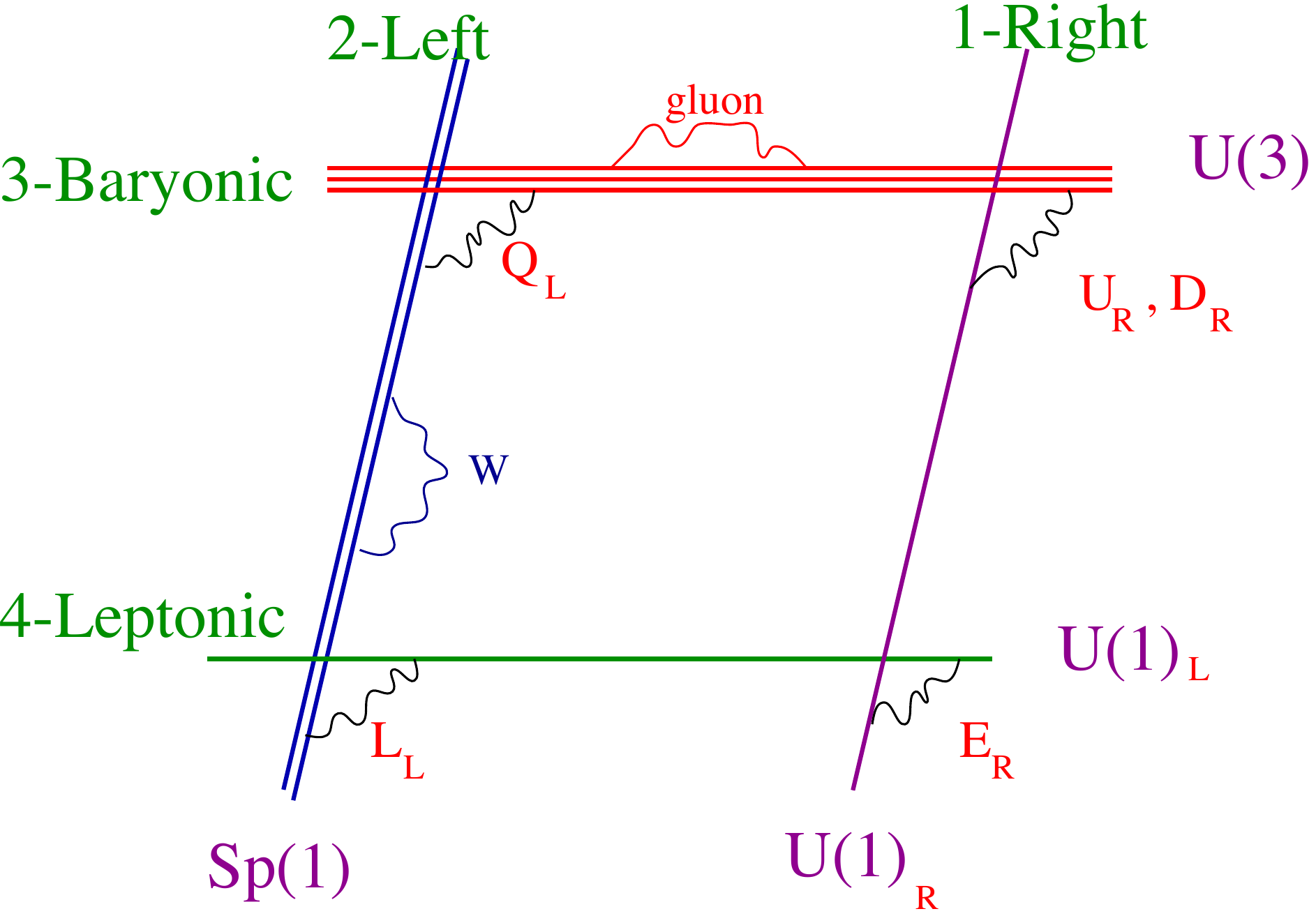}{0.8}
\caption{Pictorial representation of the $U(1)_C \times Sp(1)_L \times U(1)_L \times U(1)_R$ D-brane model.}
\label{4stacks}
\end{figure}
\begin{table}
\caption{Chiral fermion spectrum of the $U(3)_C \times Sp(1)_L \times U(1)_L \times U(1)_R$ D-brane model.}
\begin{center}
\begin{tabular}{c|ccccc}
\hline
\hline
 Name &~~Representation~~& ~$Q_3$~& ~$ Q_{1L}$~ & ~$Q_{1R}$~ & ~~$Q_{Y}$~~ \\
\hline
~~$U_i$~~ & $({\bar 3},1)$ &    $-1$ & $\phantom{-}0$ & $-1$ & $-\frac{2}{3}$  \\[1mm]
~~$D_i$~~ &  $({\bar 3},1)$&    $-1$ & $\phantom{-}0$ & $\phantom{-} 1$ & $\phantom{-}\frac{1}{3}$  \\[1mm]
~~$L_i$~~ & $(1,2)$&    $\phantom{-}0$ &  $\phantom{-}1$ & $\phantom{-}0$ & $-\frac 1 2$  \\[1mm]
~~$E_i$~~ &  $(1,1)$&   $\phantom{-}0$ & $-1$ &  $\phantom{-} 1$ & $\phantom{-} 1$ \\[1mm]
~~$Q_i$~~ & $(3,2)$& $\phantom{-}1$ & $\phantom{-}0 $ & $\phantom{-} 0$ & $\phantom{-}\frac{1}{6}$   \\[1mm]
\hline
\hline
\end{tabular}
\end{center}
\label{t:spectrum}
\end{table}

The SM embedding in four D-brane stacks leads to many more models that
have been classified in Refs.~\cite{Antoniadis:2002qm,
  Anastasopoulos:2006da}. To make a phenomenologically
interesting choice, we focus on models where $U(2)$ can be reduced to
$Sp(1)$.  Besides the fact that this reduces the number of extra
$U(1)$'s, one avoids the presence of a problematic Peccei--Quinn
symmetry, associated in general with the $U(1)$ of $U(2)$ under which
Higgs doublets are charged~\cite{Antoniadis:2000ena}. We then impose
baryon and lepton number symmetries that determine completely the
model $U(3)_C \times Sp(1)_L \times U(1)_L \times U(1)_R$, as
described in~~\cite{Ibanez:2001nd,Anastasopoulos:2006da}.  A schematic
representation of the D-brane structure is shown in
Fig.~\ref{4stacks}.  The corresponding fermion quantum numbers are
given in Table~\ref{t:spectrum}. The two extra $U(1)$'s are the baryon
and lepton numbers, $B$ and $L$, respectively; they are given by the
following combinations: \ba B=Q_3/3\quad;\quad L=Q_{1L}\quad;\quad Q_Y
= c_1 Q_{1R} + c_3 Q_3 + c_4 Q_{1L} \, ,
\label{hyperchargeY4stacks}
\ea with $c_1 = 1/2,$ $c_3 = 1/6$, and $c_4 =-1/2$; or equivalently by
the inverse relations
\ba
Q_3=3B\quad;\quad Q_{1L}=L\quad;\quad Q_{1R}=2Q_Y-(B-L)\, .
\label{bb-l}
\ea
As usual, the  $U(1)$ gauge interactions arise through the covariant derivative 
\begin{equation}\el{covderi} \CD_\mu = \p_\mu - i g'_3 \, C_\mu \, Q_3 -i g'_4 \, \tilde B_\mu \, Q_{1L} -i g'_1 \, B_\mu \, Q_{1R} \, ,
\end{equation}
where $g'_1$, $g'_3$, and $g'_4$ are the gauge coupling constants.  We
can define $Y_\mu$ and two other fields $Y'{}_\mu, Y''{}_\mu$ that are
related to $C_\mu,  B_\mu, \tilde B_\mu$ by the orthogonal transformation~\cite{Anchordoqui:2011ag}
\begin{equation}
O = \left(
\begin{array}{ccc}
 C_\theta C_\psi  & -C_\phi S_\psi + S_\phi S_\theta C_\psi  & S_\phi
S_\psi +  C_\phi S_\theta C_\psi  \\
 C_\theta S_\psi  & C_\phi C_\psi +  S_\phi S_\theta S_\psi  & - S_\phi
C_\psi + C_\phi S_\theta S_\psi  \\
 - S_\theta  & S_\phi C_\theta  & C_\phi C_\theta
\end{array}
\right) \,,
\end{equation}
with Euler angles $\theta$, $\psi,$ and $\phi$. Equation~(\ref{covderi}) can be rewritten in terms of $Y_\mu$, $Y'_\mu$, and
$Y''_\mu$ as follows
\begin{eqnarray}
\CD_\mu & = & \partial_\mu -i Y_\mu \left(-S_\xt g'_1 Q_{1R} + C_\theta S_\psi  g'_4  Q_{1L} +  C_\theta C_\psi g'_3 Q_3 \right) \nonumber \\
 & - & i Y'_\mu \left[ C_\theta S_\phi  g'_1 Q_{1R} +\left( C_\phi C_\psi + S_\theta S_\phi S_\psi \right)  g'_4 Q_{1L} +  (C_\psi S_\theta S_\phi - C_\phi S_\psi) g'_3 Q_3 \right] \label{linda} \\
& - & i Y''_\mu \left[ C_\theta C_\phi g'_1 Q_{1R} +  \left(-C_\psi S_\phi + C_\phi S_\theta S_\psi \right)  g'_4  Q_{1L} + \left( C_\phi C_\psi S_\theta + S_\phi S_\psi\right) g'_3 Q_3 \right]   \, .  \nonumber
\end{eqnarray}
Now, by demanding that $Y_\mu$ has the
hypercharge $Q_Y$ given in Eq.~\er{hyperchargeY4stacks}  we  fix the first column of the rotation matrix $O$
\begin{equation}
\label{othogaugefield}
\left(
\begin{array}{c} C_\mu \\ \tilde B_\mu \\ B_\mu
\end{array} \right) = \left(
\begin{array}{lr}
  Y_\mu \, c_3g_Y /g'_3& \dots \\
  Y_\mu \, c_4 g_Y/g'_4 & \dots\\
   Y_\mu \, c_1g_Y/g'_1 & \dots
\end{array}
\right) \, ,
\end{equation}
and we determine the value of the two associated Euler angles
\begin{equation}
\theta = {\rm -arcsin} [c_1 g_Y/g'_1]
\label{theta}
\end{equation}
and
\begin{equation}
\psi = {\rm arcsin}  [c_4 g_Y/ (g'_4 \, C_\theta)] \, .
\label{psi}
\end{equation}
The couplings $g'_1$ and $g'_4$ are related through the orthogonality condition,
\begin{equation}
 \left(\frac{c_4}{ g' _4} \right)^2  = \frac{1}{g_Y^2} - \left(\frac{c_3}{g'_3} \right)^2  - \left(\frac{c_1}{g'_1}\right)^2  \, ,
\end{equation}
with $g'_3$ fixed by the relation $g_3 (M_s) = \sqrt{6} \, g'_3
(M_s)$~\cite{Anchordoqui:2011eg}. The field $Y_\mu$ then appears in
the covariant derivative with the desired $Q_Y$. The ratio of the
coefficients in Eq.~\er{othogaugefield} is determined by the form of
Eqs.~\er{hyperchargeY4stacks} and \er{covderi}.  The value of $g_Y$
is determined so that the coefficients in Eq.~\er{othogaugefield} are
components of a normalized vector so that they can be a row vector of
$O$. The rest of the transformation (the ellipsis part) involving
$Y',Y''$ is not necessary for our calculation. The point is that we
now know the first row of the matrix $O$, and hence we can get the
first column of $O^T$, which gives the expression of $Y_\mu$ in terms
of $C_\mu, B_\mu, \tilde B_\mu$, \be Y_\mu = \frac {c_3 g_Y} {g'_3}
C_\mu + \frac {c_1 g_Y} {g'_1} B_\mu +\frac {c_4 g_Y} {g'_4} \tilde
B_\mu.\ee This is all we need when we calculate the interaction
involving $Y_\mu$; the rest of $O$, which tells us the expression of
$Y', Y''$ in terms of $C,X,B$, is not necessary.  For later
convenience, we define $\xk, \eta, \xi$ as \be Y_\mu = \xk C_\mu +
\eta B_\mu +\xi \tilde B_\mu\,;\ee therefore \be \xk = \frac {c_3 g_Y}
{g'_3},\quad \eta = \frac {c_1 g_Y} {g'_1},\quad \xi = \frac {c_4 g_Y}
{g'_4}\,.\label{consts}\ee The expression for the $C-Y$ mixing
parameter $\kappa$ is the same as that of the $U(3) \times Sp(1)
\times U(1)$ minimal D-brane model.

Note that with the ``canonical'' charges of the right-handed neutrino
$Q_{1L}=Q_{1R}=-1$, the combination $B-L$ is anomaly free, while for
$Q_{1L}=Q_{1R}=+1$, both $B$ and $B-L$ are anomalous.\footnote{We
  noted elsewhere~\cite{Anchordoqui:2011nh} that such right-handed
  neutrinos would have left their imprint on the photons of the cosmic
  microwave background.}  As mentioned already, anomalous $U(1)$'s
become massive necessarily due to the GS anomaly cancellation, but
nonanomalous $U(1)$'s can also acquire masses due to effective
six-dimensional anomalies associated, for instance, to sectors
preserving ${\cal N}=2$
SUSY~\cite{Antoniadis:2002cs,Anastasopoulos:2003aj}.\footnote{In fact,
  also the hypercharge gauge boson of $U(1)_Y$ can acquire a mass
  through this mechanism.  To keep it massless, certain topological
  constraints on the compact space have to be met.} These
two-dimensional ``bulk'' masses become therefore larger than the
localized masses associated to four-dimensional anomalies, in the
large volume limit of the two extra dimensions. Specifically for
D$p$-branes with $(p-3)$-longitudinal compact dimensions the masses of
the anomalous and, respectively, the non-anomalous $U(1)$ gauge bosons
have the following generic scale behavior: \ba
{\rm anomalous}~U(1)_a:~~~M_{Z'}&=&g'_aM_s\, ,\nonumber\\
{\rm nonanomalous}~U(1)_a:~~~M_{Z''}&=&g'_aM_s^3\, V_2\, .  \ea Here,
$g'_a$ is the gauge coupling constant associated to the group
$U(1)_a$, given by $g'_a\propto g_s/\sqrt{V_\parallel}$, where $g_s$
is the string coupling and $V_\parallel$ is the internal D-brane world
volume along the $(p-3)$ compact extra dimensions, up to an order 1
proportionality constant. Moreover, $V_2$ is the internal
two-dimensional volume associated to the effective six-dimensional
anomalies giving mass to the nonanomalous $U(1)_a$.\footnote{It should
  be noted that in spite of the proportionality of the $U(1)_a$ masses
  to the string scale, these are not string excitations but zero
  modes.  The proportionality to the string scale appears because the
  mass is generated from anomalies, via an analog of the GS anomaly
  cancellations: either four-dimensional anomalies, in which case the
  GS term is equivalent to a St\"uckelberg mechanism, or from
  effective six-dimensional anomalies, in which case the mass term is
  extended in two more (internal) dimensions. The nonanomalous
  $U(1)_a$ can also grow a mass through a Higgs mechanism. The
  advantage of the anomaly mechanism vs an explicit VEV of a scalar
  field is that the global symmetry survives in perturbation theory,
  which is a desired property for the baryon and lepton number,
  protecting proton stability and small neutrino masses.}  For
example, for the case of D5-branes, for which the common intersection
locus is just four-dimensional Minkowski space, $V_\parallel=V_2$ denotes
the volume of the longitudinal, two-dimensional space along the two
internal D5-brane directions.  Since internal volumes are bigger than
one in string units to have effective field theory description, the
masses of nonanomalous $U(1)$ gauge bosons are generically larger
than the masses of the anomalous gauge bosons.

In principle, in addition to the orthogonal field mixing induced by
identifying anomalous and nonanomalous $U(1)$ sectors, there may be
kinetic mixing between these sectors.  In all the D-brane models
discussed in this section, however, since there is only one $U(1)$ per
stack of D-branes, the relevant kinetic mixing is between $U(1)$'s on
different stacks and hence involves loops with fermions at brane
intersection.  Such loop terms are typically down by $g_i^2/16 \pi^2
\sim 0.01$~\cite{Dienes:1996zr}.\footnote{See
  also Ref.~\cite{Honecker:2012qr,Honecker:2013mya}} Generally, the major
effect of the kinetic mixing is in communicating SUSY breaking from a
hidden $U(1)$ sector to the visible sector, generally in modification
of soft scalar masses.  Stability of the weak scale in various models
of SUSY breaking requires the mixing to be orders of magnitude below
these values~\cite{Dienes:1996zr}.  For a comprehensive review of
experimental limits on the mixing, see Ref.~\cite{Abel:2008ai}.  Moreover,
none of the D-brane constructions discussed above have a hidden sector
-- all the $U(1)$'s (including the anomalous ones) couple to the
visible sector.  In summary, kinetic mixing between the nonanomalous
and the anomalous $U(1)$'s in every basic model discussed in this
paper will be small because the fermions in the loop are all in the
visible sector. In the absence of electroweak symmetry breaking, the
mixing vanishes.

\section{Lowest massive Regge excitations of open strings} 
\label{s3}

The most direct way to compute the amplitude for the scattering of
four gauge bosons is to consider the case of polarized particles
because all nonvanishing contributions can be then generated from a
single, maximally helicity violating (MHV), amplitude -- the so-called
{\it partial\/} MHV amplitude~\cite{Parke:1986gb}.  Assume that two
vector bosons, with the momenta $k_1$ and $k_2$, in the $U(N)$ gauge
group states corresponding to the generators $T^{a_1}$ and $T^{a_2}$
(here in the fundamental representation), carry negative helicities
while the other two, with the momenta $k_3$ and $k_4$ and gauge group
states $T^{a_3}$ and $T^{a_4}$, respectively, carry positive
helicities. (All momenta are incoming.)  Then the partial amplitude
for such an MHV configuration is given
by~\cite{Stieberger:2006bh,Stieberger:2006te}
\begin{equation}
  \label{ampl}
  {\cal A}(A_1^-,A_2^-,A_3^+,A_4^+) ~=~ 4\, g^2\, {\rm Tr}
  \, (\, T^{a_1}T^{a_2}T^{a_3}T^{a_4}) {\langle 12\rangle^4\over
    \langle 12\rangle\langle 23\rangle\langle 34\rangle\langle
    41\rangle}V(k_1,k_2,k_3,k_4)\ ,
\end{equation}
where $g$ is the $U(N)$ coupling constant, $\langle ij\rangle$ are the
standard spinor products written in the notation of
Refs.~\cite{Mangano:1990by,Dixon:1996wi}, and the Veneziano form factor,
\begin{equation}
V(k_1,k_2,k_3,k_4) = V(  s,   t,   u)= \frac{s\,u}{tM_s^2}B(-s/M_s^2,-u/M_s^2)={\Gamma(1-   s/M_s^2)\ \Gamma(1-   u/M_s^2)\over
    \Gamma(1+   t/M_s^2)} \label{formf}
\end{equation}
is the function of Mandelstam variables,
$s=2k_1k_2$, $t=2  k_1k_3$, $u=2 k_1k_4$; $s+t+u=0$.  (For simplicity we drop carets for the parton subprocess.)
The physical content of the form factor becomes clear after using the
well-known expansion in terms of $s$-channel resonances~\cite{Veneziano:1968yb},
\begin{equation}
B(-s/M_s^2,-u/M_s^2)=-\sum_{n=0}^{\infty}\frac{M_s^{2-2n}}{n!}\frac{1}{s-nM_s^2}
\Bigg[\prod_{J=1}^n(u+M^2_sJ)\Bigg],\label{bexp}
\end{equation}
which exhibits $s$-channel poles associated to the propagation of
virtual Regge excitations with masses $\sqrt{n}M_s$. Thus near the
$n$th-level pole $(s\to nM^2_s)$,
\begin{equation}\qquad
V(  s,   t,   u)\approx \frac{1}{s-nM^2_s}\times\frac{M_s^{2-2n}}{(n-1)!}\prod_{J=0}^{n-1}(u+M^2_sJ)\ .\label{nthpole}
\end{equation}
In specific amplitudes, the residues combine with the remaining
kinematic factors, reflecting the spin content of particles exchanged
in the $s$ channel, ranging from $J=0$ to $J=n+1$. The low-energy
expansion reads
\begin{equation}
\label{vexp}
V(s,t,u)\approx 1-{\pi^2\over 6} \frac{s\,
    u}{M_s^4}-\zeta(3)\, \frac{s\, t\, u}{M_s^6}+\dots \, .
\end{equation}

Interestingly, because of the proximity of the eight gluons and the photon
on the color stack of D-branes, the gluon fusion into $\gamma$ + jet
couples at tree level~\cite{Anchordoqui:2007da}.  This implies that
there is an order $g_3^2$ contribution in string theory, whereas this
process is not occurring until order $g_3^4$ (loop level) in field
theory. One can write down the total amplitude for this process
projecting the gamma ray onto the hypercharge,
\begin{equation}
  {\cal M} (gg \to \gamma g) = \cos \theta_W \, \, {\cal M} (gg \to Y
  g) = \kappa \, \, \cos \theta_W \, \, {\cal M} (gg \to C g) \, , 
\label{collectableG}
\end{equation}
where $\kappa$ is the (model-dependent) $C$-$Y$ mixing coefficient.

Consider the amplitude involving three $SU(N)$
gluons $g_1,~g_2,~g_3$ and one $U(1)$ gauge boson $\gamma_4$
associated to the same $U(N)$ stack:
\begin{equation}
\label{gens}
T^{a_1}=T^a \ ,~ \ T^{a_2}=T^b\ ,~ \
  T^{a_3}=T^c \ ,~ \ T^{a_4}=Q \mathbb{I}\ ,
\end{equation}
where $\mathbb{I}$ is the $N{\times}N$ identity matrix and $Q$ is the
$U(1)$ charge of the fundamental representation. 
The color
factor \begin{equation}\label{colf}{\rm
    Tr}(T^{a_1}T^{a_2}T^{a_3}T^{a_4})=Q(d^{abc}+{i\over 4}f^{abc})\ ,
\end{equation}
where the totally symmetric symbol $d^{abc}$ is the symmetrized trace
while $f^{abc}$ is the totally antisymmetric structure constant (see
Appendix~\ref{App1}).

The full MHV amplitude can be
obtained~\cite{Stieberger:2006bh,Stieberger:2006te} by summing the
partial amplitudes (\ref{ampl}) with the indices permuted in as \begin{equation}
\label{afull} {\cal M}(g^-_1,g^-_2,g^+_3,\gamma^+_4)
  =4\,g_3^{2}\langle 12\rangle^4 \sum_{\sigma } { {\rm Tr} \, (\,
    T^{a_{1_{\sigma}}}T^{a_{2_{\sigma}}}T^{a_{3_{\sigma}}}T^{a_{4}})\
    V(k_{1_{\sigma}},k_{2_{\sigma}},k_{3_{\sigma}},k_{4})\over\langle
    1_{\sigma}2_{\sigma} \rangle\langle
    2_{\sigma}3_{\sigma}\rangle\langle 3_{\sigma}4\rangle \langle
    41_{\sigma}\rangle }\ ,
\end{equation}
where the sum runs over all six permutations $\sigma$ of $\{1,2,3\}$ and
$i_{\sigma}\equiv\sigma(i)$, $N=3$. Note that in the effective field theory
of gauge bosons there are no Yang--Mills interactions that could
generate this scattering process at the tree level. Indeed, $V=1$ at
the leading order of Eq.~(\ref{vexp}), and the amplitude vanishes
due to the following identity:
\begin{equation}\label{ymlimit}
{1\over\langle 12\rangle\langle
      23\rangle\langle 34\rangle\langle
      41\rangle}+{1\over\langle 23\rangle\langle
      31\rangle\langle 14\rangle\langle 42\rangle}+{1\over\langle 31\rangle\langle
      12\rangle\langle 24\rangle\langle 43\rangle} ~=~0\ .
\end{equation}
Similarly,
the antisymmetric part of
the color factor (\ref{colf}) cancels out in the full amplitude (\ref{afull}). As a result,
one obtains
\begin{equation}\label{mhva}
{\cal
    M}(g^-_1,g^-_2,g^+_3,\gamma^+_4)=8\, Q\, d^{abc}g_3^{2}\langle
  12\rangle^4\left({\mu(s,t,u)\over\langle 12\rangle\langle
      23\rangle\langle 34\rangle\langle
      41\rangle}+{\mu(s,u,t)\over\langle 12\rangle\langle
      24\rangle\langle 13\rangle\langle 34\rangle}\right),
\end{equation}
 where
\begin{equation}
\label{mudef}
\mu(s,t,u)= \Gamma(1-u/M_s^2)\left( {\Gamma(1-s/M_s^2)\over
      \Gamma(1+t/M_s^2)}-{\Gamma(1-t/M_s^2)\over \Gamma(1+s/M_s^2)}\right) .
\end{equation}
All nonvanishing amplitudes can be obtained in a similar way. In particular,
\begin{equation}
\label{mhvb}
{\cal M}(g^-_1,g^+_2,g^-_3,\gamma^+_4)=8\, Q\,
  d^{abc}g_3^{2}\langle 13\rangle^4\left({\mu(t,s,u)\over\langle
      13\rangle\langle 24\rangle\langle 14\rangle\langle
      23\rangle}+{\mu(t,u,s)\over\langle 13\rangle\langle
      24\rangle\langle 12\rangle\langle 34\rangle}\right),
\end{equation}
and the remaining ones can be obtained either by appropriate
permutations or by complex conjugation.

To obtain the cross section for the (unpolarized) partonic
subprocess $gg\to g\gamma$, we take the squared moduli of individual
amplitudes, sum over final polarizations and colors, and average over
initial polarizations and colors. As an example, the modulus square of
the amplitude (\ref{afull}) is:
\begin{equation}
\label{mhvsq}
|{\cal
    M}(g^-_1,g^-_2,g^+_3,\gamma^+_4)|^2=64\, Q^2\, d^{abc}d^{abc}g_3^{4}
  \left|{s\mu(s,t,u)\over u}+{s\mu(s,u,t)\over t} \right|^2 \, .
\end{equation}
 Taking
into account all $4(N^2-1)^2$ possible initial polarization/color
configurations and the formula~\cite{vanRitbergen:1998pn}
\begin{equation}
\label{dsq}
\sum_{a,b,c}d^{abc}d^{abc}={(N^2-1)(N^2-4)\over 16 N},
\end{equation}
 we
obtain the average squared amplitude~\cite{Anchordoqui:2007da}
\begin{equation}
\label{mhvav}
|{\cal M}(gg\to
  g\gamma)|^2=g_3^4Q^2C(N)\left\{ \left|{s\mu(s,t,u)\over
        u}+{s\mu(s,u,t)\over t} \right|^2+(s\leftrightarrow
    t)+(s\leftrightarrow u)\right\},
\end{equation}
 where
\begin{equation}\label{cnn}
C(N)={2(N^2-4)\over N(N^2-1)} \, .
\end{equation}
Before proceeding, we need to make precise the
value of $Q$. If we were considering the process $gg\rightarrow C g,$ then $Q =
\sqrt{1/6}$ due to the $U(N)$ normalization condition~\cite{Antoniadis:2000ena}. However,
for $gg\rightarrow \gamma g$ there are two additional projections given in (\ref{collectableG}):
from $C_\mu$ to the hypercharge boson $Y_\mu$, yielding a mixing factor
$\kappa$; and from $Y_\mu$ onto a photon, providing an additional
factor $\cos\theta_W$. This gives
\begin{equation}
Q = \sqrt{\tfrac{1}{6}} \ \kappa \ \cos \theta_W \, .
\end{equation}

The two most interesting energy regimes of $gg\to g\gamma$
scattering are far below the string mass scale $M_s$
and near the threshold for the production of massive string
excitations. At low energies, Eq.~(\ref{mhvav}) becomes
\begin{equation}
\label{mhvlow}
|{\cal M}(gg\to
  g\gamma)|^2\approx g_3^4Q^2C(N){\pi^4\over 4 M_s^8}(s^4+t^4+u^4)\qquad
  (s,t,u\ll M_s^2) \, .
\end{equation}
The absence of massless poles, at $s=0$, {\it etc.}, translated
into the terms of effective field theory, confirms that there are
no exchanges of massless particles contributing to this process.
On the other hand, near the string threshold $s\approx M_s^2$,
\begin{equation}
\label{mhvlow3}
|{\cal M}(gg\to g\gamma)|^2\approx
4g_3^4Q^2C(N){M_s^8+t^4+u^4\over M_s^4(s-M_s^2)^2}
\qquad (s\approx M_s^2) \, .
\end{equation}

The general form of (\ref{afull})  for any given four external gauge bosons reads
 \begin{eqnarray}
{\cal M}(A_{1}^{-},A_{2}^{-},A_{3}^{+},A_{4}^{+}) & = & 4\, g^{2}\langle12\rangle^{4}\bigg[
\frac{V_{t}}{\langle12\rangle\langle23\rangle\langle34\rangle\langle41\rangle}
\makebox{Tr}(T^{a_{1}}T^{a_{2}}T^{a_{3}}T^{a_{4}}+T^{a_{2}}T^{a_{1}}T^{a_{4}}
T^{a_{3}})\nonumber \\
 & + & \frac{V_{u}}{\langle13\rangle\langle34\rangle\langle42\rangle\langle21\rangle}
 \makebox{Tr}(T^{a_{2}}T^{a_{1}}T^{a_{3}}T^{a_{4}}+T^{a_{1}}T^{a_{2}}T^{a_{4}}
 T^{a_{3}})\nonumber \\
 & + & \frac{V_{s}}{\langle14\rangle\langle42\rangle\langle23\rangle\langle31
 \rangle}\makebox{Tr}(T^{a_{1}}T^{a_{3}}T^{a_{2}}T^{a_{4}}+T^{a_{3}}T^{a_{1}}
 T^{a_{4}}T^{a_{2}})\bigg] , \label{mhv}
\end{eqnarray}
where
\begin{equation}
V_t =V(  s,  t,  u) ~,\qquad V_u=V(  t,  u,  s) ~,\qquad  V_s=V(  u,   s,   t) \, .
\end{equation}
The modulus square of the four-gluon amplitude, summed over final polarizations and colors, and averaged over all $4 (N^2 -1)^2$ possible initial polarization/color configurations follows from (\ref{mhv}) and is given by~\cite{Lust:2008qc} 
\begin{eqnarray}
|{\cal M} (gg \to gg)| ^2 & = & g_3^4 \left(\frac{1}{  s^2} + \frac{1}{  t^2} + \frac{1}{  u^2} \right) \left[ \frac{2 N^2}{N^2-1} \, (  s^2 \, V_s^2 +   t^2 \, V_t^2 +   u^2 \,  V_u^2) \right. \nonumber \\
 & + & \left. \frac{4 (3 - N^2)}{N^2 (N^2-1)} \, (  s \, V_s +   t \, V_t +   u \, V_u)^2 \right] \, .
\label{gggg}
\end{eqnarray}
The average square amplitudes for two gluons and two quarks are given by 
\begin{equation}
|{\cal M} (gg \to q \bar q)|^2 =
g_3^4 N_f \frac{  t^2 +   u^2}{  s^2} \left[\frac{1}{2N} \frac{1}{  u \,   t} (  t \, V_t +   u \, V_u)^2 - \frac{N}{N^2 -1} \, V_t \, V_u \right] \,\,,
\label{ggqbarq}
\end{equation}
\begin{equation}
|{\cal M} (q \bar q \to gg)|^2 =
g_3^4 \ \frac{  t^2 +   u^2}{  s^2} \ \left[\frac{(N^2 -1)^2}{2 N^3} \ \frac{1}{  u   t} \ (  t \, V_t +   u \, V_u)^2 - \frac{N^2 -1}{N} \, V_t \, V_u \right] \, \,,
\label{qbarqgg}
\end{equation}
and 
\begin{equation}
|{\cal M}(qg \to qg)|^2 =
g_3^4 \ \frac{  s^2 +   u^2}{  t^2} \left[V_s \, V_u - \frac{N^2 -1}{2 N^2} \ \ \frac{1}{  s   u} \ (  s\, V_s +   u \, V_u)^2 \right] \,\, .
\label{qgqg}
\end{equation}

The amplitudes for the four-fermion processes like quark-antiquark
scattering are more complicated because the respective form factors
describe not only the exchanges of Regge states but also of heavy
Kaluza--Klein (KK) and winding states with a model-dependent spectrum
determined by the geometry of extra dimensions. Fortunately, they are
suppressed, for two reasons: {\it (i)}~the QCD $SU(3)$ color group
factors favor gluons over quarks in the initial state, and {\it (ii)}~the
parton luminosities in proton-proton collisions at the LHC, at the
parton center-of-mass energies above~1 TeV, are significantly lower
for quark-antiquark subprocesses than for gluon-gluon and
gluon-quark~\cite{Anchordoqui:2009mm}.  The collisions of valence
quarks occur at higher luminosity; however, there are no Regge
recurrences appearing in the $s$ channel of quark-quark
scattering~\cite{Lust:2008qc}.

In the following we isolate the contribution from the first resonant
state in Eqs.~(\ref{gggg}) -- (\ref{qgqg}). For partonic center-of-mass
energies $\sqrt{s}<M_s$, contributions from the Veneziano functions
are strongly suppressed, as $\sim (\sqrt{s}/M_s)^8$, over SM
processes; see Eq.~(\ref{mhvlow}). [Corrections to SM processes at
$\sqrt{s} \ll M_s$ are of order $(\sqrt{s}/M_s)^4$; see
Eq.~(\ref{vexp}).] To factorize amplitudes on the poles due
to the lowest massive string states, it is sufficient to consider
$s=M_s^2$. In this limit, $V_s$ is regular while
\begin{equation}
V_t \to \frac{  u}{  s-M_s^2}~,\qquad V_u \to \frac{  t}{  s-M_s^2} \, .
\end{equation}
Thus the $s$-channel pole term of the average square amplitude
(\ref{gggg}) can be rewritten as
\begin{equation}
|{\cal M} (gg \to gg)| ^2  =  2 \ 
\frac{g_3^4}{M_s^4}\ \left(\frac{N^2-4+(12/N^2)}{N^2-1}\right) 
 \ \frac{M_s^8+  t^4 +   u^4}{(  s - M_s^2)^2} \, .
\label{ggggpole}
\end{equation}
Note that the contributions of single poles to the cross section are
antisymmetric about the position of the resonance, and vanish in any
integration over the resonance.\footnote{As an illustration, consider the
  amplitude $a +b/D$ in the vicinity of the pole, where $a$ and $b$
  are real, $D = x+i\epsilon,$ $x=s-M_s^2,$ and $\epsilon = \Gamma
  M_s.$ Then, since Re$(1/D) = x/|D|^2$, the cross section becomes
  $\sigma \propto a^2 + b^2/|D|^2 + 2 \, a \, b \, x/|D|^2 \simeq a^2
  + b^2 \, \pi \, \delta(x)/\epsilon + 2ab \, \pi \, x \
  \delta(x)/\epsilon$. Integrating over the width of the resonance,
  one obtains $a^2 \epsilon + b^2 \pi/\epsilon \simeq b \pi$, because
  $b \propto \epsilon$, $a \propto g^2$ and $\epsilon \propto g^2$.}

Before proceeding, we pause to present our notation. The first Regge
excitations of the gluon $g$, the color singlet $C$, and quarks $q$
will be denoted by $G^{(1)},\ C^{(1)},$ and $Q^{(1)}$,
respectively. Recall that $C_\mu$ has an anomalous mass in general
lower than the string scale by an order of magnitude. If that is the
case, and if the mass of the $C^{(1)}$ is composed (approximately) of
the anomalous mass of the $C_\mu$ and $M_s$ added in quadrature, we
would expect only a minor error in our results by taking the $C^{(1)}$
to be degenerate with the other resonances.  The singularity at
$s=M_s^2$ needs softening to a Breit--Wigner form, reflecting the
finite decay widths of resonances propagating in the $s$
channel. Because of averaging over initial polarizations,
Eq.(\ref{ggggpole}) contains additive contributions from both
spin-$J=0$ and spin-$J=2$ $U(3)$ bosonic Regge excitations ($G^{(1)}$
and $C^{(1)}$), created by the incident gluons in the helicity
configurations ($\pm \pm$) and ($\pm \mp$), respectively.  The $M_s^8$
term in Eq.~(\ref{ggggpole}) originates from $J=0$, and the $ t^4+
u^4$ piece reflects $J=2$ activity. Since the resonance widths depend
on the spin and on the identity of the intermediate state ($G^{(1)}$,
$C^{(1)}$), the pole term (\ref{ggggpole}) should be smeared
as~\cite{Anchordoqui:2008di}
\begin{eqnarray}
\label{gggg2}
|{\cal M} (gg \to gg)| ^2 & = & 2\ 
\frac{g_3^4}{M_s^4}\ \left(\frac{N^2-4+(12/N^2)}{N^2-1}\right)  \\
 & \times & \left\{ W_{G^{(1)}}^{gg \to gg} \, \left[\frac{M_s^8}{(  s-M_s^2)^2 
+ (\Gamma_{G^{(1)}}^{J=0}\ M_s)^2} \right. \right.
\left. +\frac{  t^4+   u^4}{(  s-M_s^2)^2 + (\Gamma_{G^{(1)}}^{J=2}\ M_s)^2}\right] \nonumber \\
   & + &
W_{C^{(1)}}^{gg \to gg} \, \left. \left[\frac{M_s^8}{(  s-M_s^2)^2 + (\Gamma_{C^{(1)}}^{J=0}\ M_s)^2} \right.
\left. +\frac{  t^4+  u^4}{(  s-M_s^2)^2 + (\Gamma_{C^{(1)}}^{J=2}\ M_s)^2}\right] \right\}, \nonumber
\end{eqnarray}
where $\Gamma_{G^{(1)}}^{J=0} = 75\, (M_s/{\rm TeV})~{\rm GeV}$,
$\Gamma_{C^{(1)}}^{J=0} = 150 \, (M_s/{\rm TeV})~{\rm GeV}$,
$\Gamma_{G^{(1)}}^{J=2} = 45 \, (M_s/{\rm TeV})~{\rm GeV}$, and
$\Gamma_{C^{(1)}}^{J=2} = 75 \, (M_s/{\rm TeV})~{\rm GeV}$ are the
total decay widths for intermediate states $G^{(1)}$ and $C^{(1)}$, with
angular momentum $J$~\cite{Anchordoqui:2008hi}.  The associated weights of these intermediate states are given in terms of the probabilities for the various entrance and exit channels
\begin{eqnarray} 
\el{totalcrossdecom} \frac{N^2-4+12/N^2}{N^2-1} & = & \frac {16} {(N^2-1)^2}\left[\left(N^2-1\right)\left(\frac{N^2-4}{ 4N}\right)^2+
  \left(\frac{N^2-1}{2N}\right)^2\right] \nonumber \\ & \propto & \frac {16} {(N^2-1)^2}
\left[(N^2-1)(\Gamma_{G^{(1)}\to gg})^2 + (\Gamma_{C^{(1)}\to gg})^2\right] \,,
\label{LCuno}
\end{eqnarray}
yielding
\begin{equation}
W_{G^{(1)}}^{gg \to gg} = \frac{8 (\Gamma_{G^{(1)} \to gg})^2}{8(\Gamma_{G^{(1)} \to gg})^2 +
(\Gamma_{C^{(1)} \to gg})^2} = 0.44, 
\label{w1}
\end{equation}
and
\begin{equation}
W_{C^{(1)}}^{gg \to gg} = \frac{(\Gamma_{C^{(1)}
  \to gg})^2}{8(\Gamma_{G^{(1)} \to gg})^2 + (\Gamma_{C^{(1)} \to gg})^2} =0.56  \, .
\label{w2}
\end{equation}
A similar calculation transforms Eq.~(\ref{ggqbarq}) near the pole into
\begin{eqnarray}
  |{\cal M} (gg \to q \bar q)|^2 & = & \frac{g_3^4}{M_s^4}\ N_f\ \left(\frac{N^2-2}{N(N^2-1)}\right)
  \left [W_{G^{(1)}}^{gg \to q \bar q}\, \frac{  u   t(   u^2+   t^2)}{(  s-M_s^2)^2 + (\Gamma_{G^{(1)}}^{J=2}\ M_s)^2} \right. \nonumber \\
  & + & \left. W_{C^{(1)}}^{gg \to q \bar q}\, \frac{  u   t (   u^2+   t^2)}{(  s-M_s^2)^2 + 
      (\Gamma_{C^{(1)}}^{J=2}\ M_s)^2} \right] \, ,
\label{LCdos}
\end{eqnarray}
where 
\begin{equation}
W_{G^{(1)}}^{gg \to q \bar q}  = W_{G^{(1)}}^{q \bar q \to gg} = 
\frac{8\Gamma_{G^{(1)} \to gg} \, 
\Gamma_{G^{(1)} \to q \bar q}} {8\Gamma_{G^{(1)} \to gg} \, 
\Gamma_{G^{(1)} \to q \bar q} + \Gamma_{C^{(1)} \to gg} \, 
\Gamma_{C^{(1)} \to q \bar q}}  = 0.71
\label{w3}
\end{equation}
 and 
\begin{equation}
W_{C^{(1)}}^{gg \to q \bar q} = W_{C^{(1)}}^{q \bar q \to gg}  = 
\frac{\Gamma_{C^{(1)} \to gg} \, 
\Gamma_{C^{(1)} \to q \bar q}}{8\Gamma_{G^{(1)} \to gg} \, 
\Gamma_{G^{(1)} \to q \bar q} + \Gamma_{C^{(1)} \to gg} \, 
\Gamma_{C^{(1)} \to q \bar q}}  = 0.29\, .
\label{w4}
\end{equation}
Near the $ s$ pole, Eq.~(\ref{qbarqgg}) becomes 
\begin{eqnarray}
|{\cal M} (q \bar q \to gg)|^2  & = &  \frac{g_3^4}{M_s^4}\ \left(\frac{(N^2 -2) 
(N^2-1)}{N^3}\right)
\left[ W_{G^{(1)}}^{q\bar q \to gg} \,  \frac{  u   t(   u^2+   t^2)}{(  s-M_s^2)^2 + (\Gamma_{G^{(1)}}^{J=2}\ M_s)^2} \right. \nonumber \\
 & + & \left.  W_{C^{(1)}}^{q\bar q \to gg} \, \frac{  u   t(   u^2+   t^2)}{(  s-M_s^2)^2 + (\Gamma_{C^{(1)}}^{J=2}\ M_s)^2} \right] \,\,,
\label{LCtres}
\end{eqnarray}
whereas Eq.~(\ref{qgqg}) can be rewritten as
\begin{eqnarray}
|{\cal M}(qg \to qg)|^2 & = & - \frac{g_3^4}{M_s^2}\ 
\left(\frac{N^2-1}{2N^2}\right)
\left[ \frac{M_s^4   u}{(  s-M_s^2)^2 + (\Gamma_{Q^{(1)}}^{J=1/2}\ M_s)^2}\right. \nonumber \\
 & + & \left. \frac{  u^3}{(s-M_s^2)^2 + (\Gamma_{Q^{(1)}}^{J=3/2}\ M_s)^2}\right] \, \, .
\label{qgqg2}
\end{eqnarray}
The total decay widths for the $Q^{(1)}$ excitation are:
$\Gamma_{Q^{(1)}}^{J=1/2}  = 37\, (M_s/{\rm TeV})~{\rm GeV}$ and
$\Gamma_{Q^{(1)}}^{J=3/2} = 19\, (M_s/{\rm TeV})~{\rm
  GeV}$~\cite{Anchordoqui:2008hi}.\footnote{We added a factor of
  1/2 for the spin-3/2 exited string states as noted
  in Ref.~\cite{Hashi:2011cz}.} Superscripts $J=2$ are understood to be
inserted on all the $\Gamma$'s in Eqs.~(\ref{w1}), (\ref{w2}),
(\ref{w3}), and (\ref{w4}); we have taken $N=3$ and $N_f = 6$. Equation~(\ref{gggg2}) reflects the fact that
weights for $J=0$ and $J=2$ are the same~\cite{Anchordoqui:2008hi}.  

The $s$-channel poles near the second Regge resonance can be
approximated by expanding the Veneziano form factor $V_t$ around $s = 2
M_s^2$,
\begin{equation}
V(s,t,u) \approx \frac{u (u +M_s^2)}{M_s^2 (s - 2 M_s^2)} \, .
\end{equation}
The associated scattering amplitudes and decay widths of the $n=2$
string resonances are discussed in Secs.~\ref{s4} and \ref{s5}. Roughly speaking,
the width of the Regge excitations will grow at least linearly with
energy, whereas the spacing between levels will decrease with
energy. This implies an upper limit on the domain of validity for our 
phenomenological approach~\cite{Anchordoqui:2009ja}. In particular,
for a resonance $R$ of mass $M$, the total width is given by
\begin{equation}
\Gamma_{\rm tot} \sim \frac{g^2}{4 \, \pi} \, {\cal C} \, \frac{M}{4},
\end{equation}
 where ${\cal C} > 1$ because of the growing multiplicity of decay modes~\cite{Anchordoqui:2008hi,Dong:2010jt}. On the other hand, since $\Delta (M^2) = M_s^2$ the level spacing at mass $M$ is $\Delta M \sim M_s^2/(2M)$; thus,
\begin{equation}
\frac{\Gamma_{\rm tot}}{\Delta M} \sim \frac{g^2}{8 \pi} \ {\cal C} \ \left(\frac{M}{M_s} \right)^2 = \frac{g^2}{8 \pi} \ {\cal C} \ n <1 \, .
\end{equation}
For excitation of the resonance $R$ via $a+b\rightarrow R$, the
assumption $\Gamma_{\rm tot}(R) \sim \Gamma(R\rightarrow ab)$ (which
underestimates the real width) yields a perturbative regime for $n
\lesssim 40$. This is to be compared with the $n \sim 10^4$ levels of
the string needed for black hole production.\footnote{The mass scale
  $M_{\rm BH} \sim M_s/g_s^2$, which corresponds to the onset of black
  hole production, follows from the \mbox{string $\leftrightharpoons$
    black} hole correspondence principle~\cite{Horowitz:1996nw}. For
  $g_s = 0.1$, we obtain $M_{\rm BH} \sim 100 M_s$.}

Before discussing the decay widths of the second massive-level
string states, we note that the Breit--Wigner form for gluon fusion into $\gamma$ + jet follows from (\ref{mhvlow3}) and is given by
\begin{equation}
\label{mhvlow2}
|{\cal M}(gg\to g\gamma)|^2\simeq
\frac{5g_3^4Q^2}{3M_s^4}\bigg[{M_s^8\over (s-M_s^2)^2+(\Gamma_{G^{(1)}}^{J=0} M_s)^2}+
{t^4+ u^4\over (s-M_s^2)^2+(\Gamma_{G^{(1)}}^{J=2} M_s)^2}\bigg] \,,
\end{equation}
and the dominant $s$-channel pole term of the average square
amplitude contributing to \mbox{$pp \to \gamma$ + jet}
reads 
\begin{equation}
|{\cal M}(qg \to q \gamma)|^2   =   -\frac{g_3^4 Q^2}{3M_s^2}\
\left[ \frac{M_s^4 \,  u}{( s-M_s^2)^2 + (\Gamma_{Q^{(1)}}^{J=\frac{1}{2}}\ M_s)^2}  \right.  + \left. \frac{u^3}{(s-M_s^2)^2 + (\Gamma_{Q^{(1)}}^{J=\frac{3}{2}}\ M_s)^2}\right] \, .
\label{qgqz}
\end{equation}

\section{Decay widths of the second massive-level string states}
\label{s4}
\subsection{Amplitudes and factorization}

The main goal of this section is to obtain the decay widths of the
second massive-level string states which will appear as resonances in
scattering processes $gg\to gg$, $gq\to gq$ and $gg\to q\bar{q}$ in
hadron colliders. In intersecting brane models, gluons $g$ are the
zeroth-level massless strings attaching to the $U(3)_{a}$ stack of
D-branes; left-handed quarks $q_{L}$ which participate in the weak
interactions are massless strings stretching between the $U(3)_{a}$ stack
and the $SU(2)$ stack [$U(2)$ or $Sp(1)$]; right-handed quarks $q_{R}$
could arise as either massless strings stretching between the $U(3)_{a}$
stack and another $U(1)$ stack, or massless strings attaching only to
the 
$U(3)_{a}$ stack and appearing as the antisymmetric representation of
$U(3)$.

Let us first clarify our notation on various string states in
different massive levels.  We follow the notations in
Refs.~\cite{Lust:2008qc,Lust:2009pz,Feng:2010yx,Feng:2011qc,Feng:2012bb},
and we will focus on the string states which contribute to $gg\to gg$
and $gq\to gq$ processes.  The bosonic sector of the first
massive-level consists of two universal string states: a spin-2 field
$\alpha$ and a complex scalar $\Phi$.  In addition, there is a spin-1
field $d$ for which the vertex operator involves the internal current
$\mathcal{J}$.  This vector $d$ can decay into $q \bar q$, which is a
universal property of all $\mathcal{N}=1$
compactifications~\cite{Feng:2010yx}.  As the $U(3)$ generators
decompose to the $SU(3)$ color generators plus the $U(1)$ generator
(color singlet), we have two copies of the string excitations. We will
denote the color octets by $G^{(n)}$ and the color singlets by
$C^{(n)}$, where $n$ indicates the $n$th massive level. For the
fermionic sector, the excited quark triplets $Q^{(1)}$ consists of one
spin-$\frac{3}{2}$ field $\chi$ and one spin-$\frac{1}{2}$ field $a$
(and also their opposite chirality fields $\bar \chi,\bar a$).  For
the bosonic sector of the second massive level ($G^{(2)}, C^{(2)}$),
four universal states has been determined~\cite{Feng:2011qc}: a spin-3
field $\sigma$, a spin-2 field $\pi$, and two complex vector fields
$\Xi_{1,2}$.

The total decay width of a second massive-level bosonic string state
$G^{(2)}$ consists of four contributions: $G^{(2)}$ decays into two
massless string states ($G^{(2)}\to gg$ and $G^{(2)}\to q\bar{q}$),
$G^{(2)}$ decays into one first massive-level string state plus one
massless string state ($G^{(2)}\to G^{(1)}g$ and $G^{(2)}\to
Q^{(1)}q$), $G^{(2)}$ decays into a color singlet [anomalous $U(1)$'s]
plus a massless gluon or an excited gluon ($G^{(2)}\to g A_{a}$ and
$G^{(2)}\to G^{(1)} A_{a}$), and $G^{(2)}$ decays into the excitation of
the color singlet $C^{(1)}$ plus one massless gluon. For a second
massive-level color singlet string state $C^{(2)}$, its decay width
also involves four contributions: $C^{(2)}$ decays into two massless
string states ($C^{(2)}\to gg$ and $C^{(2)}\to q\bar{q}$), $C^{(2)}$
decays into one first massive-level string state plus one massless
string state ($C^{(2)}\to G^{(1)}g$ and $C^{(2)}\to Q^{(1)}q$),
$C^{(2)}$ decays into two anomalous $U(1)$'s, and $C^{(2)}$ decays into
the excitation of the color singlet $C^{(1)}$ plus one anomalous
$U(1)$. For a second massive-level excited quark $Q^{(2)}$, its total
decay width could consist of five contributions: $Q^{(2)}$ decays into
one massless gluon plus one massless quark ($Q^{(2)}\to gq$),
$Q^{(2)}$ decays into one first massive-level string state and one
massless string state ($Q^{(2)}\to G^{(1)}q$ and $Q^{(2)}\to
Q^{(1)}g$), $Q^{(2)}$ decays into anomalous $U(1)$'s plus a massless
quark or an excited quark ($Q^{(2)}\to qA_{a}$ and $Q^{(2)}\to
Q^{(1)}A_{a}$), $Q^{(2)}$ decays into the excitation of the color
singlet $C^{(1)}$ plus one quark, and finally, for $Q^{(2)}$ which
participates in weak interactions, it could also decay into $SU(2)$
gauge bosons plus one quark.  All above decay channels of the second
massive-level string states are summarized in
Table~\ref{decaychannels}.  Most of these decay channels are universal
to all compactifications, while there are also several model-dependent
channels.  We will comment on them in Secs.~\ref{SU2}, \ref{AnU1}
and \ref{LHRH}.

\begin{table}
\begin{center}
\begin{tabular}{c|c|c|c|c}
\hline
\hline
 & 2 massless & 1 first-level string state  & involve 1 or 2  & involve 1 first-level\tabularnewline
 &  string states & plus 1 massless string state & color singlet(s) & color singlet excitation\tabularnewline
\hline
$G^{(2)}$ & $gg,q\bar{q}$ & $G^{(1)}g,Q^{(1)}\bar{q},\bar{Q}^{(1)}q$ & $gA_{a},G^{(1)}A_{a}$ & $C^{(1)}g$\tabularnewline
\hline
$C^{(2)}$ & $gg,q\bar{q}$ & $G^{(1)}g,Q^{(1)}\bar{q},\bar{Q}^{(1)}q$ & $A_{a}A_{a}$ & $C^{(1)}A_{a}$\tabularnewline
\hline
$Q^{(2)}$ & $gq$ & $G^{(1)}q,Q^{(1)}g$ & $qA_{a},Q^{(1)}A_{a}$ & $C^{(1)}q$\tabularnewline
\hline
\hline
\end{tabular}
\end{center}
\caption{Possible decay channels for the second massive-level string states $G^{(2)}, C^{(2)}, Q^{(2)}$. Excited massive quarks which participate in weak interactions can also decay into $SU(2)$ gauge bosons plus another quark.}
\label{decaychannels}
\end{table}

The partial decay widths of $G^{(2)}$ and $Q^{(2)}$ decaying into two
massless string states were already obtained in
Ref.~\cite{Dong:2010jt,Hashi:2011cz} by using factorization. However,
we realize that there are some mistakes in those results. The widths
of $G^{(2)}$ decaying into $gg$ in Ref.~\cite{Dong:2010jt} should be
reduced by one-half.  Moreover, there are in fact two distinct
$Q^{(2)}(J=3/2)$ states. They can decay into $gq$ of helicities $(+1,
+1/2)$ and $(-1, +1/2)$, respectively, and do not mix with each
other. So we need to consider their widths separately (instead of
adding them up as in Ref.~\cite{Hashi:2011cz}).  In this section, we will
obtain the partial decay widths of $G^{(2)},\, C^{(2)}$ and $Q^{(2)}$
decaying into one first massive-level string state ($G^{(1)}, \, C^{(1)}$
or $Q^{(1)}$) plus one massless string state ($g$ or $q$) using
four-point amplitudes with one leg being the first massive-level
string state obtained in Ref.~\cite{Feng:2010yx}. We will comment on other
decay channels at the end of this section.

We have seen in Sec.~\ref{s3} that four-point amplitudes
$\mathcal{A}(g,g,g,g)$ and $\mathcal{A}(g,g,q,\bar{q})$ carry the form
factor $V(s,t,u)$ which can be expanded in terms of $s$-channel
resonances.  Recasting the expansion we can reexpress the amplitudes
as sums of Wigner d matrices, and one could then obtain two
three-point amplitudes of massive string states decaying into
different final states with specific spin
combinations~\cite{Anchordoqui:2008hi}.  Using this method, one could
identify the contributions of various string states with different
spins appearing as resonances in the $s$-channel pole at a certain
massive level.  Previous works only deal with the four-point amplitude
with four massless string states, whereas in this work we consider the
factorization of four-point amplitudes, one of which has massive
external legs. More specifically, we consider four-point amplitudes
$\mathcal{A}(G^{(1)},g,g,g)$, $\mathcal{A}(G^{(1)},g,q,\bar{q})$ and
$\mathcal{A}(Q^{(1)},g,g,\bar{q})$ which were computed in
Ref.~\cite{Feng:2010yx}.  By factorizing these amplitudes and using
the known results (amplitudes that $G^{(2)},Q^{(2)}$ decaying into two
massless string states), we could obtain the partial decay widths of
one second massive-level string state decaying into a first
massive-level string state plus a massless one.

\begin{figure}[t!]
\begin{center}
\includegraphics[scale=0.55]{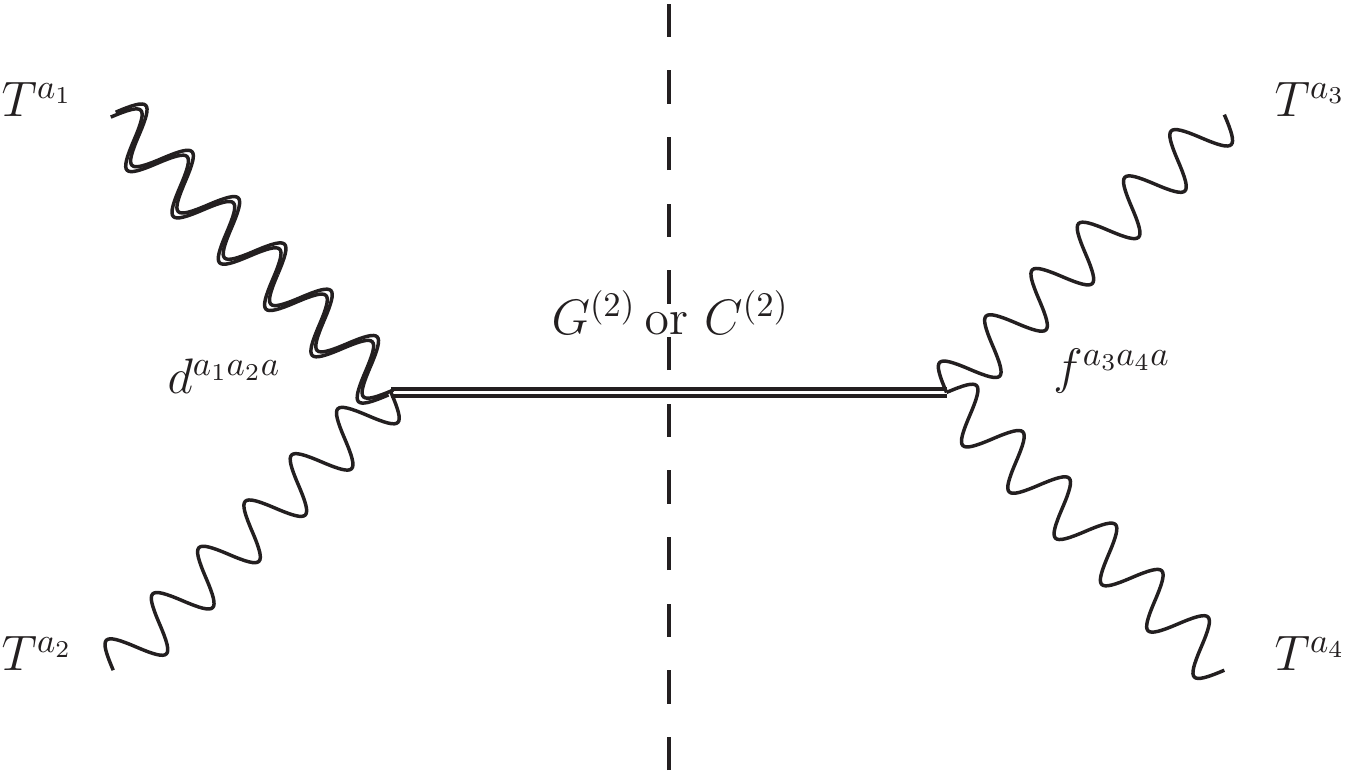}
 \caption{Factorization of the amplitude $\mathcal{A}(G^{(1)},g,g,g)$ gives different group factors on two sides.
 The doubled wavy line presents the first massive-level bosonic string state, whereas the single lines present massless bosonic string states.
 $G^{(2)}$ or $C^{(2)}$ are the second massive-level intermediate string states obtained from factorization.}
\label{factor}
\end{center}
\end{figure}

For the four bosonic string states scattering,
there is one subtlety which is the decomposition of the group factors.
The structure constant of the gauge group $f^{a_{1}a_{2}a_{3}}$ or
the total symmetric trace $d^{a_{1}a_{2}a_{3}}$ would arise when
we combine the three-point amplitudes of two different orderings $(1,2,3)$
and $(1,3,2)$ on the world sheet.
This depends on the overall world sheet parity $(-1)^{N+1}$
where $N$ is the sum of the overall massive-level number of the three scattering
string states. More specifically, the combined amplitudes have
the following group factors
\begin{align*}
{\rm Tr}(T^{a_1}[T^{a_2},T^{a_3}]) & =\tfrac{i}{2}f^{a_1 a_2 a_3}\,,\qquad N\ {\rm even}\,;\\
{\rm Tr}(T^{a_1}\{T^{a_2},T^{a_3}\}) & =2d^{a_1 a_2 a_3}\,,\qquad N\ {\rm odd}\,.
\end{align*}
When factorizing a four-point amplitude with one first massive-level
leg, on one side one gets a second massive-level string state decaying
into a first massive string state plus a zeroth-level mode, and on the
other side one gets the same second massive-level string state
decaying into two zeroth-level massless string states. Thus one would
get a group factor of $d^{a_1a_2a}$ on the left and $f^{a_3a_4a}$ on
the right; see Fig.~\ref{factor}.  Factorizing amplitudes involving
two fermions is simpler since there are only two Chan--Paton factors
involved.  Our notation on these group factors is summarized in
Appendix~\ref{App1}.

In this section all the four-point amplitudes with one first
massive-level string state are taken from Ref.~\cite{Feng:2010yx}.  In
Ref.~\cite{Feng:2010yx}, the massive string state was placed at
position 4, and the three massless ones took the positions 1, 2, and
3.  For our convenience, in this work we prefer to place the massive
string state at position 1, while the three massless string states
were placed at 2, 3, and 4.  The corresponding amplitudes can be
easily obtained by performing permutations of the original amplitudes.

The helicity wave function of a massive higher spin particle is
specified by a pair of lightlike vectors $p^\mu,q^\mu$, which is a
decomposition of the momentum of the particle $k^\mu =
p^\mu+q^\mu$.\footnote{ We will give a brief review of the massive
  helicity formalism in the next section.  Helicity formalism for
  massless fields as well as massive fermion fields is briefly
  reviewed in Appendixes \ref{AppB} and \ref{AppC}.}  The spin
quantization axis is along the direction of $\vec{q}$ in the rest
frame; here, it is most convenient to set $q^\mu = k_2^\mu$, so that
the spin axis of the first massive-level string state (at position 1)
is along the same direction as the spin axis of the massless string
state at position 2, and we denote this direction to be $+\vec{z}$.
Because of angular momentum conservation, the spin axis of the
intermediate second massive-level string state (see Fig.~\ref{factor})
should also align to $+\vec{z}$, and the corresponding helicity
amplitudes of these three states with only specific $j_z$ combinations
can survive.  The reference momenta of particle 1 are chosen to be
\begin{equation}
\label{pqmomentum}
p^\mu = \left(\frac {\sqrt s} 2, 0,0,-\frac {\sqrt s} 2\right)\,,\quad
q^\mu = k_2^\mu= \left(\frac {M_s^2}{2\sqrt s}, 0,0,\frac {M_s^2}{2\sqrt s}\right) \,.
\end{equation}
The spinor products become
\begin{equation}
\br p 2 \ke [2 p] = s/2\,,\quad \br p 3 \ke [3 p] = 2 t\,,\quad \br p 4 \ke [4 p] = 2 u\,,
\end{equation}
where $s,t,u$ are Mandelstam variables.  With this choice, we could
extract the helicity amplitudes of the second massive-level strings
decaying into a first massive-level string plus a massless one with
their spin axes all along $+\vec{z}$ (the direction of the momentum of
the massless string state), from the four-point amplitudes in
Ref.~\cite{Feng:2010yx}.  In the next section, we will focus on the
spin-3 and spin-2 universal string states from the second massive
level, computing their scattering amplitudes and their partial decay
widths, where we will also align the spins of the three interacting
states in the direction of the momentum of the massless particle.
Thus, we are expecting the helicity amplitudes we obtained from
factorization in this section to match exactly with the string
amplitudes from CFT computations in the next section.

We will discuss the factorization of the four-point amplitudes in the
following order.  We start from the amplitudes which involve the first
massive-level spin-2 field $\alpha$ and obtain the decay widths of
second massive-level string states decaying into $\alpha$ plus another
massless string state.  Then we discuss the decays which involve the
final states $d,\Phi,\chi,a$ in order, which are obtained from the
four-point amplitudes with $d,\Phi,\chi,a$ plus three other massless
string states.  The full results of decay widths for $n=2$ resonances
are summarized in Table~\ref{decayratetable1} at the end of this
section.

\subsection{$\alpha(J=2)$}

The highest spin field from the first massive level is the spin-2 boson $\alpha$
with its vertex operator given in Eq.~\eqref{Valpha}.
We will need to use the  amplitudes (all particles are incoming)~\cite{Feng:2010yx}
\ba
\mathcal{A}[\alpha_1,\epsilon_2,\epsilon_3,\epsilon_4] & = &
8\, g_3^2 \big( \, V_t \, t^{a_1a_2a_3a_4}  +  V_s \,
t^{a_2a_3a_1a_4} +  V_u \, t^{a_3a_1a_2a_4} \,\big)
\sqrt{2\alpha'}\ \mathscr{A}[\alpha_1,\epsilon_2,\epsilon_3,\epsilon_4], \quad\label{b3g} \\
\mathcal{A}[\alpha_1,u_2,\bar u_3,\epsilon_4] & = & 2\,g_3^2\ \big[V_t(T^{a_4}T^{a_1})^{\alpha_2}_{\alpha_3}+
 V_s(T^{a_1}T^{a_4})^{\alpha_2}_{\alpha_3}\big]\ \sqrt{2\alpha'}\ \mathscr{A} [\alpha_1,u_2,\bar u_3,\epsilon_4]\,, \label{bqq1}
\ea
where $\epsilon$ denotes the polarization vector of a gluon $g$, and
\begin{align}
\mathscr{A}\left[\alpha(+2),+,+,-\right] &=\frac{1}{2\sqrt{2}}\ \!\frac{\langle
p4\rangle^{4}}{\langle23\rangle\langle34\rangle
\langle42\rangle}\,,\nonumber\\
\mathscr{A}\left[\alpha(+1),+,+,-\right] &=\frac{1}{\sqrt{2}}\ \,\frac{\langle
p4\rangle^{3}\langle4 q\rangle}{\langle23\rangle\langle34\rangle
\langle42\rangle}\,,\nonumber\\
\mathscr{A}\left[\alpha(~0~),+,+,-\right] &=\frac{\sqrt{3}}{2}\ \,\frac{\langle
p4\rangle^{2}\langle4q\rangle^{2}}{\langle23\rangle\langle34\rangle
\langle42\rangle}\,,\label{b3gh}\\
\mathscr{A}\left[\alpha(-1),+,+,-\right] &=\frac{1}{\sqrt{2}}\ \,\frac{\langle
q4\rangle^{3}\langle4p\rangle}{\langle23\rangle\langle34\rangle
\langle42\rangle}\,,\nonumber\\
\mathscr{A}\left[\alpha(-2),+,+,-\right] &=\frac{1}{2\sqrt{2}}\ \!\frac{\langle
q4\rangle^{4}}{\langle23\rangle\langle34\rangle
\langle42\rangle}\,,\nonumber
\end{align}
and
\begin{align}
\mathscr{A}\left[\alpha(+2),\hpl,\hmi,+\right] &=\frac{1}{\sqrt{2}}\ \frac{\langle p2\rangle\langle
p3\rangle^3}{\langle23\rangle\langle34\rangle \langle42\rangle}\,,\nonumber\\
\mathscr{A}\left[\alpha(+1),\hpl,\hmi,+\right] &= \frac{1}{2\sqrt{2}}\,
\frac{\langle p3\rangle^2}{\langle23\rangle\langle34\rangle \langle42\rangle} \big(\langle q2\rangle\langle p3\rangle +3\langle p2\rangle\langle
q3\rangle\big)\,,\nonumber\\
\mathscr{A}\left[\alpha(~0~),\hpl,\hmi,+\right] & = \frac{\sqrt{3}}{2}\
\frac{\langle p3\rangle\langle q3\rangle}{\langle23\rangle\langle34\rangle \langle42\rangle}\ \big(\langle q2\rangle\langle p3\rangle +\langle
p2\rangle\langle q3\rangle\big)\,,\label{bqq2}\\
\mathscr{A}\left[\alpha(-1),\hpl,\hmi,+\right] & =
\frac{1}{2\sqrt{2}}\,\frac{\langle q3\rangle^2}{\langle23\rangle\langle34\rangle \langle42\rangle}\big(3\langle q2\rangle\langle p3\rangle +\langle
p2\rangle\langle q3\rangle\big) \,, \nonumber\\
\mathscr{A}\left[\alpha(-2),\hpl,\hmi,+\right] & =  \frac{1}{\sqrt{2}}\
\frac{\langle q2\rangle\langle q3\rangle^3}{\langle23\rangle\langle34\rangle \langle42\rangle}\,.\nonumber
\end{align}
The other nonvanishing amplitudes can be obtained by taking the
complex conjugate and permutation.

\subsubsection{$G^{(2)}(J=3,2)\to \alpha+g$}

We now factorize the four-point amplitudes $\mathcal{A}[\alpha,
+,+,-]$ to get the matrix elements of $G^{(2)}(J=2,3)$ decaying into
$\alpha+g^+$.  Amplitudes $\mathcal{A}[\alpha, -,-,+]$, can be
obtained via the complex conjugate, and they give the matrix elements of
the decays $G^{(2)}(J=3,2) \to \alpha+g^-$.  The factorization of
$\mathcal{A}[\alpha(+2), +,+,-]$ gives \be \el{alpggg}
\mathcal{A}[\alpha(+2), +,+,-] = \frac{g_3^2 M_s^2}{s-2M_s^2}
\frac{16}{\sqrt 3}d^3_{-3,-2}(\theta)\, f^{a_1a_2a}d^{a_3a_4a}\,, \ee
where $\theta$ is the angle between $-\vec{z}$ and the spatial
momentum of particle $3$. It is related to the Mandelstam variables
$u,t$ by
\begin{eqnarray} \label{tu}
u=-\frac{s}{2}(1+\cos \theta )\,,~~~ t=-\frac{s}{2}(1-\cos \theta )\,.
\end{eqnarray}
From \er{alpggg} we can read off the matrix elements as,
\begin{eqnarray}
F^{a, J=3}_{+2+a_1a_2}=F^{a, J=3}_{-2- a_1a_2}=8 g_3M_s d^{a_1a_2a}\,,
\label{Spin3toalphag1}
\end{eqnarray}
where we use $F^{a, J}_{\lambda_1 \lambda_2 a_1a_2}$ to denote the
amplitude of a spin-$J$ particle with angular momentum $j_z =
\lambda_1 + \lambda_2$ (and gauge index $a$) decaying into particles
1 and 2 with momenta along the $\vec{z}$ axis. $\lambda_1, \lambda_2$
are helicities of the two particles while $a_1, a_2$ are gauge
indices.  Thus, the result of Eq.~\eqref{Spin3toalphag1} presents the
decay of a second massive-level spin-3 string state with $j_z =
-3$ decaying into $\alpha_1(j_z=-2)$ and $\epsilon_2^{-}$, which is
exactly what we get in Eq.~\eqref{3to21.1} in the next section. In
Eq.~\eqref{3to21.1}, all particles are incoming and the corresponding
outgoing particles are one $\xa(-2)$ and one $\xe^-$. We would like to
remind the reader that the definition of $F^{a, J}_{\lambda_1
  \lambda_2 a_1a_2}$ is in some sense different from what is used in
the literature \cite{Dong:2010jt, Hashi:2011cz,
  Anchordoqui:2008hi}. Previously the helicity $\xl_1$ (of a massless
particle) was usually defined with its spin axis along $\vec k_1$. In
our convention the spin axis of every particle is along
$+\vec{z}$. Particle $1$ is moving along $-\vec{z}$ and its spin axis
is opposite to $\vec k_1$.

Similarly, we can do the factorization for amplitudes with other spin configurations:
\begin{gather}
\mathcal{A}[\alpha(+1), +,+,-]
=  \frac{g_3^2 M_s^2}{s-2M_s^2} \left( \frac{16}{3}d^3_{-2,-2}(\theta)-\frac{16}{3}d^2_{-2,-2}(\theta) \right)  f^{a_1a_2a}d^{a_3a_4a}\,,\\
F^{a, J=3}_{+1+a_1a_2}=F^{a, J=3}_{-1- a_1a_2}=\frac 8 {\sqrt 3} g_3M_s d^{a_1a_2a}\,,\quad F^{a, J=2}_{+1+a_1a_2}=F^{a, J=2}_{-1- a_1a_2}=4 \sqrt{\frac 2 3} g_3M_s d^{a_1a_2a}\,. \label{Spin3toalphag2}\\
\mathcal{A}[\alpha(0), +,+,-]
=  \frac{g_3^2 M_s^2}{s-2M_s^2} \left( 8\sqrt{\frac{2}{15}}d^3_{-1,-2}(\theta)-\frac{8}{\sqrt 3}d^2_{-1,-2}(\theta) \right)  f^{a_1a_2a}d^{a_3a_4a}\,,\\
F^{a, J=3}_{0+a_1a_2}=F^{a, J=3}_{0- a_1a_2}=4\sqrt{\frac{2}{5}} g_3M_s d^{a_1a_2a}\,,\quad F^{a, J=2}_{0+a_1a_2}=F^{a, J=2}_{0- a_1a_2}=2 \sqrt{2} g_3M_s d^{a_1a_2a}\,.\\
\mathcal{A}[\alpha(-1), +,+,-]
=  \frac{g_3^2 M_s^2}{s-2M_s^2} \left( 4\sqrt{\frac{2}{15}}d^3_{0,-2}(\theta)-4\sqrt{\frac{2} 3}d^2_{0,-2}(\theta) \right)  f^{a_1a_2a}d^{a_3a_4a}\,,\\
F^{a, J=3}_{-1+a_1a_2}=F^{a, J=3}_{+1- a_1a_2}=2\sqrt{\frac{2}{5}} g_3M_s d^{a_1a_2a}\,,\quad F^{a, J=2}_{-1+a_1a_2}=F^{a, J=2}_{+1- a_1a_2}=2  g_3M_s d^{a_1a_2a}\,.\\
\mathcal{A}[\alpha(-2), +,+,-]
=  \frac{g_3^2 M_s^2}{s-2M_s^2} \left( \frac{4}{3\sqrt 5}d^3_{+1,-2}(\theta)-\frac{4 \sqrt 2} 3 d^2_{+1,-2}(\theta) \right)  f^{a_1a_2a}d^{a_3a_4a}\,,\\
F^{a, J=3}_{-2+a_1a_2}=F^{a, J=3}_{+2- a_1a_2}=\frac{2}{\sqrt{15}} g_3M_s d^{a_1a_2a}\,,\quad F^{a, J=2}_{-2+a_1a_2}=F^{a, J=2}_{+2- a_1a_2}=\frac 2 {\sqrt 3} g_3M_s d^{a_1a_2a}\,.
\label{Spin3toalphag5}
\end{gather}
The decay width can be computed using (an extra factor of $1/2$ is needed if
outgoing particles are a pair of gluons) \footnote{Since the decay product includes a massive particle, the decay width is suppressed by $M_s^2/s$ compared to the width of decaying into two massless particles. The suppression is due to the difference in $|\vec k_1|/\sqrt s$, which appears in phase space integration of the final states. In the case of two outgoing massless particles this ratio is $\tfrac 1 2$ while in the current case, it is $\tfrac {M_s^2} {2s}$ [see, \textit{e.g.}, Eq.\eqref{pqmomentum}].}
\begin{eqnarray}
\Gamma^{aJ}_{\lambda_1 \lambda_2, a_1a_2} = \frac{1}{32(2J+1)\sqrt{2} \pi M_s} |F^{aJ}_{\lambda_1\lambda_2, a_1a_2}|^2\,.
\end{eqnarray}
We need to take into account both the channels into $\alpha+g^+$ and into $\alpha+g^-$ and the results are
\be
\Gamma^{J=3}_{G^{(2)}\rightarrow \alpha g} = \frac{117 g_3^2 M_s }{2240 \sqrt{2} \pi } N\,, \qquad
\Gamma^{J=2}_{G^{(2)}\rightarrow \alpha g} = \frac{3 g_3^2 M_s}{160 \sqrt{2} \pi }N\,.
\ee

\subsubsection{$G^{(2)}(J=1)\to \alpha + g$}
The spin-1 resonances arise from factorization of the 
amplitude $\mathcal{A}[\alpha, -, +,+]$,
\be
\mathcal{A}[\alpha(+2), -, +,+]
=  \frac{4g_3^2 M_s^2}{s-2M_s^2} d^1_{-1,0}(\theta)  f^{a_1a_2a}d^{a_3a_4a}\,,
\ee
and we obtain
\begin{eqnarray}
\el{FJ1J2g}
F^{a, J=1}_{+2-a_1a_2}=F^{a, J=1}_{-2+ a_1a_2}=2 g_3M_s d^{a_1a_2a}\,,
\end{eqnarray}
which corresponds to the complex vectors found in Ref.~\cite{Feng:2011qc}.
Unlike $G^{(2)}(J=3,2)$, $G^{(2)}(J=1)$ is not parity invariant, the matrix
elements in \er{FJ1J2g} are for two different particles and should not be added together.
Thus the corresponding partial decay width reads
\be
\Gamma^{J=1}_{G^{(2)}\rightarrow \alpha g} = \frac{g_3^2 M_s}{384 \sqrt{2} \pi }N\,.
\ee

\subsubsection{$Q^{(2)}(J=5/2, 3/2)\to \alpha+q$}
We could obtain the second massive-level spin-$\frac 5 2$ and spin-$\frac 3 2$ resonances from
factorizing amplitude $\mathcal{A}[\alpha, +\tfrac 1 2,-\tfrac 1 2,+]$:
\begin{gather}
\mathcal{A}\left[\alpha(+2), +\tfrac 1 2,-\tfrac 1 2,+\right]
=  \frac{g_3^2 M_s^2}{s-2M_s^2}  \frac{4}{\sqrt 5}d^{5/2}_{-5/2,+3/2}(\theta)  T^{a_1}_{\alpha_2 \alpha}T^{a_4}_{\alpha \alpha_3}\,,\\
F^{\xa, J=5/2}_{+2+\frac 1 2 a_1\xa_2}=F^{\xa, J=5/2}_{-2-\frac 1 2 a_1\xa_2}=\sqrt 2 g_3M_s T^{a_1}_{\alpha_2 \alpha}\,.\\
\mathcal{A}\left[\alpha(+1), +\tfrac 1 2,-\tfrac 1 2,+\right]
=  \frac{g_3^2 M_s^2}{s-2M_s^2} \left( \frac{3\sqrt 2}{5}d^{5/2}_{-3/2,+3/2}(\theta)-\frac{3 \sqrt 2} 5 d^{3/2}_{-3/2,+3/2}(\theta) \right)  T^{a_1}_{\alpha_2 \alpha}T^{a_4}_{\alpha \alpha_3}\,,\\
F^{\xa, J=5/2}_{+1+\frac 1 2 a_1\xa_2}=F^{\xa, J=5/2}_{-1-\frac 1 2 a_1\xa_2}=\frac{3}{2\sqrt{5}} g_3M_s T^{a_1}_{\alpha_2 \alpha}\,,\quad
F^{\xa, J=3/2}_{+1+\frac 1 2 a_1\xa_2}=F^{\xa, J=3/2}_{-1-\frac 1 2 a_1\xa_2}=\sqrt{\frac 3 { 10}} g_3M_s T^{a_1}_{\alpha_2 \alpha}\,.\\
\mathcal{A}\left[\alpha(0), +\tfrac 1 2,-\tfrac 1 2,+\right]
=  \frac{g_3^2 M_s^2}{s-2M_s^2} \left( -\frac{\sqrt 3}{5}d^{5/2}_{-1/2,+3/2}(\theta)-\frac{2 \sqrt 2} 5 d^{3/2}_{-1/2,+3/2}(\theta) \right)  T^{a_1}_{\alpha_2 \alpha}T^{a_4}_{\alpha \alpha_3}\,,\\
F^{\xa, J=5/2}_{0+\frac 1 2 a_1\xa_2}=F^{\xa, J=5/2}_{0-\frac 1 2 a_1\xa_2}=\frac{1}{2}\sqrt{\frac{3}{10}} g_3M_s T^{a_1}_{\alpha_2 \alpha}\,,\quad
F^{\xa, J=3/2}_{0+\frac 1 2 a_1\xa_2}=F^{\xa, J=3/2}_{0-\frac 1 2 a_1\xa_2}=\sqrt{\frac 2 { 15}} g_3M_s T^{a_1}_{\alpha_2 \alpha}\,.\\
\mathcal{A}\left[\alpha(-1), +\tfrac 1 2,-\tfrac 1 2,+\right]
=  \frac{g_3^2 M_s^2}{s-2M_s^2} \left(\frac{1}{10}d^{5/2}_{+1/2,+3/2}(\theta)-\frac 1 5\sqrt{\frac{3} 2} d^{3/2}_{+1/2,+3/2}(\theta) \right)  T^{a_1}_{\alpha_2 \alpha}T^{a_4}_{\alpha \alpha_3}\,,\\
F^{\xa, J=5/2}_{-1+\frac 1 2 a_1\xa_2}=F^{\xa, J=5/2}_{+1-\frac 1 2 a_1\xa_2}=\frac{1}{4 \sqrt{10}} g_3M_s T^{a_1}_{\alpha_2 \alpha}\,,\quad
F^{\xa, J=3/2}_{-1+\frac 1 2 a_1\xa_2}=F^{\xa, J=3/2}_{+1-\frac 1 2 a_1\xa_2}=\frac{1}{2 \sqrt{10}} g_3M_s T^{a_1}_{\alpha_2 \alpha}\,.
\end{gather}
Left-handed and right-handed fermions are stretching between different
branes.  As a result, left-handed excited quarks cannot decay into
right-handed quarks plus gluons.  For example, we have $F^{\xa,
  J=5/2}_{+2+\frac 1 2 a_1\alpha_2}=F^{\xa, J=5/2}_{-2-\frac 1 2
  a_1\alpha_2}$ but they are decay amplitudes for left- and
right-handed excited quarks and should not be combined. The
corresponding decay widths are \be \Gamma^{J=5/2}_{Q^{(2)}\rightarrow
  \alpha q} = \frac{27 g_3^2 M_s}{4096 \sqrt{2} \pi }N\,,\qquad
\Gamma^{J=3/2}_{Q^{(2)}\rightarrow \alpha q} = \frac{11 g_3^2
  M_s}{6144 \sqrt{2} \pi }N\,.  \ee

\subsubsection{$Q^{(2)}(J=3/2, 1/2)\to \alpha+q$}

The second massive-level spin-$\frac 3 2$ and spin-$\frac 1 2$ resonances
can be obtained from amplitude $\mathcal{A}[\alpha, -\frac 1 2,+, +\frac 1 2]$:
\begin{gather}
\mathcal{A}\left[\alpha(+2),-\tfrac 1 2,+, +\tfrac 1 2\right]
=  \frac{g_3^2 M_s^2}{s-2M_s^2}  \frac{2 }{\sqrt{3}} d^{3/2}_{-3/2,-1/2}(\theta)  T^{a_1}_{\alpha_2 \alpha}T^{a_3}_{\alpha \alpha_4}\,,\\
F^{\xa, J=3/2}_{+2-\frac 1 2 a_1\xa_2}=F^{\xa, J=3/2}_{-2+\frac 1 2 a_1\xa_2}=\frac 1 {\sqrt 2} g_3M_s T^{a_1}_{\alpha_2 \alpha}\,.\\
\mathcal{A}\left[\alpha(+1),-\tfrac 1 2,+, +\tfrac 1 2\right]
=  \frac{g_3^2 M_s^2}{s-2M_s^2} \left(\frac{1}{3\sqrt2}d^{3/2}_{-1/2,-1/2}(\theta)-\frac{1}{3\sqrt2} d^{1/2}_{-1/2,-1/2}(\theta) \right)  T^{a_1}_{\alpha_2 \alpha}T^{a_3}_{\alpha \alpha_4}\,,\\
F^{\xa, J=3/2}_{+1-\frac 1 2 a_1\xa_2}=F^{\xa, J=3/2}_{-1+\frac 1 2 a_1\xa_2}=\frac{1}{4 \sqrt{3}} g_3M_s T^{a_1}_{\alpha_2 \alpha}\,,\quad
F^{\xa, J=1/2}_{+1 -\frac 1 2 a_1\xa_2}=F^{\xa, J=1/2}_{-1+\frac 1 2 a_1\xa_2}=\frac{1}{2 \sqrt{6}} g_3M_s T^{a_1}_{\alpha_2 \alpha}\,.
\end{gather}
The spin-$\frac 3 2$ fermion $\tilde Q^{(2)}$ here is different from
the spin-$\frac 3 2$ fermion $Q^{(2)}$ we obtained from the amplitude $\mathcal{A}[\alpha, +\tfrac 1 2,-\tfrac 1 2,+]$,
as this one can decay into $(+ ,+\frac 1 2)$ (instead of $(- ,+\frac 1 2)$).
Since the amplitude $\mathcal{A}[+, +, +\frac{1}{2}, -\frac{1}{2}] =0$, these two states do not mix, and we obtain
\be
\Gamma^{J=3/2}_{\tilde Q^{(2)}\rightarrow \alpha q} = \frac{25 g_3^2 M_s}{12288 \sqrt{2} \pi }N\,, \qquad
\Gamma^{J=1/2}_{Q^{(2)}\rightarrow \alpha q} = \frac{g_3^2 M_s}{3072 \sqrt{2} \pi }N\,.
\ee

\subsection{$d(J=1)$}

The spin-1 field $d$ is different from the universal bosonic fields $\alpha,\Phi$
in that it is tied to spacetime SUSY.
Although its vertex operator contains the world sheet current $\mathcal{J}$,
the vector $d$ does give rise to universal amplitudes into a quark-antiquark pair~\cite{Feng:2010yx}.
The existence of this vector resonance is a universal property
of all $\mathcal{N}=1$ SUSY compactifications.
We will need the amplitude $\mathcal{A}[d_1,u_2,\bar u_3,\epsilon_4]$, which reads
\begin{equation}
\mathcal{A}[d_1,u_2,\bar u_3,\epsilon_4]=\sqrt{3}\,g_3^2\ \big[V_t(T^{a_4}T^{a_1})^{\alpha_2}_{\alpha_3}+
 V_s(T^{a_1}T^{a_4})^{\alpha_2}_{\alpha_3}\big]\ \mathscr{A} \left[d_1,u_2,\bar u_3,\epsilon_4\right]\,,
\end{equation}
where
\begin{align}
\mathscr{A}\left[d(+1),\hpl,\hmi,+\right] &=
\frac{\langle p3\rangle^2}{\langle24\rangle\langle34\rangle}\,,
\nonumber\\[1mm]
\mathscr{A}\left[d(~0~),\hpl,\hmi,+\right] & = \sqrt{2}\,
\frac{\langle p3\rangle\langle q3\rangle}{\langle24\rangle\langle34\rangle}\,,\\
\mathscr{A}\left[d(-1),\hpl,\hmi,+\right] & =
-\frac{\langle q3\rangle^2}{\langle24\rangle\langle34\rangle}\,. \nonumber
\end{align}
These amplitudes will give rise to two channels of the second
massive-level string resonances.\footnote{ Indeed, by factorizing
  $\mathcal{A}[d,+,-\frac 1 2, +\frac 1 2]$ amplitudes, one can get
  the second massive-level $J=2,1$ resonances where the states can
  decay into $d+g$.  These states are not the same as the
  $G^{(2)}(J=2,1)$ we have discussed above.  For $\mathcal{N}=1$
  compactification, the vertex operator of this vector $d$ involves
  internal current $\mathcal J$~\cite{Feng:2010yx}. It only couples to
  quark-antiquark pairs,  while the $G^{(2)}(J=2,1)$ states, for which
  vertex operators, cf. Ref.~\cite{Feng:2011qc}, cannot decay into $d+g$.
  Thus the vertex operators of $J=2,1$ resonances which arise from
  this channel must also contain internal components.  These $J=2,1$
  states do not couple to a pair of gluons and thus play no role in
  processes $gg \to gg$ or $gg \to q \bar q$.  Even though these
  states do couple to quark-antiquark pairs and may contribute to
  four-fermion amplitudes, we will not consider such processes as they
  are suppressed~\cite{Anchordoqui:2008di}.  Thus we will not discuss
  these states in this work.}

\subsubsection{$Q^{(2)}(J=5/2, 3/2)\to d+q$}

We could obtain the second massive-level spin-$\frac 5 2$ and spin-$\frac 3 2$ resonances from
factorizing amplitude $\mathcal{A}[d, +\frac 1 2,-\frac 1 2,+]$:
\begin{gather}
\mathcal{A}[d(+1), +\tfrac 1 2,-\tfrac 1 2,+]
=  \frac{g_3^2 M_s^2}{s-2M_s^2} \left( \frac{\sqrt 6}{5}d^{5/2}_{-3/2,+3/2}(\theta)+\frac{\sqrt 6} 5 d^{3/2}_{-3/2,+3/2}(\theta) \right)  T^{a_1}_{\alpha_2 \alpha}T^{a_4}_{\alpha \alpha_3}\,,\\
F^{\xa, J=5/2}_{+1+\frac 1 2 a_1\xa_2}=F^{\xa, J=5/2}_{-1-\frac 1 2 a_1\xa_2}=\frac 1 2 \sqrt{\frac{3}{5}} g_3M_s T^{a_1}_{\alpha_2 \alpha}\,,\quad
F^{\xa, J=3/2}_{+1+\frac 1 2 a_1\xa_2}=F^{\xa, J=3/2}_{-1-\frac 1 2 a_1\xa_2}=\frac 1 {\sqrt{ 10}} g_3M_s T^{a_1}_{\alpha_2 \alpha}\,.\\
\mathcal{A}[d(0), +\tfrac 1 2,-\tfrac 1 2,+]
=  \frac{g_3^2 M_s^2}{s-2M_s^2} \left( \frac{\sqrt 3}{5}d^{5/2}_{-1/2,+3/2}(\theta)+\frac{2\sqrt 2} 5 d^{3/2}_{-1/2,+3/2}(\theta) \right)  T^{a_1}_{\alpha_2 \alpha}T^{a_4}_{\alpha \alpha_3}\,,\\
F^{\xa, J=5/2}_{0+\frac 1 2 a_1\xa_2}=F^{\xa, J=5/2}_{0-\frac 1 2 a_1\xa_2}=\frac1 2 \sqrt{\frac{3}{10}} g_3M_s T^{a_1}_{\alpha_2 \alpha}\,,\quad
F^{\xa, J=3/2}_{0+\frac 1 2 a_1\xa_2}=F^{\xa, J=3/2}_{0-\frac 1 2 a_1\xa_2}=\sqrt{\frac{2}{15}} g_3M_s T^{a_1}_{\alpha_2 \alpha}\,.\\
\mathcal{A}[d(-1), +\tfrac 1 2,-\tfrac 1 2,+]
=  \frac{g_3^2 M_s^2}{s-2M_s^2} \left( \frac{\sqrt 3}{10}d^{5/2}_{+1/2,+3/2}(\theta)+\frac{3} {5\sqrt 2} d^{3/2}_{+1/2,+3/2}(\theta) \right)  T^{a_1}_{\alpha_2 \alpha}T^{a_4}_{\alpha \alpha_3}\,,\\
F^{\xa, J=5/2}_{-1+\frac 1 2 a_1\xa_2}=F^{\xa, J=5/2}_{+1-\frac 1 2 a_1\xa_2}=\frac 1 4 \sqrt{\frac{3}{10}} g_3M_s T^{a_1}_{\alpha_2 \alpha}\,,\quad
F^{\xa, J=3/2}_{-1+\frac 1 2 a_1\xa_2}=F^{\xa, J=3/2}_{+1-\frac 1 2 a_1\xa_2}=\frac 1 2 \sqrt{\frac{3}{10}} g_3M_s T^{a_1}_{\alpha_2 \alpha}\,.
\end{gather}
The corresponding partial decay widths read
\be
\Gamma^{J=5/2}_{Q^{(2)}\rightarrow d q} = \frac{13 g_3^2 M_s}{20480 \sqrt{2} \pi }N\,, \qquad
\Gamma^{J=3/2}_{Q^{(2)}\rightarrow d q} = \frac{37 g_3^2 M_s}{30720 \sqrt{2} \pi }N\,.
\ee

\subsubsection{$Q^{(2)}(J=3/2, 1/2)\to d+q$}

The second massive-level spin-$\frac 3 2$ and spin-$\frac 1 2$ resonances
arise from amplitude $\mathcal{A}[d, -\frac 1 2,+, +\frac 1 2]$:
\begin{gather}
\mathcal{A}[d(+1), -\frac 1 2,+, +\frac 1 2]
=  \frac{g_3^2 M_s^2}{s-2M_s^2} \left(\frac{1}{\sqrt 6}d^{3/2}_{-1/2,-1/2}(\theta)-\frac{1}{\sqrt 6} d^{1/2}_{-1/2,-1/2}(\theta) \right)  T^{a_1}_{\alpha_2 \alpha}T^{a_3}_{\alpha \alpha_4}\,,\\
F^{\xa, J=3/2}_{+1-\frac 1 2 a_1\xa_2}=F^{\xa, J=3/2}_{-1+\frac 1 2 a_1\xa_2}=\frac{1}{4} g_3M_s T^{a_1}_{\alpha_2 \alpha}\,,\quad
F^{\xa, J=1/2}_{+1-\frac 1 2 a_1\xa_2}=F^{\xa, J=1/2}_{-1+\frac 1 2 a_1\xa_2}=\frac{1}{2 \sqrt{2}} g_3M_s T^{a_1}_{\alpha_2 \alpha}\,,
\end{gather}
and the corresponding partial decay widths read
\be
\Gamma^{J=3/2}_{\tilde Q^{(2)}\rightarrow d q} = \frac{ g_3^2 M_s}{4096 \sqrt{2} \pi }N\,,\qquad
\Gamma^{J=1/2}_{Q^{(2)}\rightarrow d q} = \frac{ g_3^2 M_s}{1024 \sqrt{2} \pi }N\,.
\ee
Similar to previous case, we identify the spin-$\frac 3 2$ fermion in this channel as $\tilde Q^{(2)} (J=3/2)$.

\subsection{$\Phi_\pm(J=0)$}

$\Phi$ is a complex scalar field, which couples to only
(anti)self-dual gauge field configurations, i.e., to gluons in $(++)$ or $(--)$
helicity configurations. The vertex operator of $\Phi$ is given in Eq.~\eqref{VPhi}.
We will use the following amplitudes:
\begin{align}
\mathcal{A}[\Phi_+,+,+,-] & =
4\, g_3^2 \big( \, V_t \, t^{a_1a_2a_3a_4}  +  V_s \,
t^{a_2a_3a_1a_4}
 +  V_u \, t^{a_3a_1a_2a_4} \,\big)\sqrt{\alpha'}
\frac{[23]^4}{[23][34][42]} ,\quad
\label{f3g2} \\
\mathcal{A}[\Phi_+,+,+,+] & =
4\, g_3^2 \big( \, V_t \, t^{a_1a_2a_3a_4}  +  V_s \,
t^{a_2a_3a_1a_4}
 +  V_u \, t^{a_3a_1a_2a_4} \,\big)
\frac{(\alpha')^{-3/2}}{\langle 23\rangle \langle 34\rangle  \langle 42\rangle}
\label{f3g1}, \\
 \mathcal{A}\left[\Phi_+,+\tfrac 1 2,-\tfrac 1 2,+\right] & =  2g_3^2\big[V_t(T^{a_4}T^{a_1})^{\alpha_2}_{
\alpha_3 } +
 V_s(T^{a_1}T^{a_4})^{\alpha_2}_{\alpha_3}\big]\,\sqrt{\alpha'}\,\frac{[24]^2}{[23]} \,.
\end{align}

\subsubsection{$G^{(2)}(J=3,2)\to \Phi_{+}+g^+$}

The second massive-level spin-3 and spin-2 excitations arise
from factorization of $\mathcal{A}[\Phi_+,+,+,-]$:
\begin{gather}
\mathcal{A}[\Phi_+,+,+,-]
= \frac{g_3^2 M_s^2}{s-2M_s^2} \left( \frac{4}{3\sqrt 5}d^3_{-1,-2}(\theta)+\frac{4\sqrt 2}{3}d^2_{-1,-2}(\theta) \right)  f^{a_1a_2a}d^{a_3a_4a}\,,\\
F^{a, J=3}_{\Phi_++a_1a_2}=F^{a, J=3}_{\Phi_- - a_1a_2}=\frac 2 {\sqrt {15}} g_3M_s d^{a_1a_2a}\,,\quad
F^{a, J=2}_{\Phi_++a_1a_2}=F^{a, J=2}_{\Phi_-- a_1a_2}= \frac 2{\sqrt3} g_3M_s d^{a_1a_2a}\,.
\label{Spin3toPhig}
\end{gather}
$G^{(2)}(J=3,2)$ can decay both into $\Phi_{+}+g^+$ and $\Phi_{-}+g^-$ (from $\mathcal{A}[\Phi_-,-,-,+]$).
However, $\Phi_{+}+g^-$ is not possible since
$\mathcal{A}[\Phi_+,+,-,-] = 0$, and
neither is $\Phi_{-}+g^+$ as $\mathcal{A}[\Phi_-,-,+,+] = 0$.
These will also be confirmed in the next section.
The corresponding decay widths read
\be
\Gamma^{J=3}_{G^{(2)}\rightarrow \Phi g} = \frac{g_3^2 M_s}{6720 \sqrt{2} \pi }N\,,\qquad
\Gamma^{J=2}_{G^{(2)}\rightarrow \Phi g} = \frac{g_3^2 M_s}{960 \sqrt{2} \pi }N\,.
\ee

\subsubsection{$G^{(2)}(J=1)\to \Phi_{+}+g$}
$G^{(2)}(J=1)$ can arise from the following two channels.

$G^{(2)}(J=1)\to \Phi_{+}+g^+$:
\begin{gather}
\mathcal{A}[\Phi_+,+,+,+]
=  \frac{4g_3^2 M_s^2}{s-2M_s^2}  d^1_{-1,0}(\theta)  f^{a_1a_2a}d^{a_3a_4a}\,,\\
F^{a, J=1}_{\Phi_++a_1a_2}=F^{a, J=1}_{\Phi_- - a_1a_2}=2 g_3M_s d^{a_1a_2a}\,.
\end{gather}

$G^{(2)}(J=1)\to \Phi_{+}+g^-$:
\begin{gather}
\mathcal{A}[\Phi_+,-,+,+]
=  \frac{16g_3^2 M_s^2}{s-2M_s^2}  d^1_{+1,0}(\theta)  f^{a_1a_2a}d^{a_3a_4a}\,,\\
F^{a, J=1}_{\Phi_+-a_1a_2}=F^{a, J=1}_{\Phi_- + a_1a_2}=8 g_3M_s d^{a_1a_2a}\,.
\end{gather}

The $G^{(2)}(J=1)$ that goes into $\Phi_++g^ +$ is not parity invariant.
Instead, its partner decays into $\Phi_-+g^ -$.
On the other hand, both channels of $\Phi_++g^ +$ and $\Phi_++g^ -$ are possible and we need to add them up.
\be
\Gamma^{J=1}_{G^{(2)}\rightarrow \Phi g} = \frac{17 g_3^2 M_s}{384 \sqrt{2} \pi }N\,.
\ee

\subsubsection{$Q^{(2)}(J=5/2, 3/2)\to \Phi_{+}+q$}

The second massive-level spin-$\frac 5 2$ and spin-$\frac 3 2$ resonances arise from
\begin{gather}
\mathcal{A}\left[\Phi_+,+\tfrac 1 2,-\tfrac 1 2,+\right]
=  \frac{g_3^2 M_s^2}{s-2M_s^2} \left( \frac{\sqrt 2}{5}d^{5/2}_{-1/2,+3/2}(\theta)-
\frac{2 \sqrt 3} 5 d^{3/2}_{-1/2,+3/2}(\theta) \right)  T^{a_1}_{\alpha_2 \alpha}T^{a_4}_{\alpha \alpha_3}\,,\\
F^{\xa, J=5/2}_{\Phi_++\frac 1 2 a_1\xa_2}=F^{\xa, J=5/2}_{\Phi_--\frac 1 2 a_1\xa_2}=\frac{1}{2\sqrt{5}} g_3M_s T^{a_1}_{\alpha_2 \alpha}\,,\quad
F^{\xa, J=3/2}_{\Phi_++\frac 1 2 a_1\xa_2}=F^{\xa, J=3/2}_{\Phi_- -\frac 1 2 a_1\xa_2}=\frac 1 {\sqrt 5 } g_3M_s T^{a_1}_{\alpha_2 \alpha}\,.
\end{gather}
The corresponding partial decay widths read
\be
\Gamma^{J=5/2}_{Q^{(2)}\rightarrow \Phi q} = \frac{g_3^2 M_s}{7680 \sqrt{2} \pi }N\,,\qquad
\Gamma^{J=3/2}_{Q^{(2)}\rightarrow \Phi q} = \frac{ g_3^2 M_s}{1280 \sqrt{2} \pi }N\,.
\ee

\subsubsection{$Q^{(2)}(J=3/2, 1/2)\to \Phi_{+}+q$}

The second massive-level spin-$\frac 3 2$ and spin-$\frac 1 2$ resonances arise from
\begin{gather}
\mathcal{A}\left[\Phi_+,-\tfrac 1 2,+,+\tfrac 1 2\right]
=  \frac{g_3^2 M_s^2}{s-2M_s^2} \left( \frac{4}{3}d^{3/2}_{+1/2,-1/2}(\theta)+\frac{4} 3 d^{1/2}_{+1/2,-1/2}(\theta) \right)  T^{a_1}_{\alpha_2 \alpha}T^{a_3}_{\alpha \alpha_4}\,,\\
F^{\xa, J=3/2}_{\Phi_++\frac 1 2 a_1\xa_2}=F^{\xa, J=3/2}_{\Phi_- -\frac 1 2 a_1\xa_2}=\sqrt{\frac{2}{3}} g_3M_s T^{a_1}_{\alpha_2 \alpha}\,,\quad
F^{\xa, J=1/2}_{\Phi_+ +\frac 1 2 a_1\xa_2}=F^{\xa, J=1/2}_{\Phi_- -\frac 1 2 a_1\xa_2}=\frac 2 {\sqrt 3 } g_3M_s T^{a_1}_{\alpha_2 \alpha}\,.
\end{gather}
The corresponding partial decay widths read
\be
\Gamma^{J=3/2}_{\tilde Q^{(2)}\rightarrow \Phi q} = \frac{g_3^2 M_s}{384 \sqrt{2} \pi }N, ~~~~
\Gamma^{J=1/2}_{Q^{(2)}\rightarrow \Phi q} = \frac{g_3^2 M_s}{96 \sqrt{2} \pi }N \ .
\ee
Similar to previous cases, we identify the spin-$\frac 3 2$ fermion to be $\tilde Q^{(2)}(J=3/2)$.

\subsection{$\chi (J=3/2)$}

The vertex operator of the spin-$\frac 3 2$ fermion $\chi$
is given in Eq.~\eqref{Vchi}.
We will need to use the following amplitudes:
\begin{equation}
\mathcal{A}[\chi_1,\epsilon_2,\epsilon_3,u_4] ~=~ 2\,g_3^2\,
[ \, V_t \, (T^{a_2}  T^{a_3})^{\alpha_4} _{\alpha_{1}} \
- \ V_s \, (T^{a_3}  T^{a_2})_{\alpha_1}^{\alpha_4}\, ]\,
\mathscr{A}[\chi_1,\epsilon_2,\epsilon_3,u_4]\ ,
\end{equation}
where
\begin{align}
\mathscr{A}\left[\chi(-\tfrac 3 2),-,-,\hpl \right]&
=~\frac{[4q]^{3}}{[23][34][42]}\,,\notag\\
\mathscr{A}\left[\chi(-\tfrac 1 2),-,-,\hpl \right]&
=\sqrt{3}\frac{[4q]^{2}[p4]}{[23][34][42]}\,,\notag\\
\mathscr{A}\left[\chi(-\tfrac 1 2),-,-,\hpl \right]&
=\sqrt{3}\frac{[4p]^{2}[q4]}{[23][34][42]}\,,\\
\mathscr{A}\left[\chi(-\tfrac 3 2),-,-,\hpl \right]&
=~\frac{[4p]^{3}}{[23][34][42]}\,,\notag
\end{align}
and
\begin{align}
\mathscr{A}\left[\chi(-\tfrac 3 2),+,-,\hpl\right]&
=~~\sqrt{\al'}\frac{\langle p3\rangle^{3}}{\langle 23\rangle\langle
24\rangle}\,,\notag\\
\mathscr{A}\left[\chi(-\tfrac 1 2),+,-,\hpl\right]&
=~\sqrt{3\al'}\,\frac{\langle p3\rangle^{2}\langle q3\rangle}{\langle 23\rangle\langle
24\rangle}\,,\notag\\
\mathscr{A}\left[\chi(-\tfrac 1 2),+,-,\hpl\right]&
=-\sqrt{3\al'}\frac{\langle q3\rangle^{2}\langle p3\rangle}{\langle 23\rangle\langle
24\rangle}\,,\\
\mathscr{A}\left[\chi(-\tfrac 3 2),+,-,\hpl\right]&
=~{-}\sqrt{\al'}\frac{\langle q3\rangle^{3}}{\langle 23\rangle\langle
24\rangle}\notag\,.
\end{align}

\subsubsection{$G^{(2)}(J=3,2)\to \chi +\bar q$}

The second massive level spin-3 and spin-2 excitations arise
from factorization of $\mathcal{A}[\chi,+\frac{1}{2},-,+]$:
\begin{gather}
\mathcal{A}\left[\chi(+\tfrac 3 2), +\tfrac 1 2,-,+\right]
=  \frac{g_3^2 M_s^2}{s-2M_s^2} \lb \frac 2 3 d^3_{-2,+2}(\theta) + \frac 2 3
d^2_{-2,+2}(\theta) \rb T^a_{\xa_1\xa_2}f^{a_3a_4a}\,,\\
F^{a, J=3}_{+\frac 3 2 +\frac 1 2\xa_1\xa_2}=F^{a, J=3}_{-\frac 3 2 -\frac 1 2 \xa_1\xa_2}=\frac 1 {\sqrt 3} g_3M_s T^a_{\xa_1\xa_2}\,, \quad
F^{a, J=2}_{+\frac 3 2 +\frac 1 2\xa_1\xa_2}=F^{a, J=2}_{-\frac 3 2 -\frac 1 2 \xa_1\xa_2}=\frac 1 {\sqrt 6} g_3M_s T^a_{\xa_1\xa_2}\,. \label{Spin3tochiq1}\\
\mathcal{A}\left[\chi(+\tfrac 1 2), +\tfrac 1 2,-,+\right]
=  \frac{g_3^2 M_s^2}{s-2M_s^2} \lb \frac 2 {\sqrt{15}} d^3_{-1,+2}(\theta) + \sqrt {\frac 2 3} d^2_{-1,+2}(\theta) \rb
T^a_{\xa_1\xa_2}f^{a_3a_4a}\,,\\
F^{a, J=3}_{+\frac 1 2 +\frac 1 2\xa_1\xa_2}=F^{a, J=3}_{-\frac 1 2 -\frac 1 2 \xa_1\xa_2}=\frac 1 {\sqrt 5} g_3M_s T^a_{\xa_1\xa_2}\,, \quad
F^{a, J=2}_{+\frac 1 2 +\frac 1 2\xa_1\xa_2}=F^{a, J=2}_{-\frac 1 2 -\frac 1 2 \xa_1\xa_2}=\frac 1 {2} g_3M_s T^a_{\xa_1\xa_2}\,.\\
\mathcal{A}\left[\chi(-\tfrac 1 2), +\tfrac 1 2,-,+\right]
=  \frac{g_3^2 M_s^2}{s-2M_s^2} \lb \frac 1 {\sqrt{10}} d^3_{0,+2}(\theta) +  \frac 1 {\sqrt 2 } d^2_{0,+2}(\theta) \rb
T^a_{\xa_1\xa_2}f^{a_3a_4a}\,,\\
F^{a, J=3}_{-\frac 1 2 +\frac 1 2\xa_1\xa_2}=F^{a, J=3}_{+\frac 1 2 -\frac 1 2 \xa_1\xa_2}=\frac 1 2 \sqrt{\frac 3 { 10}} g_3M_s T^a_{\xa_1\xa_2}\,, \quad
F^{a, J=2}_{-\frac 1 2 +\frac 1 2\xa_1\xa_2}=F^{a, J=2}_{+\frac 1 2 -\frac 1 2 \xa_1\xa_2}=\frac 1 {4} \sqrt 3 g_3M_s T^a_{\xa_1\xa_2}\,.\\
\mathcal{A}\left[\chi(-\tfrac 3 2), +\tfrac 1 2,-,+\right]
=  \frac{g_3^2 M_s^2}{s-2M_s^2} \lb \frac 1 {3\sqrt{5}} d^3_{+1,+2}(\theta) +  \frac {\sqrt 2} {3 } d^2_{+1,+2}(\theta) \rb
T^a_{\xa_1\xa_2}f^{a_3a_4a}\,,\\
F^{a, J=3}_{-\frac 3 2 +\frac 1 2\xa_1\xa_2}=F^{a, J=3}_{+\frac 3 2 -\frac 1 2 \xa_1\xa_2}=\frac 1 { 2\sqrt{15}} g_3M_s T^a_{\xa_1\xa_2}\ , \quad F^{a, J=2}_{-\frac 3 2 +\frac 1 2\xa_1\xa_2}=F^{a, J=2}_{+\frac 3 2 -\frac 1 2 \xa_1\xa_2}=\frac 1 {2
\sqrt 3}  g_3M_s T^a_{\xa_1\xa_2}\,. \label{Spin3tochiq4}
\end{gather}
The corresponding decay widths read
\be
\Gamma^{J=3}_{(G^{(2)}\rightarrow \chi \bar q)+(G^{(2)}\rightarrow  \bar\chi q)} = \frac{5 g_3^2 M_s  N_f}{896 \sqrt{2} \pi }\,,\qquad
\Gamma^{J=2}_{(G^{(2)}\rightarrow \chi \bar q)+(G^{(2)}\rightarrow  \bar\chi q)} = \frac{11 g_3^2 M_s  N_f}{1280 \sqrt{2} \pi }\,.
\ee

\subsubsection{$G^{(2)}(J=1)\to \chi +\bar q$}

The second massive-level spin-1 excitations arise from factorization of
\begin{gather}
\mathcal{A}\left[\chi(+\tfrac 3 2), +\tfrac 1 2,-,-\right]
=  \frac{g_3^2 M_s^2}{s-2M_s^2}  d^1_{+1,0}(\theta)  T^a_{\xa_1\xa_2}f^{a_3a_4a}\,,\\
F^{a, J=1}_{+\frac 3 2 +\frac 1 2\xa_1\xa_2}=F^{a, J=1}_{-\frac 3 2 -\frac 1 2 \xa_1\xa_2}=\frac 1 2 g_3M_s T^a_{\xa_1\xa_2}\,.
\end{gather}
We also need to take into account the channel of $G^{(2)}(J=1) \to \bar \chi + q$. The sum of the decay widths reads
\be
\Gamma^{J=1}_{(G^{(2)}\rightarrow \chi \bar q)+(G^{(2)}\rightarrow \bar\chi q)} = \frac{g_3^2 M_s  N_f}{384 \sqrt{2} \pi }\ .
\ee

\subsubsection{$Q^{(2)}(J=5/2, 3/2)\to \chi +g$}
$Q^{(2)}(J=5/2, 3/2)\to \chi +g^-$ can be obtained from:
\begin{gather}
\mathcal{A}\left[\chi(+\tfrac 3 2),-,-,+\tfrac 1 2\right]
=  \frac{g_3^2 M_s^2}{s-2M_s^2} \left( \frac{1}{5}d^{5/2}_{-1/2,+3/2}(\theta)-\frac{\sqrt 6} 5 d^{3/2}_{-1/2,+3/2}(\theta) \right)  T^{a_2}_{\alpha_1 \alpha}T^{a_3}_{\alpha \alpha_4}\,,\\
F^{\xa, J=5/2}_{+\frac 3 2 - \xa_1a_2}=F^{\xa, J=5/2}_{-\frac 3 2 + \xa_1a_2}=\frac{1}{2\sqrt{10}} g_3M_s T^{a_2}_{\alpha_1 \alpha}\,,\quad
F^{\xa, J=3/2}_{+\frac 3 2 - \xa_1a_2}=F^{\xa, J=3/2}_{-\frac 3 2 + \xa_1a_2}=\frac 1 {\sqrt {10} } g_3M_s T^{a_2}_{\alpha_1 \alpha}\,.\\
\mathcal{A}\left[\chi(+\tfrac 1 2),-,-,+\tfrac 1 2\right]
=  \frac{g_3^2 M_s^2}{s-2M_s^2} \left( \frac{\sqrt 6}{5}d^{5/2}_{+1/2,+3/2}(\theta)-\frac{4} 5 d^{3/2}_{+1/2,+3/2}(\theta) \right)  T^{a_2}_{\alpha_1 \alpha}T^{a_3}_{\alpha \alpha_4}\,,\\
F^{\xa, J=5/2}_{+\frac 1 2 - \xa_1a_2}=F^{\xa, J=5/2}_{-\frac 1 2 + \xa_1a_2}=\frac1 2 \sqrt{\frac{3}{5}} g_3M_s T^{a_2}_{\alpha_1 \alpha}\,,\quad
F^{\xa, J=3/2}_{+\frac 1 2 - \xa_1a_2}=F^{\xa, J=3/2}_{-\frac 1 2 + \xa_1a_2}=\frac 2 {\sqrt {15} } g_3M_s T^{a_2}_{\alpha_1 \alpha}\,.\\
\mathcal{A}\left[\chi(-\tfrac 1 2),-,-,+\tfrac 1 2\right]
=  \frac{g_3^2 M_s^2}{s-2M_s^2} \left( \frac{2\sqrt 6}{5}d^{5/2}_{+3/2,+3/2}(\theta)-\frac{2\sqrt 6} 5 d^{3/2}_{+3/2,+3/2}(\theta) \right)  T^{a_2}_{\alpha_1 \alpha}T^{a_3}_{\alpha \alpha_4}\,,\\
F^{\xa, J=5/2}_{-\frac 1 2 - \xa_1a_2}=F^{\xa, J=5/2}_{+\frac 1 2 + \xa_1a_2}=\sqrt{\frac{3}{5}} g_3M_s T^{a_2}_{\alpha_1 \alpha}\,,\quad
F^{\xa, J=3/2}_{-\frac 1 2 - \xa_1a_2}=F^{\xa, J=3/2}_{+\frac 1 2 + \xa_1a_2}=\sqrt {\frac 2 {5}} g_3M_s T^{a_2}_{\alpha_1 \alpha}\,.\\
\mathcal{A}\left[\chi(-\tfrac 3 2),-,-,+\tfrac 1 2\right]
=  \frac{g_3^2 M_s^2}{s-2M_s^2}  \frac{4}{\sqrt 5}d^{5/2}_{+5/2,+3/2}(\theta)  T^{a_2}_{\alpha_1 \alpha}T^{a_3}_{\alpha \alpha_4}\,,\\
F^{\xa, J=5/2}_{-\frac 3 2 - \xa_1a_2}=F^{\xa, J=5/2}_{+\frac 3 2 + \xa_1a_2}=\sqrt{2} g_3M_s T^{a_2}_{\alpha_1 \alpha}\,.
\end{gather}
$Q^{(2)}(J=5/2, 3/2)\to \chi +g^+$ can be obtained from:
\begin{gather}
\mathcal{A}\left[\chi(+\tfrac 3 2),+,-,+\tfrac 1 2\right]
=  \frac{g_3^2 M_s^2}{s-2M_s^2} 4\sqrt{\frac{2}{5}}d^{5/2}_{-5/2,+3/2}(\theta) T^{a_2}_{\alpha_1 \alpha}T^{a_3}_{\alpha \alpha_4}\,,\\
F^{\xa, J=5/2}_{+\frac 3 2 + \xa_1a_2}=F^{\xa, J=5/2}_{-\frac 3 2 - \xa_1a_2}= 2 g_3M_s T^{a_2}_{a_1 a}\,.\\
\mathcal{A}\left[\chi(+\tfrac 1 2),+,-,+\tfrac 1 2\right]
=  \frac{g_3^2 M_s^2}{s-2M_s^2} \left( \frac{4\sqrt{3}}{5}d^{5/2}_{-3/2,+3/2}(\theta)+\frac{4\sqrt{3}}{5} d^{3/2}_{-3/2,+3/2}(\theta) \right)  T^{a_2}_{\alpha_1 \alpha}T^{a_3}_{\alpha \alpha_4}\,,\\
F^{\xa, J=5/2}_{+\frac 1 2 + \xa_1a_2}=F^{\xa, J=5/2}_{-\frac 1 2 - \xa_1a_2}= \sqrt{\frac{6}{5}}g_3M_s T^{a_2}_{\alpha_1 \alpha}\,,\quad
F^{\xa, J=3/2}_{+\frac 1 2 + \xa_1a_2}=F^{\xa, J=3/2}_{-\frac 1 2 - \xa_1a_2}=\frac 2 {\sqrt 5 } g_3M_s T^{a_2}_{\alpha_1 \alpha}\,.\\
\mathcal{A}\left[\chi(-\tfrac 1 2),+,-,+\tfrac 1 2\right]
=  \frac{g_3^2 M_s^2}{s-2M_s^2} \left( \frac{2\sqrt{3}}{5}d^{5/2}_{-1/2,+3/2}(\theta)+\frac{4\sqrt{2}}{5} d^{3/2}_{-1/2,+3/2}(\theta) \right)  T^{a_2}_{\alpha_1 \alpha}T^{a_3}_{\alpha \alpha_4}\,,\\
F^{\xa, J=5/2}_{-\frac 1 2 + \xa_1a_2}=F^{\xa, J=5/2}_{+\frac 1 2 - \xa_1a_2}= \sqrt{\frac{3}{10}}g_3M_s T^{a_2}_{\alpha_1 \alpha}\,,\quad
F^{\xa, J=3/2}_{-\frac 1 2 + \xa_1a_2}=F^{\xa, J=3/2}_{+\frac 1 2 - \xa_1a_2}=2\sqrt {\frac 2 {15} } g_3M_s T^{a_2}_{\alpha_1 \alpha}\,.\\
\mathcal{A}\left[\chi(-\tfrac 3 2),+,-,+\tfrac 1 2\right]
=  \frac{g_3^2 M_s^2}{s-2M_s^2} \left( \frac{\sqrt{2}}{5}d^{5/2}_{+1/2,+3/2}(\theta)+\frac{2\sqrt{3}}{5} d^{3/2}_{+1/2,+3/2}(\theta) \right)  T^{a_2}_{\alpha_1 \alpha}T^{a_3}_{\alpha \alpha_4}\,,\\
F^{\xa, J=5/2}_{-\frac 3 2 + \xa_1a_2}=F^{\xa, J=5/2}_{+\frac 3 2 - \xa_1a_2}= \frac{1}{2\sqrt{5}}g_3M_s T^{a_2}_{\alpha_1 \alpha}\,,\quad
F^{\xa, J=3/2}_{-\frac 3 2 + \xa_1a_2}=F^{\xa, J=3/2}_{+\frac 3 2 - \xa_1a_2}=\frac 1 {\sqrt {5}} g_3M_s T^{a_2}_{\alpha_1 \alpha}\,.
\end{gather}
The corresponding decay widths read
\ba
\Gamma^{J=5/2}_{Q^{(2)}\rightarrow \chi g} = \frac{111 g_3^2 M_s}{5120 \sqrt{2} \pi }N\,,\qquad
\Gamma^{J=3/2}_{Q^{(2)}\rightarrow \chi g} = \frac{23 g_3^2 M_s}{2560 \sqrt{2} \pi }N\,.
\ea

\subsubsection{$Q^{(2)}(J=3/2, 1/2)\to \chi +g^-$}

The second massive-level spin-$\frac 3 2$ and spin-$\frac 1 2$ resonances arise from
\begin{gather}
\mathcal{A}\left[\chi(+\tfrac 3 2),-,+\tfrac 1 2,+\right]
=  \frac{g_3^2 M_s^2}{s-2M_s^2} \left( \frac{\sqrt{2}}{3}d^{3/2}_{-1/2,+1/2}(\theta)+\frac{\sqrt{2}}{3} d^{1/2}_{-1/2,+1/2}(\theta) \right)  T^{a_2}_{\alpha_1 \alpha}T^{a_4}_{\alpha \alpha_3}\,,\\
F^{\xa, J=3/2}_{+\frac 3 2 - \xa_1a_2}=F^{\xa, J=3/2}_{-\frac 3 2 + \xa_1a_2}= \frac{1}{2\sqrt{3}}g_3M_s T^{a_2}_{\alpha_1 \alpha}\,,\quad
F^{\xa, J=1/2}_{+\frac 3 2 - \xa_1a_2}=F^{\xa, J=1/2}_{-\frac 3 2 + \xa_1a_2}=\frac 1 {\sqrt 6 } g_3M_s T^{a_2}_{\alpha_1 \alpha}\,.
\end{gather}
Channels to $Q^{(2)}(J=3/2, 1/2) \to \chi g^-$ are not possible since $\mathcal{A}[\chi ,+,+\frac 1 2,+] = 0$.
\be
\Gamma^{J=3/2}_{\tilde Q^{(2)}\rightarrow \chi g^-} = \frac{g_3^2 M_s}{3072 \sqrt{2} \pi }N\,,\qquad
\Gamma^{J=1/2}_{Q^{(2)}\rightarrow \chi g^-} = \frac{g_3^2 M_s}{768 \sqrt{2} \pi }N\,.
\ee

\subsection{$ a (J=1/2)$}

The vertex operator of the spin-$\frac 1 2$ fermion $a$
is given in Eq.~\eqref{Va}.
We will use the following amplitudes:
\begin{equation}
\mathcal{A}[a_1,\epsilon_2,\epsilon_3,u_4] ~=~ 2g_3^2\,(\alpha')^{-1}
[ \, V_t \, (T^{a_2}  T^{a_3})^{\alpha_4} _{\alpha_1} \
- \ V_s \, (T^{a_3}  T^{a_2})_{\alpha_1}^{\alpha_4}\, ]\,
\mathscr{A}[a_1,\epsilon_2,\epsilon_3,u_4]\,,
\end{equation}
where
\begin{align}
\mathscr{A}\left[a(\hpl),+,+,\hpl\right]&
=\frac{\langle p4\rangle}{\langle 23\rangle\langle
34\rangle\langle 42\rangle}\,,\notag\\
\mathscr{A}\left[a(\hmi),+,+,\hpl\right]&
=\frac{\langle q4\rangle}{\langle 23\rangle\langle
34\rangle\langle 42\rangle}\,,
\end{align}
and
\begin{align}
\mathscr{A}\left[a(\hpl),+,-,\hpl\right]&
=~{\al'}^{3/2}\,\frac{[q2][24]^{2}}{[23][34]}\,,\notag\\
\mathscr{A}\left[a(\hmi),+,-,\hpl\right]&
={\al'}^{3/2}\frac{[p2][24]^{2}}{[23][34]}\,.
\end{align}

\subsubsection{$G^{(2)}(J=3,2)\to a +\bar q$}

The second massive level spin-3 and spin-2 resonances arise from
\begin{gather}
\mathcal{A}\left[a(+\tfrac 1 2), +\tfrac 1 2,-,+\right]
=  \frac{g_3^2 M_s^2}{s-2M_s^2} \lb \frac 1 {3\sqrt 5} d^3_{-1,+2}(\theta) - \frac{\sqrt 2}{3}
d^2_{-1,+2}(\theta) \rb T^a_{\xa_1\xa_2}f^{a_3a_4a}\,,\\
F^{a, J=3}_{+\frac 1 2 +\frac 1 2\xa_1\xa_2}=F^{a, J=3}_{-\frac 1 2 -\frac 1 2 \xa_1\xa_2}=\frac 1 {2\sqrt {15}} g_3M_s T^a_{\xa_1\xa_2}\,, \quad
F^{a, J=2}_{+\frac 1 2 +\frac 1 2\xa_1\xa_2}=F^{a, J=2}_{-\frac 1 2 -\frac 1 2 \xa_1\xa_2}=\frac 1 {2\sqrt 3} g_3M_s T^a_{\xa_1\xa_2}\,. \label{Spin3toaq1}\\
\mathcal{A}\left[a(-\tfrac 1 2) +\tfrac 1 2,-,+\right]
=  \frac{g_3^2 M_s^2}{s-2M_s^2} \lb \frac 1 {\sqrt {30}} d^3_{0,+2}(\theta) - \frac 1 {\sqrt 6}
d^2_{0,+2}(\theta) \rb T^a_{\xa_1\xa_2}f^{a_3a_4a}\,,\\
F^{a, J=3}_{-\frac 1 2 +\frac 1 2\xa_1\xa_2}=F^{a, J=3}_{+\frac 1 2 -\frac 1 2 \xa_1\xa_2}=\frac 1 {2\sqrt {10}} g_3M_s T^a_{\xa_1\xa_2}\,, \quad
F^{a, J=2}_{-\frac 1 2 +\frac 1 2\xa_1\xa_2}=F^{a, J=2}_{+\frac 1 2 -\frac 1 2 \xa_1\xa_2}=\frac 1 {4} g_3M_s T^a_{\xa_1\xa_2}\,. \label{Spin3toaq2}
\end{gather}
The corresponding decay widths read
\be
\Gamma^{J=3}_{(G^{(2)}\rightarrow a \bar q)+(G^{(2)}\rightarrow  \bar a q)} = \frac{g_3^2 M_s  N_f}{2688 \sqrt{2} \pi }\,,\qquad
\Gamma^{J=2}_{(G^{(2)}\rightarrow a \bar q)+(G^{(2)}\rightarrow  \bar a q)} = \frac{7 g_3^2 M_s  N_f}{3840 \sqrt{2} \pi }\,.
\ee

\subsubsection{$G^{(2)}(J=1)\to a +\bar q$}

The second massive-level spin-1 resonances arise from
\begin{gather}
\mathcal{A}\left[a(+\tfrac 1 2), +\tfrac 1 2,+,+\right]
=  \frac{g_3^2 M_s^2}{s-2M_s^2}  d^1_{-1,0}(\theta)
T^a_{\xa_1\xa_2}f^{a_3a_4a}\,,\\
F^{a, J=1}_{+\frac 1 2 +\frac 1 2\xa_1\xa_2}=F^{a, J=1}_{-\frac 1 2 -\frac 1 2 \xa_1\xa_2}=\frac 1 {2} g_3M_s T^a_{\xa_1\xa_2}\,.
\end{gather}
The corresponding decay width reads
\be
\Gamma^{J=1}_{(G^{(2)}\rightarrow a \bar q)+(G^{(2)}\rightarrow  \bar a q)} = \frac{g_3^2 M_s  N_f}{384 \sqrt{2} \pi }\,.
\ee

\subsubsection{$Q^{(2)}(J=5/2, 3/2)\to a +g^+$}

We could obtain the second massive-level spin-$\frac 5 2$ and spin-$\frac 3 2$ resonances from
\begin{gather}
\mathcal{A}\left[a(-\tfrac 1 2),+,-,+\tfrac 1 2\right]
=  \frac{g_3^2 M_s^2}{s-2M_s^2} \left( \frac{1}{5}d^{5/2}_{-1/2,+3/2}(\theta)-\frac{\sqrt 6}{5} d^{3/2}_{-1/2,+3/2}(\theta) \right)  T^{a_2}_{\alpha_1 \alpha}T^{a_3}_{\alpha \alpha_4}\,,\\
F^{\xa, J=5/2}_{-\frac 1 2 + \xa_1a_2}=F^{\xa, J=5/2}_{+\frac 1 2 - \xa_1a_2}= \frac{1}{2\sqrt{10}}g_3M_s T^{a_2}_{\alpha_1 \alpha}\,,\quad
F^{\xa, J=3/2}_{-\frac 1 2 + \xa_1a_2}=F^{\xa, J=3/2}_{+\frac 1 2 - \xa_1a_2}=\frac{1}{\sqrt{10}} g_3M_s T^{a_2}_{\alpha_1 \alpha}\,.
\end{gather}
Again, decaying into $a +g^-$ is not possible since $\mathcal{A}[a(+\frac 1 2),-,-,+\frac 1 2] =0$.
The decay widths read
\ba
\Gamma^{J=5/2}_{Q^{(2)}\rightarrow a g} = \frac{g_3^2 M_s}{15360 \sqrt{2} \pi }N\,,\qquad
\Gamma^{J=3/2}_{Q^{(2)}\rightarrow a g} = \frac{g_3^2 M_s}{2560 \sqrt{2} \pi }N\,.
\ea

\subsubsection{$Q^{(2)}(J=3/2, 1/2)\to a + g$}

$Q^{(2)}(J=3/2, 1/2)\to a +g^+$ can be obtained from
\begin{gather}
\mathcal{A}\left[a(+\tfrac 1 2),+,+\tfrac 1 2,+\right]
=  \frac{g_3^2 M_s^2}{s-2M_s^2} \frac{2}{\sqrt{3}}d^{3/2}_{-3/2,+1/2}(\theta)  T^{a_2}_{\alpha_1 \alpha}T^{a_4}_{\alpha \alpha_3}\,,\\
F^{\xa, J=3/2}_{+\frac 1 2 + \xa_1a_2}=F^{\xa, J=3/2}_{-\frac 1 2 - \xa_1a_2}= \frac{1}{\sqrt{2}}g_3M_s T^{a_2}_{\alpha_1 \alpha}\,.\\
\mathcal{A}\left[a(-\tfrac 1 2),+,+\tfrac 1 2,+\right]
=  \frac{g_3^2 M_s^2}{s-2M_s^2} \left( \frac{\sqrt{2}}{3}d^{3/2}_{-1/2,+1/2}(\theta)+\frac{\sqrt{2}}{3}d^{1/2}_{-1/2,+1/2}(\theta) \right)  T^{a_2}_{\alpha_1 \alpha}T^{a_4}_{\alpha \alpha_3}\,,\\
F^{\xa, J=3/2}_{-\frac 1 2 + \xa_1a_2}=F^{\xa, J=3/2}_{+\frac 1 2 - \xa_1a_2}= \frac{1}{2\sqrt{3}}g_3M_s T^{a_2}_{\alpha_1 \alpha}\,,\quad
F^{\xa, J=1/2}_{-\frac 1 2 + \xa_1a_2}=F^{\xa, J=1/2}_{+\frac 1 2 - \xa_1a_2}=\frac 1 {\sqrt 6 } g_3M_s T^{a_2}_{\alpha_1 \alpha}\,.
\end{gather}
$Q^{(2)}(J=3/2, 1/2)\to a +g^-$ can be obtained from
\begin{gather}
\mathcal{A}\left[a(+\tfrac 1 2),-,+\tfrac 1 2,+\right]
=  \frac{g_3^2 M_s^2}{s-2M_s^2} \left( \frac{4}{3}d^{3/2}_{+1/2,+1/2}(\theta)+\frac{4}{3} d^{1/2}_{+1/2,+1/2}(\theta) \right)  T^{a_2}_{\alpha_1 \alpha}T^{a_4}_{\alpha \alpha_3}\,,\\
F^{\xa, J=3/2}_{+\frac 1 2 - \xa_1a_2}=F^{\xa, J=3/2}_{-\frac 1 2 + \xa_1a_2}= \sqrt \frac{2}{3}g_3M_s T^{a_2}_{\alpha_1 \alpha}\,,\quad
F^{\xa, J=1/2}_{+\frac 1 2 - \xa_1a_2}=F^{\xa, J=1/2}_{-\frac 1 2 + \xa_1a_2}=\frac 2 {\sqrt 3 } g_3M_s T^{a_2}_{\alpha_1 \alpha}\,.\\
\mathcal{A}\left[a(-\tfrac 1 2),-,+\tfrac 1 2,+\right]
=  \frac{g_3^2 M_s^2}{s-2M_s^2} 4\sqrt{\frac{2}{3}}d^{3/2}_{+3/2,+1/2}(\theta) T^{a_2}_{\alpha_1 \alpha}T^{a_4}_{\alpha \alpha_3}\,,\\
F^{\xa, J=3/2}_{-\frac 1 2 - \xa_1a_2}=F^{\xa, J=3/2}_{+\frac 1 2 + \xa_1a_2}= 2g_3M_s T^{a_2}_{\alpha_1 \alpha}\,.
\end{gather}
The corresponding decay widths read
\be
\Gamma^{J=3/2}_{\tilde Q^{(2)}\rightarrow a g} = \frac{21 g_3^2 M_s}{1024 \sqrt{2} \pi }N\,,\qquad
\Gamma^{J=1/2}_{Q^{(2)}\rightarrow a g} = \frac{3 g_3^2 M_s}{256 \sqrt{2} \pi }N\,.
\ee

\subsection{Excited quarks decay to $SU(2)$ gauge bosons}\label{SU2}

For excited quarks which arise from the intersection of the $U(3)$ stack
and $U(2)$ [or $Sp(1)$] stack, it is easy to see that the massive
quarks could decay into a $SU(2)$ gauge boson plus a massless quark.
One could obtain the total decay width of the massive quark decaying
into $SU(2)$ gauge bosons $A^{a}$ by performing a factorization
of the amplitude $\mathcal{A}(q,A^{a},\bar{q},g)$ which was obtained
in Ref.~\cite{Lust:2008qc},
while in the broken electroweak symmetry, $W$ and
$Z$ bosons are produced. Hence we need to translate the decay widths
of the massive quarks to $A^{a}$ into the decay width of $W$ and
$Z$ bosons.

For illustration, let us focus on the higher-level excited quark $u^{(n)}$.
Effectively, its couplings can be written as
\begin{align}
\mathcal{L}_{{\rm int}} & =\frac{1}{2}g_{2}\bar{u}_{L}^{(n)}\gamma^{\mu}d_{L}(A_{\mu}^{1}-iA_{\mu}^{2})+\frac{1}{2}g_{2}\bar{u}_{L}^{(n)}\gamma^{\mu}u_{L}A_{\mu}^{3}+\frac{1}{6}g_{Y}\bar{u}_{L}^{(n)}\gamma^{\mu}u_{L}Y_{\mu}\nonumber \\
 & \to\frac{1}{\sqrt{2}}g_{2}\bar{u}_{L}^{(n)}\gamma^{\mu}d_{L}W_{\mu}^{+}+\frac{g_{2}}{c_{W}}\big(\frac{1}{2}-\frac{2}{3}s_{W}^{2}\big)\bar{u}_{L}^{(n)}\gamma^{\mu}u_{L}Z_{\mu}+\big(\frac{2}{3}e\big)\bar{u}_{L}^{(n)}\gamma^{\mu}u_{L}A_{\mu}^{\gamma}\,,
\end{align}
where $c_{W}\equiv\cos\theta_{W},s_{W}\equiv\sin\theta_{W}$, $e=g_{2}g_{Y}/\sqrt{g_{2}^{2}+g_{Y}^{2}}$
and
\begin{equation}
W^{+}=\frac{1}{\sqrt{2}}(A^{1}-iA^{2})\,,\quad Z=c_{W}A^{3}-s_{W}Y\,,\quad A^{\gamma}=s_{W}A^{3}+c_{W}Y\,.
\end{equation}
Since $u^{(n)}$ is very massive ($\sim\sqrt{n}M_{s}$), we can
simply treat all the gauge bosons after the electroweak symmetry breaking
as massless. A simple calculation shows
\begin{gather}
\Gamma(u_{L}^{(n)}\to W^{+}+d_{L})=2\Gamma(u_{L}^{(n)}\to A^{1}+d_{L})=2\Gamma(u_{L}^{(n)}\to A^{2}+d_{L})=2\Gamma(u_{L}^{(n)}\to A^{3}+u_{L})\,,
\end{gather}
and
\begin{equation}
\Gamma(u_{L}^{(n)}\to Z+u_{L})=\frac{2}{c_{W}^{2}}\big(\frac{1}{2}-\frac{2}{3}s_{W}^{2}\big)^{2}\ \Gamma(u_{L}^{(n)}\to W^{+}+d_{L})\,.
\end{equation}
At 10 --100 TeV, we have
$\frac{2}{c_{W}^{2}}\big(\frac{1}{2}-\frac{2}{3}s_{W}^{2}\big)^{2}\approx0.28$.
Thus, we conclude, the decay widths of the massive quark $u_{L}^{(n)}$
that decay into $W^{+}$ and $Z$ are approximately
\begin{equation}
\Gamma(u_{L}^{(n)}\to W^{+}+d_{L})+\Gamma(u_{L}^{(n)}\to Z+u_{L})\approx 0.86 \times \sum_{a=1,2,3}\Gamma(u_{L}^{(n)}\to A^{a}+\cdots)\,.
\end{equation}
Since $g_{3}$ is not much greater than $g_{2}$ at 10 --100 TeV,
we should also include these contributions to the total decay widths
of the massive quark excitations.

For the second massive-level excited quarks, the decay channels
$Q^{(2)}\to A^{a}+Q^{(1)}$ also exist. A similar analysis gives the
same front factor
\begin{equation}
\sum\Gamma(Q^{(2)}\to Q^{(1)}+W/Z)\approx 0.86 \times \sum_{a=1,2,3}\Gamma(Q^{(2)}\to A^{a}+Q^{(1)})\,.
\end{equation}
For the massive string states decaying into photon plus other string
states, see the discussion of the next subsection on massive string states
decaying to anomalous $U(1)$'s.

\subsection{Massive string states decaying to anomalous $U(1)$'s}\label{AnU1}

We have seen that for intersecting D-brane brane models the SM gauge group
must be extended with new $U(1)$ symmetries. These $U(1)$'s are in general anomalous.
They couple to RR axions and   would obtain a string scale mass~\cite{Antoniadis:2002cs}.
These $U(1)$'s would mix with each other through the $U(1)$ mass-squared matrix.
The mass mixing effects have been discussed in Sec.~\ref{massmixing}.
Massive string excitations carry the SM gauge charges and thus they
could decay into anomalous $U(1)$'s if kinetically allowed. In this
subsection, we will briefly study the possible
decay channels of massive string excitations.

Let us first focus on the amplitude $\mathcal{A}(g,g,g,A_{a})$, where $A_{a}$
denotes the $U(1)$ from the $U(3)_{a}$ stack. Factorization gives
rise to the resonances of excited massive gluons, and we have
\begin{equation}
G^{(n)} \to g+A_{a}\,.
\label{GntogA}
\end{equation}
Similarly, the factorization of amplitude $\mathcal{A}(g,g, A_{a},A_{a})$
gives rise to a massive color singlet that 
\begin{equation}
C^{(n)}\to A_{a}+A_{a}\,,
\end{equation}
and we also need to write this decay in terms of mass eigenfields.
We can also consider amplitudes $\mathcal{A}(G^{(1)},g,g,A_{a})$ and $\mathcal{A}(C^{(1)},g,g,A_{a})$,
for which
factorization could give the following decay channels 
\begin{gather}
G^{(n)} \to G^{(1)} + A_a\,,\qquad G^{(n)}\to C^{(1)}+g \,, \\
C^{(n)}\to C^{(1)}+A_{a}\,.
\end{gather}
Additionally, the factorization of the amplitude $\mathcal{A}(g,q,\bar{q},A^{a})$ gives
rise to higher-level excited massive quarks decaying into anomalous $U(1)$'s:
\begin{equation}
Q^{(n)}\to q+A_a\,,
\end{equation}
if kinetically allowed.
Also, factorization of the amplitudes $\mathcal{A}(g,q,\bar{q},C^{(1)})$ and $\mathcal{A}(Q^{(1)},g,\bar{q},A_a)$ gives
\begin{equation}
Q^{(n)}\to C^{(1)} + q\,,\qquad Q^{(n)}\to Q^{(1)}+A_a\,.
\end{equation}

Since $A_{a}$ is not in the physical eigenbasis,
we need to write it in terms of physical fields (fields in the mass eigenbasis).
Using Eq.~\eqref{AinMassB}, we rewrite Eq.~\eqref{GntogA} as
\begin{align}
G^{(n)} & \to g+A_{a}\nonumber \\
 & =g+O_{a1}A_{1}^{(m)}+O_{a2}A_{2}^{(m)}+\cdots\nonumber \\
 & =g+O_{a1}B_{\mu}+O_{a2}Z'+\cdots
\end{align}
and similarly for other decay channels.
As long as kinetically allowed, the massive string excitations can
decay also into heavier massive anomalous $U(1)$'s.
This is a model-dependent issue,
since the transformation matrix $O$ depends on the details of the model building.
Unless we know an explicit model construction,
we cannot perform further studies for these decay channels.

In this work, we follow the treatment of
Ref.~\cite{Anchordoqui:2008hi} that we consider $A_a$ [the anomalous
$U(1)$ from the $U(3)_a$ stack] as massless and do not consider the
mass mixing effect of this $U(1)$ with others (this field was referred
as $C^0$ in Ref.~\cite{Anchordoqui:2008hi}).  The cases involving the
excitation of the color singlet fields $C^{(1)}$ (as a decay product) is
simpler.  It has a mass $M_s$, and we expect they do not couple to RR
axions.

\subsection{Comments on how to realize right-handed quarks in intersecting brane models}\label{LHRH}

In intersecting brane models, right-handed quarks can be realized as
either open string stretching between the $U(3)_{a}$ stack and another
$U(1)$ stack (let us label this stack as $c$ stack) or open string
stretching between the $U(3)_{a}$ stack and its orientifold image.  In the
former case, right-handed quarks are bifundamental representations
under $U(3)_{a}$ and $U(1)_c$; whereas in the latter case,
right-handed quarks are the antisymmetric representation of $U(3)$.

For the former case, $U(1)_{B}$ is a symmetry remaining unbroken at
the perturbative level in the low-energy effective
theory~\cite{Ibanez:1999it}, but it can be broken by nonperturbative
effects, which are in principle sufficient to suppress proton decay.
For the latter case that (one of the two) right-handed quarks are
realized as an antisymmetric representation of $U(3)$, $U(1)_{B}$ is not
a symmetry.  This is problematic since the leftover global $U(1)$ of
$U(3)$ allows for baryon number violating couplings already at the
lowest order. However, this might be cured by the implementation of
discrete gauge
symmetries~\cite{BerasaluceGonzalez:2011wy,Anastasopoulos:2012zu,Honecker:2013hda}
to forbid the unwanted couplings.

The difference between these two realizations is that
we can have the scattering process $\mathcal{A}(g,q_R,\bar{q}_R,A_c)$
for the former case, but this process is absent for the latter case.
Thus, compared to the latter case, from factorization we know that
the second massive-level right-handed quark excitations
have several more decay channels $Q^{(2)} \to q + A_c$, $Q^{(2)} \to Q^{(1)} + A_c$ and $Q^{(2)} \to A^{(1)}_c + q$.
However as we discussed in the previous subsection,
$A_c$ is not in the physical eigenbasis and we need to rewrite it in
terms of physical mass eigenfields.\footnote{Note that in the
  four-stack SM D-brane construct of  Sec.~\ref{SMDb},
  $A_c$ can 
  either be $B$ or $\tilde B$, the $U(1)_L$ or $U(1)_R$ gauge fields, respectively.}
These are all model-dependent issues.
Unless we focus on a specific D-brane model,
we cannot make any general statements on them.

Similarly for the left-handed quarks, if one uses $Sp(1)$ type
construction, there is no additional $U(1)$ coming from this stack.
Thus, compared to the $U(2)$ type constructions, decay channels
$Q^{(2)} \to q + A_b$, $Q^{(2)} \to Q^{(1)} + A_b$ and $Q^{(2)} \to
A^{(1)}_b + q$ do not exist, since the amplitude
$\mathcal{A}(g,q_R,\bar{q}_R,A_b)$ is absent for $Sp(1)$ cases.

\subsection{Summary of the results}

\begin{table}
\begin{center}
  \begin{tabular}{c|c|c|c|c|c|c|c}
\hline      \hline
    Channel  & $\Gamma_{G^{(2)}}^{ J=3}$ & $\Gamma_{G^{(2)}}^{ J=2}$ &  $\Gamma_{G^{(2)}}^{J=1}$  &  $\Gamma_{Q^{(2)}}^{ J=5/2}$ &  $\Gamma_{Q^{(2)}}^{ J=3/2}$ & $\Gamma_{\tilde Q^{(2)}}^{ J=3/2}$ &  $\Gamma_{Q^{(2)}}^{J=1/2}$ \\ \hline
    $gg$ & $ \frac{N}{21 \sqrt{2}} $  & $\frac{\sqrt{2}N}{15} $  & $ \frac{N}{6 \sqrt{2}} $ & - & - & - & -\\ \hline
    $\xa g$ &  $\frac{117 N}{560 \sqrt{2}}$ &  $\frac{3N}{40 \sqrt{2}}$ &  $\frac{N}{96 \sqrt{2}}$ & - & - & - & -\\ \hline
    $\Phi_\pm g$ &  $\frac{N}{1680 \sqrt{2}}$ &  $\frac{N}{240 \sqrt{2}}$ &  $\frac{17N}{96 \sqrt{2}}$ & - & - & - & -\\ \hline
     $q\bar{q}$ & $\frac{\sqrt{2} N_f}{105} $ &  $\frac{N_f}{120 \sqrt{2}}$ &  0 & - & - & - & - \\ \hline
     $\chi \bar q+\bar \chi q$ &  $\frac{5 N_f}{224 \sqrt{2}}$ &  $\frac{11 N_f}{320 \sqrt{2}}$ &  $\frac{N_f}{96 \sqrt{2}}$ & - & - & - & -\\ \hline
     $a \bar q+\bar a q$ &  $\frac{N_f}{672 \sqrt{2}}$ &  $\frac{7 N_f}{960 \sqrt{2}}$ &  $\frac{N_f}{96 \sqrt{2}}$ & - & - & - & -\\ \hline
  $gq$ &  - & - & - & $\frac{N}{30 \sqrt{2}}$ &  $\frac{3N}{40 \sqrt{2}}$ &  $\frac{N}{12 \sqrt{2}}$ &  $\frac{N}{12 \sqrt{2}}$ \\ \hline
  $\xa q$ &  - & - & - &  $\frac{27N}{1024 \sqrt{2}}$ &  $\frac{11N}{1536\sqrt{2}}$ &  $\frac{25N}{3072 \sqrt{2}}$ &  $\frac{N}{768 \sqrt{2}}$ \\ \hline
  $\Phi_\pm q$ &  - & - & - &  $\frac{N}{1920 \sqrt{2}}$ &  $\frac{N}{320 \sqrt{2}}$ &  $\frac{N}{96 \sqrt{2}}$ &  $\frac{N}{24 \sqrt{2}}$ \\ \hline
  $d q$ &  - & - & - &  $\frac{13N}{5120 \sqrt{2}}$ &  $\frac{37N}{7680  \sqrt{2}}$ &  $\frac{N}{1024 \sqrt{2}}$ &  $\frac{N}{256 \sqrt{2}}$ \\ \hline
  $g \chi$ &  - & - & - &  $\frac{111N}{1280 \sqrt{2}}$ &  $\frac{23N}{640 \sqrt{2}}$ &  $\frac{N}{768 \sqrt{2}}$ &  $\frac{N}{192 \sqrt{2}}$ \\ \hline
  $g a$ &  - & - & - &  $\frac{N}{3840 \sqrt{2}}$ &  $\frac{N}{640 \sqrt{2}}$ &  $\frac{21N}{256 \sqrt{2}}$ &  $\frac{3N}{64 \sqrt{2}}$ \\ \hline
  total &  $\frac{3(6 N+ N_f)}{70 \sqrt{2}}$ &  $\frac{17 N+4 N_f}{80
    \sqrt{2}}$ &  $\frac{17 N+ N_f}{48 \sqrt{2}}$ &  $\frac{115N}{768
    \sqrt{2}}$ &  $\frac{49 N}{384 \sqrt{2}}$ &  $\frac{143N}{768
    \sqrt{2}}$ &  $\frac{35N}{192 \sqrt{2}}$ \\ \hline \hline
    \end{tabular}
    \end{center}
    \caption{The decay widths of $n=2$ string resonances. All of them are to be multiplied by the factor $\frac{g_3^2}{4 \pi}M_s$. For the widths of $G^{(2)}$, we have $N=3$, $N_f=6$. On the other hand, $Q^{(2)}$ can decay into bosons on different stacks. For example, the decay product $G^{(1)}$ of a left-handed $Q^{(2)}$ in \eqref{2nddecayQ2} can be either an $SU(3)$ or an $SU(2)$ boson, but for each channel the width is of the same form (with different coupling constant and $N$). So the widths $\Gamma_{Q^{(2)}}$ in the table should be understood as only for a particular channel, and we need to sum over all possible channels to get the total widths.} \label{decayratetable1}
\end{table}

Using factorization, for the second massive-level bosonic string
states, we have identified a spin-3 field, a spin-2 field, complex
vector fields, which contribute to scattering processes $gg\to gg$ and
$gg\to q\bar{q}$.  For the second massive-level fermionic states, we
have identified a spin-$\frac 5 2$ field, two spin-$\frac 3 2$ fields,
and a spin-$\frac 1 2$ field, which contribute to scattering process
$gq\to gq$.

For a second massive-level color octet, its total decay width includes
\begin{align}
\Gamma_{G^{(2)}} &=
\Gamma(G^{(2)}\to gg) + \Gamma(G^{(2)}\to q\bar q)
+\Gamma(G^{(2)}\to G^{(1)} g) + \Gamma(G^{(2)}\to Q^{(1)}\bar q, \bar{Q}^{(1)} q) \nonumber\\
&+\Gamma(G^{(2)}\to C g) + \Gamma(G^{(2)}\to G^{(1)} C)  + \Gamma(G^{(2)}\to C^{(1)} g)\,.
\end{align}
For the second massive-level color singlets, we have
\begin{align}
\Gamma_{C^{(2)}} &=
\Gamma(C^{(2)}\to gg) + \Gamma(C^{(2)}\to q\bar q)
+\Gamma(C^{(2)}\to G^{(1)} g) + \Gamma(C^{(2)}\to Q^{(1)}\bar q, \bar{Q}^{(1)} q) \nonumber\\
&+\Gamma(C^{(2)}\to C C) + \Gamma(C^{(2)}\to C^{(1)} C)\,.
\end{align}
For the second massive-level excited quarks, we have
\begin{align}
\label{2nddecayQ2}
\Gamma_{Q^{(2)}} &=
\Gamma(Q^{(2)}\to gq) + \Gamma(Q^{(2)}\to G^{(1)} q) + \Gamma(Q^{(2)}\to Q^{(1)} g) \nonumber \\
&+\Gamma(Q^{(2)}\to C q) + \Gamma(Q^{(2)}\to C^{(1)} q) + \cdots \,,
\end{align}
where ``$\cdots$'' denotes model-dependent decay channels for left- or
right-handed excited quarks.  In general left- and right-handed
excited quarks have different decay channels and therefore different
widths. We note that among the amplitudes contributing to the dijet
signal, $Q^{(2)}_L$ only appears as the intermediate state in the
channel of $gq_L\to gq_L$ and similarly $Q^{(2)}_R$ only appears in
$gq_R\to gq_R$. In the phenomenology analysis, we will take the
average of $|\CM(gq_L\to gq_L)|^2$ and $|\CM(gq_R\to gq_R)|^2$ since
the incoming quark is equally likely to be left or right handed.

The total decay widths of the second massive-level string states are
summarized in Table~\ref{decayratetable1}.

\section{String computation of partial decay widths}
\label{s5}

In this section, we will focus on two second massive-level universal
string states: the spin-3 field $\sigma_{\mu\nu\rho}$ and the
spin-2 field $\pi_{\mu\nu}$, computing their decays in various channels.

$N$-point tree-level string amplitudes are obtained by calculating the
$N$-point correlation functions\footnote{ The relevant world sheet
  fields correlation functions can be found
  in Refs.~\cite{Lust:2008qc,Lust:2009pz}.}  of associate vertex operators
inserted on the boundary of the disk world sheet, which read
\begin{equation}
\mathcal{A}=\sum V_{CKG}^{-1}\int(\prod_{i=3}^{N}\mathrm{d}z_{i})\langle V(\mathbf{1})V(\mathbf{2})V(\mathbf{3})\cdots V(\mathbf{N})\rangle,
\end{equation}
where the sum runs over all the cyclic ordering of the $N$ ($N\ge3$)
vertices on the boundary of the disk. The corresponding string vertex
operators are constructed from the fields of the underlying superconformal
field theory and contain explicit Chan--Paton factors. To
cancel the total background ghost charge $-2$ on the disk, we should
choose the vertex operators in the correlator in appropriate ghost
``pictures'' which makes the total ghost number to be $-2$. In
addition, the factor $V_{CKG}$ is defined to be the volume of the
conformal Killing group of the disk after choosing the conformal
gauge, which would be canceled by fixing three vertices and introducing
respective $c$-ghost fields into the vertex operators. Then we integrate
over other $N-3$ points and get the amplitude.

To obtain the decay widths of the second massive-level string states,
we only need to compute the three-point amplitudes, in which all the positions
of the vertex operators on the disk boundary are fixed.

\subsection{Vertex operators of the second massive-level universal string states}

Before we compute the amplitudes, we summarize all the relevant vertex
operators of the zeroth to the second massive-level string states.
For the zeroth-level string, the vertex operator for massless gluon
$g$ (with the polarization vector $\epsilon_{\mu}$) in the $-1$
and 0 ghost picture read, respectively,
\begin{align}
V_{\epsilon^{a}}^{(-1)} & =[T^{a}]_{\alpha_{2}}^{\alpha_{1}}\sqrt{2\alpha'}g_3\epsilon_{\mu}\psi^{\mu}{\rm e}^{-\phi}{\rm e}^{ikX}\,,\\
V_{\epsilon^{a}}^{(0)} & =[T^{a}]_{\alpha_{2}}^{\alpha_{1}}g_3\epsilon_{\mu}(i\partial X^{\mu}+2\alpha'k\cdot\psi\psi^{\mu}){\rm e}^{ikX}\,,
\end{align}
where $\epsilon_{\mu}\cdot k^{\mu}=k^{2}=0$. The Chan--Paton factor
$T^{a}$ indicates the vertex operator is inserted on the segment
of disk boundary on stack $a$, and $\alpha_{1},\alpha_{2}$ represent
the two string ends. Massless quarks originated from brane intersections
are given by
\begin{align}
V_{u_{\beta}^{\alpha}}^{(-\frac{1}{2})} & =[T_{\beta}^{\alpha}]_{\alpha_{1}}^{\beta_{1}}\sqrt{2}\alpha^{\prime\frac{3}{4}}{\rm e}^{\phi_{10}/2}u^{a}S_{a}\Xi^{a\cap b}{\rm e}^{-\phi/2}{\rm e}^{ikX}\,, \label{Valpha}\\
V_{\bar{u}_{\alpha}^{\beta}}^{(-\frac{1}{2})} & =[T_{\alpha}^{\beta}]_{\beta_{1}}^{\alpha_{1}}\sqrt{2}\alpha^{\prime\frac{3}{4}}{\rm e}^{\phi_{10}/2}\bar{u}_{\dot{a}}S^{\dot{a}}\overline{\Xi}^{a\cap b}{\rm e}^{-\phi/2}{\rm e}^{ikX}\,, \label{VPhi}
\end{align}
where the $u^{a},\bar{u}_{\dot{a}}$ satisfy the Dirac equation $u^{a}\slashed k_{a\dot{a}}=\bar{u}_{\dot{a}}\slashed k^{\dot{a}a}=0$,
and $\Xi^{a\cap b}$ is the boundary changing operator~\cite{Lust:2008qc}. These vertex
operators connect two segments of disk boundary, associate to two
stacks of D-branes, with the indices $\alpha_{1}$ and $\beta_{1}$
representing the string ends on the respective stacks.

The first massive-level string states and their properties were
comprehensively studied in Refs.~\cite{Feng:2010yx,Feng:2012bb}.  For the
bosonic sector, we only need the spin-2 field $\alpha_{\mu\nu}$ and
the complex scalar $\Phi_{\pm}$:
\begin{align}
V_{\alpha^{a}}^{(-1)} & =[T^{a}]_{\alpha_{2}}^{\alpha_{1}}g_3\,\alpha_{\mu\nu}i\partial X^{\mu}\psi^{\nu}{\rm e}^{-\phi}{\rm e}^{ikX}\,,\\
V_{\Phi_{\pm}^{a}}^{(-1)} & =[T^{a}]_{\alpha_{2}}^{\alpha_{1}}\frac{g_3}{2}\Big\{\big[(\eta_{\mu\nu}+2\alpha'k_{\mu}k_{\nu})i\partial X^{\mu}\psi^{\nu}+2\alpha'k_{\nu}\partial\psi^{\nu}\big]\nonumber \\
 & \qquad\qquad\pm\frac{i}{6}2\alpha'\varepsilon_{\mu\nu\rho\sigma}\psi^{\mu}\psi^{\nu}\psi^{\rho}k^{\sigma}\Big\}{\rm e}^{-\phi}{\rm e}^{ikX}\,,
\end{align}
where $\alpha_{\mu \nu}$ is symmetric, transverse, and traceless.

The fermionic sector contains spin-$\frac{3}{2}$ and spin-$\frac{1}{2}$
fields which read
\begin{align}
V_{\chi_{\beta}^{\alpha}}^{(-\frac{1}{2})} & =[T_{\beta}^{\alpha}]_{\alpha_{1}}^{\beta_{1}}\alpha^{\prime\frac{1}{4}}{\rm e}^{\phi_{10}/2}\chi_{\mu}^{a}(i\partial X^{\mu}S_{a}-\sqrt 2 \alpha'\slashed k_{a\dot{a}}S^{\mu\dot{a}})\Xi^{a\cap b}{\rm e}^{-\phi/2}{\rm e}^{ikX}\,, \label{Vchi}\\
V_{a_{\beta}^{\alpha}}^{(-\frac{1}{2})} & =[T_{\beta}^{\alpha}]_{\alpha_{1}}^{\beta_{1}}\frac{\alpha^{\prime\frac{3}{4}}}{\sqrt{2}}{\rm e}^{\phi_{10}/2}a^{b}\big[(\sigma_{\mu}\slashed k)_{b}^{\phantom{b}c}i\partial X^{\mu}S_{c}-4\partial S_{b}\big]\Xi^{a\cap b}{\rm e}^{-\phi/2}{\rm e}^{ikX}\,, \label{Va}
\end{align}
which involve the excited spin field $S^{\mu}$ and the derivative
of the standard spin field, cf. Ref.~\cite{Feng:2012bb} for their OPEs.
The spin-$\frac 3 2$ field satisfies $\chi_\mu^a k^\mu =\chi_\mu^a \sigma_{a\dot{a}}^\mu =0$.

Here, all the normalization factors for the vertex operators listed
above were fixed by factorization as worked out in Ref.~\cite{Feng:2010yx} and have also been checked
from supersymmetry transformations in Ref.~\cite{Feng:2012bb}.

For the second massive level, we will focus on two bosonic universal
states $\sigma,\pi$, for which the vertex operators were obtained in Ref.~\cite{Feng:2011qc}
\begin{align}
V_{\sigma^{a}}^{(-1)} & =[T^{a}]_{\alpha_{2}}^{\alpha_{1}}C_{\sigma}\sigma_{\mu\nu\rho}i\partial X^{\mu}i\partial X^{\nu}\psi^{\rho}{\rm e}^{-\phi}{\rm e}^{ikX}\,, \label{Vsigma}\\
V_{\pi^{a}}^{(-1)} & =[T^{a}]_{\alpha_{2}}^{\alpha_{1}}C_{\pi}k^{\lambda}\varepsilon_{\lambda(\mu|\rho\gamma|}\pi_{\phantom{\gamma}\nu)}^{\gamma}(i\partial X^{\mu}i\partial X^{\nu}\psi^{\rho}-4\alpha'\partial\psi^{\mu}\psi^{\nu}\psi^{\rho}){\rm e}^{-\phi}{\rm e}^{ikX}\,, \label{Vpi}
\end{align}
where in $V_{\pi^{a}}^{(-1)}$ we symmetrize only $\mu,\nu$ indices.
$\sigma_{\mu \nu \rho},\pi_{\mu \nu}$ are spin-3 and spin-2 bosonic fields, respectively,
which are both symmetric, transverse, and traceless.
The normalization $C_{\sigma},C_{\pi}$ will be fixed later. Before
we carry out the scattering amplitudes and obtain the partial decay widths of
various channels, we pause and present the construction of helicity
wave functions for higher spin massive bosonic fields.

\subsection{Helicity wave functions for higher spin massive fields}

In this subsection, we first review the helicity wave functions for
spin-1 and spin-2 bosonic fields.  Then we construct the helicity
wave functions for higher spin massive bosonic fields.  The helicity
formalism for massless fields as well as massive fermions is briefly
reviewed in Appendixes~\ref{AppB} and~\ref{AppC}.

\subsubsection{Review of helicity wave functions for spin one and spin two bosonic fields}

\paragraph{Massive spin-1 boson}

A spin-$J$ particle contains $2J+1$ spin degrees of freedom associated
to the eigenstates of $J_{z}$. The choice of the quantization axis
$\vec{z}$ can be handled in an elegant way by decomposing the momentum
$k^{\mu}$ into two arbitrary lightlike reference momenta $p$ and
$q$:
\begin{equation}
k^{\mu}=p^{\mu}+q^{\mu}\,,\qquad k^{2}=-m^{2}=2pq\,,\qquad p^{2}=q^{2}=0\,.
\end{equation}
Then the spin quantization axis is chosen as the direction of $\vec{q}$
in the rest frame. The $2J+1$ spin wave functions depend on $p$
and $q$, while this dependence would drop out in the squared amplitudes
summing over all spin directions.

The massive spin-1 wave functions $\xi_{\mu}$ (transverse, i.e.,
$\xi_{\mu}k^{\mu}=0$) are given by the following polarization vectors
(up to a phase factor) \cite{Novaes:1991ft}:
\begin{flalign}
\text{\ensuremath{\xi}}_{+}^{\mu}(k) & =\frac{1}{\sqrt{2}m}p_{\dot{a}}^{*}\bar{\sigma}^{\mu\dot{a}a}q_{a}\,,\\
\text{\ensuremath{\xi}}_{0}^{\mu}(k) & =\frac{1}{2m}\bar{\sigma}^{\mu\dot{a}a}(p_{\dot{a}}^{*}p_{a}-q_{\dot{a}}^{*}q_{a})\,,\\
\text{\ensuremath{\xi}}_{-}^{\mu}(k) & =-\frac{1}{\sqrt{2}m}q_{\dot{a}}^{*}\bar{\sigma}^{\mu\dot{a}a}p_{a}\,.
\end{flalign}

\paragraph{Massive spin-2 boson}

The wave function (polarization tensor) of massive spin-2 boson
$\text{\ensuremath{\alpha}}^{\mu\nu}$ satisfies the following relations
(symmetric, transverse, traceless), which read
\begin{gather}
\text{\ensuremath{\alpha}}^{\mu\nu}(k,\lambda)=\text{\ensuremath{\alpha}}^{\nu\mu}(k,\lambda)\,,\\
k_{\mu}\text{\ensuremath{\alpha}}^{\mu\nu}(k,\lambda)=0\,,\\
g_{\mu\nu}\text{\ensuremath{\alpha}}^{\mu\nu}(k,\lambda)=0\,,
\end{gather}
where $\lambda$ denotes the helicity of $\text{\ensuremath{\alpha}}^{\mu\nu}$.

An arbitrary four by four tensor has 16 degrees of freedom. The first
condition above reduces the degree of freedom to 10, and the second
and third conditions would further reduce the degrees of freedom 4 and 1,
respectively. Thus, we are left with 5 physical degrees of freedom
as expected. Different helicity states of the spin-2 massive boson
satisfy the relation
\begin{equation}
\text{\ensuremath{\alpha}}^{\mu\nu}(k,+\lambda)=[\text{\ensuremath{\alpha}}^{\mu\nu}(k,-\lambda)]^{\dagger}\,.
\end{equation}

The spin-2 boson helicity wave functions are constructed in Ref.~\cite{Spehler:1991yw},
up to a phase factor,
\begin{flalign}
\text{\ensuremath{\alpha}}^{\mu\nu}(k,+2) & =\frac{1}{2m^{2}}\bar{\sigma}^{\mu\dot{a}a}\bar{\sigma}^{\nu\dot{b}b}p_{\dot{a}}^{*}q_{a}p_{\dot{b}}^{*}q_{b}\,,\nonumber \\
\text{\ensuremath{\alpha}}^{\mu\nu}(k,+1) & =\frac{1}{4m^{2}}\bar{\sigma}^{\mu\dot{a}a}\bar{\sigma}^{\nu\dot{b}b}\left[(p_{\dot{a}}^{*}p_{a}-q_{\dot{a}}^{*}q_{a})p_{\dot{b}}^{*}q_{b}+p_{\dot{a}}^{*}q_{a}(p_{\dot{b}}^{*}p_{b}-q_{\dot{b}}^{*}q_{b})\right]\,,\nonumber \\
\text{\ensuremath{\alpha}}^{\mu\nu}(k,\ 0\ ) & =\frac{1}{2m^{2}\sqrt{6}}\bar{\sigma}^{\mu\dot{a}a}\bar{\sigma}^{\nu\dot{b}b}\left[(p_{\dot{a}}^{*}p_{a}-q_{\dot{a}}^{*}q_{a})(p_{\dot{b}}^{*}p_{b}-q_{\dot{b}}^{*}q_{b})-p_{\dot{a}}^{*}q_{a}q_{\dot{b}}^{*}p_{b}-q_{\dot{a}}^{*}p_{a}p_{\dot{b}}^{*}q_{b}\right]\,,\label{SpinTwoHW}\\
\text{\ensuremath{\alpha}}^{\mu\nu}(k,-1) & =-\frac{1}{4m^{2}}\bar{\sigma}^{\mu\dot{a}a}\bar{\sigma}^{\nu\dot{b}b}\left[(p_{\dot{a}}^{*}p_{a}-q_{\dot{a}}^{*}q_{a})q_{\dot{b}}^{*}p_{b}+q_{\dot{a}}^{*}p_{a}(p_{\dot{b}}^{*}p_{b}-q_{\dot{b}}^{*}q_{b})\right]\,,\nonumber \\
\text{\ensuremath{\alpha}}^{\mu\nu}(k,-2) & =\frac{1}{2m^{2}}\bar{\sigma}^{\mu\dot{a}a}\bar{\sigma}^{\nu\dot{b}b}q_{\dot{a}}^{*}p_{a}q_{\dot{b}}^{*}p_{b}\,.\nonumber
\end{flalign}

\subsubsection{Building helicity wave functions for higher spin massive bosons}

This spin-$n$ massive boson $\Phi_{n}^{\mu_{1}\mu_{2}\cdots\mu_{n}}$
satisfies the following physical state conditions:
\begin{flalign}
\Phi_{n}^{\mu_{1}\mu_{2}\cdots\mu_{n}} & =\frac{1}{n!}\Phi_{n}^{(\mu_{1}\mu_{2}\cdots\mu_{n})}\,,\label{PhiPSC1}\\
k_{\mu_{i}}\Phi_{n}^{\mu_{1}\mu_{2}\cdots\mu_{n}} & =0\,,\label{PhiPSC2}\\
\eta_{\mu_{i}\mu_{j}}\Phi_{n}^{\mu_{1}\mu_{2}\cdots\mu_{n}} & =0\,.\label{PhiPSC3}
\end{flalign}
In four dimensions, the first symmetric condition brings down the
degrees of freedom from $4^{n}$ to $\left(\begin{array}{c}
4+n-1\\
n
\end{array}\right)$, and the transversality and tracelessness eliminate further $\left(\begin{array}{c}
4+n-2\\
n-1
\end{array}\right)$ and $\left(\begin{array}{c}
n\\
2
\end{array}\right)$ conditions. Thus, the $\Phi_{n}^{\mu_{1}\mu_{2}\cdots\mu_{n}}$ has
\[
\left(\begin{array}{c}
4+n-1\\
n
\end{array}\right)-\left(\begin{array}{c}
4+n-2\\
n-1
\end{array}\right)-\left(\begin{array}{c}
n\\
2
\end{array}\right)=2n+1
\]
degrees of freedom.

Thus, the helicity wave function of the highest helicity $j_{z}=+n$
of a spin-$n$ massive boson $\Phi_{n}^{\mu_{1}\mu_{2}\cdots\mu_{n}}$
can be written as, up a phase factor,
\begin{flalign*}
\Phi_{n}^{\mu_{1}\mu_{2}\cdots\mu_{n}}(n,n) & =\frac{1}{(\sqrt{2}m)^{n}}(p_{\dot{a}_{1}}^{*}\bar{\sigma}^{\mu_{1}\dot{a}_{1}a_{1}}q_{a_{1}})(p_{\dot{a}_{2}}^{*}\bar{\sigma}^{\mu_{2}\dot{a}_{2}a_{2}}q_{a_{2}})\cdots(p_{\dot{a}_{n}}^{*}\bar{\sigma}^{\mu_{n}\dot{a}_{n}a_{n}}q_{a_{n}})\,,
\end{flalign*}
and as always, $p^{\mu}+q^{\mu}=k^{\mu}$. Now to obtain all the helicity
wave functions of a spin-$n$ boson $\Phi_{n}^{\mu_{1}\mu_{2}\cdots\mu_{n}}$,
we can make use of angular momentum ladder operators $J_{-}$. By
acting $J_{-}$ on the the highest $J_{z}$ state successively, one
can obtain all the helicity wave functions of $\Phi_{n}^{\mu_{1}\mu_{2}\cdots\mu_{n}}$
using the formula $J_{-}|j,m\rangle=\sqrt{(j+m)(j-m+1)}|j,m-1\rangle$.
Based on spin-1 gauge boson wave functions, we have
\begin{gather}
J_{-}(p_{\dot{a}}^{*}\bar{\sigma}^{\mu\dot{a}a}q_{a})=(p_{\dot{a}}^{*}\bar{\sigma}^{\mu\dot{a}a}p_{a}-q_{\dot{a}}^{*}\bar{\sigma}^{\mu\dot{a}a}q_{a})\,,\\
J_{-}(p_{\dot{a}}^{*}\bar{\sigma}^{\mu\dot{a}a}p_{a}-q_{\dot{a}}^{*}\bar{\sigma}^{\mu\dot{a}a}q_{a})=-2q_{\dot{a}}^{*}\bar{\sigma}^{\mu\dot{a}a}p_{a}\,.
\end{gather}
More specifically, we have the following relations:
\begin{gather}
J_{-}p_{\dot{a}}^{*}=-q_{\dot{a}}^{*}\,,\qquad J_{-}p_{a}=0\,,\\
J_{-}q_{\dot{a}}^{*}=0\,,\qquad J_{-}q_{a}=p_{a}\,.
\end{gather}
One could write these relations in a simpler form as
\begin{equation}
J_{-}=p_{a}\frac{\partial}{\partial q_{a}}-q_{\dot{a}}^{*}\frac{\partial}{\partial p_{\dot{a}}^{*}}\,.
\end{equation}
These formulas allow us to get all the wave functions of an arbitrary
spin massive boson. By applying the $J_{-}$ operator on $\Phi_{n}^{\mu_{1}\mu_{2}\cdots\mu_{n}}(n,n)$
successively, one can obtain wave functions of all the helicities.

Indeed, this $J_{-}$ operator is extremely useful in the computation
of the helicity amplitudes involving massive states. Since the wave
function of the highest helicity state
$\Phi_{n}^{\mu_{1}\mu_{2}\cdots\mu_{n}}(n,n)$ has the simplest form,
one could relatively easily obtain the helicity amplitude
$\mathcal{A}[\Phi_{n}(n,n),\cdots]$ that
$\Phi_{n}^{\mu_{1}\mu_{2}\cdots\mu_{n}}(n,n)$ interacts with other
states, and it is usually in a simple form.  One could then apply
$J_{-}$ successively to the amplitude
$\mathcal{A}[\Phi_{n}(n,n),\cdots]$ to obtain all the helicity
amplitudes $\mathcal{A}[\Phi_{n}(n,m),\cdots]$, which is much simpler
than plugging in explicit forms of the $\Phi_{n}$ helicity wave
functions of lower $j_z$.\footnote{ As a simple example, we consider
  the amplitudes Eqs.~\eqref{b3gh} obtained in
  Ref.~\cite{Feng:2010yx}.  We have
\begin{equation*}
J_- \mathscr{A}\left[ \alpha(2,+2), +,+,- \right] = \sqrt{(2+2)(2-2+1)} \mathscr{A}\left[ \alpha(2,+1), +,+,- \right]\,,
\end{equation*}
and thus
\begin{align*}
\mathscr{A}\left[ \alpha(2,+1), +,+,- \right] &= \frac{1}{2} J_- \mathscr{A}\left[ \alpha(2,+2), +,+,- \right] \\
& = \frac{1}{2} \times \frac{4}{2 \sqrt 2} \frac{\langle p4\rangle^3 \langle 4q\rangle  }{\langle 23\rangle\langle 34\rangle\langle 42\rangle}\,,
\end{align*}
which just reproduce the desired result.
Using this method, one could then check all the results in Ref.~\cite{Feng:2010yx},
where all the helicity amplitudes were computed using the explicit forms of
the helicity wave functions in different $j_z$, for example, Eqs.~\eqref{SpinTwoHW}.}

There is another way of constructing the helicity wave functions of
a spin-$n$ massive boson, that we can treat the spin-$n$ boson as
a spin-$(n-1)$ and a spin-1 boson coupling. Thus, given the helicity
wave function of a spin-$(n-1)$ boson, one can write down an arbitrary
$J_{z}=m$ state of the spin-$n$ boson as
\begin{flalign}
\Phi_{n}^{\mu_{1}\mu_{2}\cdots\mu_{n}}(n,m) & =\langle n-1,m-1;1,+1|n,m\rangle\Phi_{n-1}^{\mu_{1}\mu_{2}\cdots\mu_{n-1}}(n-1,m-1)\text{\ensuremath{\xi}}_{+}^{\mu_{n}}\nonumber \\
 & \,+\langle n-1,m+1;1,-1|n,m\rangle\Phi_{n-1}^{\mu_{1}\mu_{2}\cdots\mu_{n-1}}(n-1,m+1)\text{\ensuremath{\xi}}_{-}^{\mu_{n}}\nonumber \\
 & \,+\langle n-1,m;1,0|n,m\rangle\Phi_{n-1}^{\mu_{1}\mu_{2}\cdots\mu_{n-1}}(n-1,m)\text{\ensuremath{\xi}}_{0}^{\mu_{n}}\,,\label{n1coupling}
\end{flalign}
where the CG coefficients read
\begin{equation}
\begin{cases}
\langle n-1,m-1;1,+1|n,m\rangle & =\sqrt{\frac{(n+m)(n+m+1)}{(2n+1)(2n+2)}}\,,\\
\quad\ \langle n-1,m;1,0|n,m\rangle & =\sqrt{\frac{(n-m+1)(n+m+1)}{(n+1)(2n+1)}}\\
\langle n-1,m+1;1,-1|n,m\rangle & =\sqrt{\frac{(n-m)(n-m+1)}{(2n+1)(2n+2)}}\,.
\end{cases}\,,\label{CGcoef}
\end{equation}
Thus Eq.~\eqref{n1coupling} can be written as
\begin{align}
\Phi_{n}^{\mu_{1}\mu_{2}\cdots\mu_{n}}(n,m) & =\sqrt{\tfrac{(n+m)(n+m+1)}{(2n+1)(2n+2)}}\Phi_{n-1}^{\mu_{1}\mu_{2}\cdots\mu_{n-1}}(n-1,m-1)\text{\ensuremath{\xi}}_{+}^{\mu_{n}}\nonumber \\
 & \,+\sqrt{\tfrac{(n-m)(n-m+1)}{(2n+1)(2n+2)}}\Phi_{n-1}^{\mu_{1}\mu_{2}\cdots\mu_{n-1}}(n-1,m+1)\text{\ensuremath{\xi}}_{-}^{\mu_{n}}\nonumber \\
 & \,+\sqrt{\tfrac{(n-m+1)(n+m+1)}{(n+1)(2n+1)}}\Phi_{n-1}^{\mu_{1}\mu_{2}\cdots\mu_{n-1}}(n-1,m)\text{\ensuremath{\xi}}_{0}^{\mu_{n}}\,.
\end{align}

Indeed, the helicity wave function of an arbitrary $j_{z}$ state
of $\Phi_{n}$ can be written in a general form
\begin{align}
\Phi_{n}^{\mu_{1}\mu_{2}\cdots\mu_{n}}(n,m) & =\Big[\sum_{\alpha}\frac{2^{n-m-2\alpha}\cdot n!}{\alpha!(m+\alpha)!(n-2\alpha-m)!}(2m^{2})^{n}\Big]^{-\frac{1}{2}}\times\nonumber \\
 & \sum_{\alpha}\Big\{\prod_{_{i}}(p^{*}\bar{\sigma}^{(\underline{\mu_{i}}}q)^{m+\alpha}\prod_{_{j}}(-q^{*}\bar{\sigma}^{\underline{\mu_{j}}}p)^{\alpha}\prod_{k}[\bar{\sigma}^{\underline{\mu_{k}})}(p^{*}p-q^{*}q)]^{n-m-2\alpha}\Big\}\,,
\end{align}
where $m\geq0$, the sum over $\alpha$ is over such values that the
factorials are non-negative, and we symmetrize all the spacetime indices
$\mu_{i},\mu_{j},\mu_{k}$. We have omitted all the spinor indices,
e.g., $p^{*}\bar{\sigma}^{\mu}q\equiv p_{\dot{a}}^{*}\bar{\sigma}^{\mu\dot{a}a}q_{a}$.
These wave functions satisfy physical state conditions (symmetric,
transverse and traceless) Eqs.~\eqref{PhiPSC1} --\eqref{PhiPSC3}.
The helicity wave functions of $\Phi_{n}^{\mu_{1}\mu_{2}\cdots\mu_{n}}(n,-m)$
can be easily obtained by
\begin{equation}
\Phi_{n}^{\mu_{1}\mu_{2}\cdots\mu_{n}}(n,-m)=\Phi_{n}^{\mu_{1}\mu_{2}\cdots\mu_{n}}(n,m)^{\dagger}\,.
\end{equation}

We now write down the helicity wave functions for the massive spin-3
boson, which we will need for further calculations:
\begin{flalign}
\Phi_{3}^{\mu\nu\rho}(k,+3) & =\frac{1}{(\sqrt{2}m)^{3}}\bar{\sigma}^{\mu\dot{a}a}\bar{\sigma}^{\nu\dot{b}b}\bar{\sigma}^{\rho\dot{c}c}p_{\dot{a}}^{*}q_{a}p_{\dot{b}}^{*}q_{b}p_{\dot{c}}^{*}q_{c}\,, \label{Spin3HW}\\
\Phi_{3}^{\mu\nu\rho}(k,+2) & =\frac{\bar{\sigma}^{\mu\dot{a}a}\bar{\sigma}^{\nu\dot{b}b}\bar{\sigma}^{\rho\dot{c}c}}{\sqrt{6}(\sqrt{2}m)^{3}}\Big[p_{\dot{a}}^{*}q_{a}p_{\dot{b}}^{*}q_{b}(p_{\dot{c}}^{*}p_{c}-q_{\dot{c}}^{*}q_{c})+p_{\dot{a}}^{*}q_{a}(p_{\dot{b}}^{*}p_{b}-q_{\dot{b}}^{*}q_{b})p_{\dot{c}}^{*}q_{c}+(p_{\dot{a}}^{*}p_{a}-q_{\dot{a}}^{*}q_{a})p_{\dot{b}}^{*}q_{b}p_{\dot{c}}^{*}q_{c}\Big]\,,\nonumber \\
\Phi_{3}^{\mu\nu\rho}(k,+1) & =\frac{\bar{\sigma}^{\mu\dot{a}a}\bar{\sigma}^{\nu\dot{b}b}\bar{\sigma}^{\rho\dot{c}c}}{\sqrt{15}(\sqrt{2}m)^{3}}\Big[p_{\dot{a}}^{*}q_{a}(p_{\dot{b}}^{*}p_{b}-q_{\dot{b}}^{*}q_{b})(p_{\dot{c}}^{*}p_{c}-q_{\dot{c}}^{*}q_{c})+(p_{\dot{a}}^{*}p_{a}-q_{\dot{a}}^{*}q_{a})p_{\dot{b}}^{*}q_{b}(p_{\dot{c}}^{*}p_{c}-q_{\dot{c}}^{*}q_{c})\nonumber \\
 & \quad+(p_{\dot{a}}^{*}p_{a}-q_{\dot{a}}^{*}q_{a})(p_{\dot{b}}^{*}p_{b}-q_{\dot{b}}^{*}q_{b})p_{\dot{c}}^{*}q_{c}-p_{\dot{a}}^{*}q_{a}p_{\dot{b}}^{*}q_{b}q_{\dot{c}}^{*}p_{c}-p_{\dot{a}}^{*}q_{a}q_{\dot{b}}^{*}p_{b}p_{\dot{c}}^{*}q_{c}-q_{\dot{a}}^{*}p_{a}p_{\dot{b}}^{*}q_{b}p_{\dot{c}}^{*}q_{c}\Big]\,,\nonumber \\
\Phi_{3}^{\mu\nu\rho}(k,\ 0\ ) & =\frac{\bar{\sigma}^{\mu\dot{a}a}\bar{\sigma}^{\nu\dot{b}b}\bar{\sigma}^{\rho\dot{c}c}}{2\sqrt{5}(\sqrt{2}m)^{3}}\Big[(p_{\dot{a}}^{*}p_{a}-q_{\dot{a}}^{*}q_{a})(p_{\dot{b}}^{*}p_{b}-q_{\dot{b}}^{*}q_{b})(p_{\dot{c}}^{*}p_{c}-q_{\dot{c}}^{*}q_{c})-p_{\dot{a}}^{*}q_{a}q_{\dot{b}}^{*}p_{b}(p_{\dot{c}}^{*}p_{c}-q_{\dot{c}}^{*}q_{c})\nonumber \\
 & \qquad\qquad\qquad-q_{\dot{a}}^{*}p_{a}p_{\dot{b}}^{*}q_{b}(p_{\dot{c}}^{*}p_{c}-q_{\dot{c}}^{*}q_{c})-p_{\dot{a}}^{*}q_{a}(p_{\dot{b}}^{*}p_{b}-q_{\dot{b}}^{*}q_{b})q_{\dot{c}}^{*}p_{c}-q_{\dot{a}}^{*}p_{a}(p_{\dot{b}}^{*}p_{b}-q_{\dot{b}}^{*}q_{b})p_{\dot{c}}^{*}q_{c}\nonumber \\
 & \qquad\qquad\qquad-(p_{\dot{a}}^{*}p_{a}-q_{\dot{a}}^{*}q_{a})p_{\dot{b}}^{*}q_{b}q_{\dot{c}}^{*}p_{c}-(p_{\dot{a}}^{*}p_{a}-q_{\dot{a}}^{*}q_{a})q_{\dot{b}}^{*}p_{b}p_{\dot{c}}^{*}q_{c}\Big]\,,\nonumber \\
\Phi_{3}^{\mu\nu\rho}(k,-1) & =-\frac{\bar{\sigma}^{\mu\dot{a}a}\bar{\sigma}^{\nu\dot{b}b}\bar{\sigma}^{\rho\dot{c}c}}{\sqrt{15}(\sqrt{2}m)^{3}}\Big[(p_{\dot{a}}^{*}p_{a}-q_{\dot{a}}^{*}q_{a})(p_{\dot{b}}^{*}p_{b}-q_{\dot{b}}^{*}q_{b})q_{\dot{c}}^{*}p_{c}+(p_{\dot{a}}^{*}p_{a}-q_{\dot{a}}^{*}q_{a})q_{\dot{b}}^{*}p_{b}(p_{\dot{c}}^{*}p_{c}-q_{\dot{c}}^{*}q_{c})\nonumber \\
 & \quad+q_{\dot{a}}^{*}p_{a}(p_{\dot{b}}^{*}p_{b}-q_{\dot{b}}^{*}q_{b})(p_{\dot{c}}^{*}p_{c}-q_{\dot{c}}^{*}q_{c})-p_{\dot{a}}^{*}q_{a}q_{\dot{b}}^{*}p_{b}q_{\dot{c}}^{*}p_{c}-q_{\dot{a}}^{*}p_{a}p_{\dot{b}}^{*}q_{b}q_{\dot{c}}^{*}p_{c}-q_{\dot{a}}^{*}p_{a}q_{\dot{b}}^{*}p_{b}p_{\dot{c}}^{*}q_{c}\Big]\,,\nonumber \\
\Phi_{3}^{\mu\nu\rho}(k,-2) & =\frac{\bar{\sigma}^{\mu\dot{a}a}\bar{\sigma}^{\nu\dot{b}b}\bar{\sigma}^{\rho\dot{c}c}}{\sqrt{6}(\sqrt{2}m)^{3}}\Big[(p_{\dot{a}}^{*}p_{a}-q_{\dot{a}}^{*}q_{a})q_{\dot{b}}^{*}p_{b}q_{\dot{c}}^{*}p_{c}+q_{\dot{a}}^{*}p_{a}(p_{\dot{b}}^{*}p_{b}-q_{\dot{b}}^{*}q_{b})q_{\dot{c}}^{*}p_{c}+q_{\dot{a}}^{*}p_{a}q_{\dot{b}}^{*}p_{b}(p_{\dot{c}}^{*}p_{c}-q_{\dot{c}}^{*}q_{c})\Big]\,,\nonumber \\
\Phi_{3}^{\mu\nu\rho}(k,-3) & =-\frac{1}{(\sqrt{2}m)^{3}}\bar{\sigma}^{\mu\dot{a}a}\bar{\sigma}^{\nu\dot{b}b}\bar{\sigma}^{\rho\dot{c}c}q_{\dot{a}}^{*}p_{a}q_{\dot{b}}^{*}p_{b}q_{\dot{c}}^{*}p_{c}\,.\nonumber
\end{flalign}

\subsection{Decay of the second massive-level string states}

We need to first fix the normalization of vertex operators for $\sigma_{\mu\nu\rho}$
and $\pi_{\mu\nu}$. To this end, we compute the amplitude that $\sigma_{\mu\nu\rho},\pi_{\mu\nu}$
decay into two massless gluons, and the result reads
\begin{align}
\mathcal{A}(\sigma_{1},\epsilon_{2},\epsilon_{3}) & ={\rm Tr}(T^{a_{1}}[T^{a_{2}},T^{a_{3}}])C_{\sigma}g_3^{2}C_{D_{2}}(2\alpha')^{\frac{7}{2}}\sigma_{\mu\nu\rho}\Big[\frac{1}{\alpha'}\epsilon_{2}^{\mu}\epsilon_{3}^{\nu}k_{2}^{\rho}+k_{2}^{\mu}k_{2}^{\nu}k_{2}^{\rho}(\epsilon_{2}\cdot\epsilon_{3})\nonumber \\
 & \qquad\qquad\qquad+k_{2}^{\mu}k_{2}^{\nu}\epsilon_{3}^{\rho}(\epsilon_{2}\cdot k_{3})-k_{2}^{\mu}k_{2}^{\nu}\epsilon_{2}^{\rho}(\epsilon_{3}\cdot k_{2})\Big]\,.
\end{align}
Applying the helicity formalism, we obtain
\begin{equation}
\mathcal{A}\big[\sigma_{1}(+2),\epsilon_{2}^{+},\epsilon_{3}^{-}\big]=\frac{8}{\sqrt{3}}C_{\sigma}{\rm Tr}(T^{a_{1}}[T^{a_{2}},T^{a_{3}}])\,.
\end{equation}
Extracting the second-level pole information from the Veneziano amplitude $\mathcal{A}(g,g,g,g)$,
we obtain (up to a phase factor)
\begin{equation}
\mathcal{A}(\sigma_{1},\epsilon_{2}^{+},\epsilon_{3}^{-})=\frac{2g_3}{\sqrt{3\alpha'}}f_{a_{1}a_{2}a_{3}}\,.
\end{equation}
Thus we obtain $C_{\sigma}=g_3/2\sqrt{\alpha'}$, where we have used
$C_{D_{2}}=1/(g_3^{2}\alpha^{\prime2})$ and Eq.~\eqref{fabc}.

For $\pi_{\mu\nu}$ decay to two massless gluons, we have
\begin{align}
\mathcal{A}(\pi_{1},\epsilon_{2},\epsilon_{3}) & ={\rm Tr}(T^{a_{1}}[T^{a_{2}},T^{a_{3}}])C_{\pi}g_3^{2}C_{D_{2}}(2\alpha')^{\frac{3}{2}}k_{1}^{\lambda}\varepsilon_{\lambda(\mu|\rho\gamma|}\pi_{\phantom{\gamma}\nu)}^{\gamma}\Big[2\epsilon_{2}^{\mu}k_{2}^{\nu}\epsilon_{3}^{\rho}+2\epsilon_{3}^{\mu}k_{2}^{\nu}\epsilon_{2}^{\rho}-2\epsilon_{2}^{\mu}\epsilon_{3}^{\nu}k_{2}^{\rho}\nonumber \\
 & \qquad\quad-2\alpha'k_{2}^{\mu}k_{2}^{\nu}\epsilon_{2}^{\rho}(\epsilon_{3}\cdot k_{2})+2\alpha'k_{2}^{\mu}k_{2}^{\nu}\epsilon_{3}^{\rho}(\epsilon_{2}\cdot k_{3})-2\alpha'k_{2}^{\mu}k_{2}^{\nu}k_{3}^{\rho}(\epsilon_{2}\cdot\epsilon_{3})\Big]\,.
\end{align}
Similarly, by applying the helicity formalism, we match the helicity
amplitude with the amplitude we extract from Veneziano amplitude, and
we obtain $C_{\pi}=g_3/4\sqrt{3}$.

The partial decay widths of second massive-level string states to two
massless string states were already obtained in
Refs.~\cite{Dong:2010jt,Hashi:2011cz}.  We are now the most interested
in computing the partial decay widths of a second massive-level string
states decay into one first massive-level string state plus a massless
one.

\subsubsection{Partial decay widths of the spin-3 state $\sigma_{\mu\nu\rho}$}

We now focus on the spin-3 bosonic string state $\sigma_{\mu\nu\rho}$.
It has four possible decay channels for which the final states consist of
one first massive-level string state and one massless string state,
which read $\sigma\to\alpha+g,\sigma\to\Phi_{\pm}+g,\sigma\to\bar{\chi}+u,\sigma\to\bar{a}+u$
(the decay widths of $\sigma\to\chi+\bar{u},\sigma\to a+\bar{u}$
are the same as the last two channels). Straightforward computation
gives
\begin{align}
\mathcal{A}(\sigma_{1},\alpha_{2},\epsilon_{3}) & ={\rm Tr}(T^{a_{1}}\{T^{a_{2}},T^{a_{3}}\})\frac{2g_3}{\sqrt{\alpha'}}\sigma_{\mu\nu\rho}\Big\{(2\alpha')^{2}\big[k_{3}^{\mu}k_{3}^{\nu}\epsilon_{3}^{\rho}\alpha_{\gamma\zeta}k_{3}^{\gamma}k_{3}^{\zeta}-k_{3}^{\mu}k_{3}^{\nu}k_{3}^{\rho}\alpha_{\gamma\zeta}\epsilon_{3}^{\gamma}k_{3}^{\zeta}\nonumber \\
 & \quad\qquad-k_{3}^{\mu}k_{3}^{\nu}\alpha^{\rho\gamma}k_{3\gamma}(\epsilon_{3}\cdot k_{2})\big]+(2\alpha')\big[3k_{3}^{\mu}k_{3}^{\nu}\alpha^{\rho\gamma}\epsilon_{3\gamma}-4k_{3}^{\mu}\epsilon_{3}^{\nu}\alpha^{\rho\gamma}k_{3\gamma}\nonumber \\
 & \quad\qquad+2k_{3}^{\mu}\alpha^{\nu\rho}(\epsilon_{3}\cdot k_{2})\big]+2\alpha^{\mu\nu}\epsilon_{3}^{\rho}\Big\}\,,\\
\mathcal{A}(\sigma_{1},\Phi_{2\pm},\epsilon_{3}) & ={\rm Tr}(T^{a_{1}}\{T^{a_{2}},T^{a_{3}}\})2g_3\sqrt{\alpha}\sigma_{\mu\nu\rho}\big[-2\alpha'k_{3}^{\mu}k_{3}^{\nu}k_{3}^{\rho}(\epsilon_{3}\cdot k_{2})-k_{3}^{\mu}k_{3}^{\nu}\epsilon_{3}^{\rho}\nonumber \\
 & \qquad\qquad\qquad\qquad\qquad\qquad\qquad\pm i2\alpha'k_{3}^{\mu}k_{3}^{\nu}\varepsilon^{\rho\gamma\zeta\lambda}\epsilon_{3\gamma}k_{3\zeta}k_{2\lambda}\big]\,.
\end{align}
We place the second massive-level string state, the first
massive-level string state, and the massless string at positions 1, 2,
and 3
with corresponding momentum $k_{1},$ $k_{2},$ and $k_{3}$, and thus we have
\begin{equation}
k_1^2 = -\frac{2}{\alpha'}\,,\qquad k_2^2 = -\frac{1}{\alpha'}\,,\qquad k_3^2=0\,. \label{MSC}
\end{equation}
To obtain the partial decay widths of the above channels, again we apply the helicity formalism.
In principle, by plugging in directly the helicity wave functions of the fields
participating in the processes, e.g. Eqs.~\eqref{SpinTwoHW} and \eqref{Spin3HW},
we could obtain the helicity amplitudes.
Then by summing over their squares we can achieve the final results.
However, special treatment is needed here.
For example, for the amplitude $\mathcal{A}(\sigma_{1},\alpha_{2},\epsilon_{3})$,
$\sigma$ has 7 degrees of freedom, $\alpha$ has 5, and $\epsilon$
has 2. Thus we need to compute total $7\times5\times2=70$ helicity
amplitudes, and the computation would be very tedious. First of all, we observe that
\begin{equation}
\Gamma(\sigma_{1}\to\alpha_{2}+\epsilon_{3}^{+})=\Gamma(\sigma_{1}\to\alpha_{2}+\epsilon_{3}^{-})\,,
\end{equation}
since
\begin{equation}
\mathcal{A}\big[\sigma_{1}(-n),\alpha_{2}(-m),\epsilon_{3}^{-}\big]=\mathcal{A}\big[\sigma_{1}(n),\alpha_{2}(m),\epsilon_{3}^{+}\big]^\dagger \,.
\end{equation}
This would reduce the total number of the amplitudes we need to
compute by half.  In addition, as we mentioned, the helicity wave
functions of massive bosonic fields are built by decomposing their
momentum into two lightlike momenta $k^{\mu}\to p^{\mu}+q^{\mu}$,
and the spin axis of the field aligns to the $\vec{q}$
direction. Hence if we align the spin axes of all the scattering
fields to one same direction, we only need to compute very few
helicity amplitudes, and the others should vanish automatically because of
the angular momentum conservation.

The most clever choice of reference momenta read\footnote{
This choice can be easily generated to more general cases: (1) Assuming
the three particles are all incoming ($k_{1}+k_{2}+k_{3}=0$) with
corresponding momentum $k_{1}^{2}=-M_{1}^{2},k_{2}^{2}=-M_{2}^{2},k_{3}^{2}=0$,
we can choose the  reference momenta
\[
p_{1}^{\mu}=-r\,,\quad q_{1}=\frac{-M_{1}^{2}}{M_{1}^{2}-M_{2}^{2}}k_{3}\,,\quad p_{2}=r\,,\quad q_{2}=\frac{M_{2}^{2}}{M_{1}^{2}-M_{2}^{2}}k_{3}\,,
\]
where $r^{2}=0$ and $r\cdot k_{3}=(M_{2}^{2}-M_{1}^{2})/2$; (2)
if all the three incoming particles are massive with corresponding
momentum $k_{1}^{2}=-M_{1}^{2},k_{2}^{2}=-M_{2}^{2},k_{3}^{2}=-M_{3}^{2}$,
we can choose the reference momenta
\begin{equation}
p_{2}=\alpha p_{1}\,,\quad q_{2}=\beta q_{1}\,,\quad p_{3}=(-\alpha-1)p_{1}\,,\quad q_{3}=(-\beta-1)q_{1}\,,
\end{equation}
where $p_{1}\cdot q_{1}=-M_{1}^{2}/2$, and the coefficients
\begin{align*}
\alpha & =\frac{M_{3}^{2}-M_{1}^{2}-M_{2}^{2}\pm\sqrt{(M_{1}^{2}+M_{2}^{2}-M_{3}^{2})^{2}-4M_{1}^{2}M_{2}^{2}}}{2M_{1}^{2}}\,,\\
\beta & =\frac{2M_{2}^{2}}{M_{3}^{2}-M_{1}^{2}-M_{2}^{2}\pm\sqrt{(M_{1}^{2}+M_{2}^{2}-M_{3}^{2})^{2}-4M_{1}^{2}M_{2}^{2}}}\,.
\end{align*}
With these choices, the spin axes of the three particles align to
the same direction, and thus the computation of helicity amplitude
will be dramatically simplified.
}
\begin{equation}
p_{1}^{\mu}=-r^{\mu}\,,\quad q_{1}^{\mu}=-2k_{3}^{\mu}\,,\quad p_{2}^{\mu}=r^{\mu}\,,\quad q_{2}^{\mu}=k_{3}^{\mu}\,,
\label{refpq}
\end{equation}
where $r$ is the reference momentum for the massless gluon $\epsilon_{3}(k_{3})$
with $r^{2}=0$. It can be easily verified that
\begin{gather}
k_{1}+k_{2}+k_{3}=p_{1}+q_{1}+p_{2}+q_{2}+k_{3}=0\,,\\
(p_{1}+q_{1})^{2}=2(p_{2}+q_{2})^{2}\,.
\end{gather}
Then by using the mass shell condition Eq.~\eqref{MSC}, we fix the
reference momentum $r$ as $r\cdot k_{3}=-1/(2\alpha')$.  This
particular choice of reference momenta not only simplifies the
computation dramatically but also aligns the spins of all the
interacting particles in one same direction (the direction of
$\vec{k}_3$), and thus we are expecting the results we obtained from
this section to match exactly with the results we obtained in the
previous section using factorization.

Using massive helicity wave functions and the above choice of
reference momenta, we compute the helicity amplitudes of
$\mathcal{A}(\sigma_{1},\alpha_{2},\epsilon_{3}^{+})$.  Only five
survive, which read
\begin{align}
\mathcal{A}\big[\sigma_{1}(-3),\alpha_{2}(+2),\epsilon_{3}^{+}\big] & =\frac{8g_3}{\sqrt{\alpha'}}d_{a_{1}a_{2}a_{3}}\,, \label{3to21.1}\\
\mathcal{A}\big[\sigma_{1}(-2),\alpha_{2}(+1),\epsilon_{3}^{+}\big] & =\frac{8g_3}{\sqrt{3\alpha'}}d_{a_{1}a_{2}a_{3}}\,,\\
\mathcal{A}\big[\sigma_{1}(-1),\alpha_{2}(\ 0\ ),\epsilon_{3}^{+}\big] & =\frac{4\sqrt{2}g_3}{\sqrt{5\alpha'}}d_{a_{1}a_{2}a_{3}}\,,\\
\mathcal{A}\big[\sigma_{1}(\ 0\ ),\alpha_{2}(-1),\epsilon_{3}^{+}\big] & =\frac{4g_3}{\sqrt{10\alpha'}}d_{a_{1}a_{2}a_{3}}\,,\\
\mathcal{A}\big[\sigma_{1}(+1),\alpha_{2}(-2),\epsilon_{3}^{+}\big] & =\frac{2g_3}{\sqrt{15\alpha'}}d_{a_{1}a_{2}a_{3}}\,.
\end{align}
All other helicity amplitudes are checked to vanish. These results
match exactly with the results obtained from factorization Eqs.~\eqref{Spin3toalphag1}-\eqref{Spin3toalphag5}, as expected.

With the same choice of the reference momenta, for
$\mathcal{A}(\sigma_{1},\Phi_{2\pm},\epsilon_{3})$, we obtain
\begin{align}
\mathcal{A}\big[\sigma_{1}(-1),\Phi_{2+},\epsilon_{3}^{+}\big] & =\frac{2g_3}{\sqrt{15\alpha'}}d_{a_{1}a_{2}a_{3}}\,,\\
\mathcal{A}\big[\sigma_{1}(-1),\Phi_{2-},\epsilon_{3}^{+}\big] & =0\,,
\end{align}
which match Eq.~\eqref{Spin3toPhig} exactly.

For the decay channels that final states being fermions. The scattering
amplitudes read,
\begin{align}
\mathcal{A}(\sigma_{1},\bar{\chi}_{2},u_{3}) & =T_{\alpha_{2}\alpha_{3}}^{a}g_3\sqrt{\alpha'}\sigma_{\mu\nu\rho}\big[2\alpha'k_{3}^{\mu}k_{3}^{\nu}(u_{3}^{a}\sigma_{a\dot{a}}^{\rho}\bar{\chi}_{2}^{\lambda\dot{a}}k_{1\lambda}-u_{3}^{a}\slashed k_{1a\dot{a}}\bar{\chi}_{2}^{\rho\dot{a}})+2k_{3}^{\mu}u_{3}^{a}\sigma_{a\dot{a}}^{\nu}\bar{\chi}_{2}^{\rho\dot{a}}\big]\,,\\
\mathcal{A}(\sigma_{1},\bar{a}_{2},u_{3}) & =T_{\alpha_{2}\alpha_{3}}^{a}\sqrt{2}g_3\alpha'\sigma_{\mu\nu\rho} (-2\alpha'k_{3}^{\mu}k_{3}^{\nu}k_{3}^{\rho}u_{3}^{b}\slashed k_{2b\dot{b}}\bar{a}_{2}^{\dot{b}}+k_{3}^{\mu}k_{3}^{\nu}u_{3}^{b}\sigma_{b\dot{b}}^{\rho}\bar{a}_{2}^{\dot{b}} )\,.
\end{align}
For scattering amplitudes involving two fermionic fields, a factor of $\tilde{C}_{D_{2}}$ would appear and
we have used $\tilde{C}_{D_{2}}={\rm e}^{-\phi_{10}}/(2\alpha^{\prime2})$~\cite{Feng:2010yx}.

For the fermionic decay channels, again we align the spin axes of the
three interacting states into the direction of $\vec{k}_3$.  We will
use exactly the same reference momenta Eq.~\eqref{refpq} as we did
for the bosonic decay channels.  Here we also need to introduce an
additional reference momentum $r$ with $r\cdot k_{3}=-1/(2\alpha')$.
Using the massive fermion helicity wave functions summarized in
Appendix~\ref{AppC}, we obtain the following helicity amplitudes:
\begin{align}
\mathcal{A}\big[\sigma_{1}(+2),\bar{\chi}_{2}(-\tfrac{3}{2}),u_{3}(-\tfrac{1}{2})\big] & =\frac{g_3}{\sqrt{3\alpha'}}T_{\alpha_{2}\alpha_{3}}^{a}\,,\\
\mathcal{A}\big[\sigma_{1}(+1),\bar{\chi}_{2}(-\tfrac{1}{2}),u_{3}(-\tfrac{1}{2})\big] & =\frac{g_3}{\sqrt{5\alpha'}}T_{\alpha_{2}\alpha_{3}}^{a}\,,\\
\mathcal{A}\big[\sigma_{1}(\ 0\ ),\bar{\chi}_{2}(+\tfrac{1}{2}),u_{3}(-\tfrac{1}{2})\big] & =\frac{\sqrt{3}g_3}{2\sqrt{10\alpha'}}T_{\alpha_{2}\alpha_{3}}^{a}\,,\\
\mathcal{A}\big[\sigma_{1}(-1),\bar{\chi}_{2}(+\tfrac{3}{2}),u_{3}(-\tfrac{1}{2})\big] & =\frac{g_3}{2\sqrt{15\alpha'}}T_{\alpha_{2}\alpha_{3}}^{a}\,,
\end{align}
and
\begin{align}
\mathcal{A}\big[\sigma_{1}(+1),\bar{a}_{2}(-\tfrac{1}{2}),u_{3}(-\tfrac{1}{2})\big] & =\frac{g_3}{2\sqrt{15\alpha'}}T_{\alpha_{2}\alpha_{3}}^{a}\,,\\
\mathcal{A}\big[\sigma_{1}(\ 0\ ),\bar{a}_{2}(+\tfrac{1}{2}),u_{3}(-\tfrac{1}{2})\big] & =\frac{g_3}{2\sqrt{10\alpha'}}T_{\alpha_{2}\alpha_{3}}^{a}\,,
\end{align}
which match exactly with the results of Eqs.~\eqref{Spin3tochiq1}-\eqref{Spin3tochiq4}, and Eqs.~\eqref{Spin3toaq1} and \eqref{Spin3toaq2} respectively.
In addition, we also have the contributions
\begin{equation}
\Gamma(\sigma_{1}\to\chi_{2}+\bar{u}_{3})=\Gamma(\sigma_{1}\to\bar{\chi}_{2}+u_{3})\,.
\end{equation}
Thus, the partial decay widths of the spin-3 field $\sigma$ match
exactly the results we obtain from factorization.

\subsubsection{Partial decay width of the spin-2 state $\pi_{\mu\nu}$}

We now turn to the decay of the spin-2 field $\pi_{\mu\nu}$. For the
decay channels
$\pi\to\alpha+g,\pi\to\Phi_{\pm}+g,\pi\to\bar{\chi}+u,\pi\to\bar{a}+u$, 
we obtain
\begin{align}
\mathcal{A}(\pi_{1},\alpha_{2},\epsilon_{3}) & ={\rm Tr}(T^{1}\{T^{2},T^{3}\})\frac{g_3}{\sqrt{3}}k_{1}^{\lambda}\varepsilon_{\lambda(\mu|\rho\gamma|}\pi_{\phantom{\gamma}\nu)}^{\gamma}\Big\{(2\alpha')^{2}\big[k_{3}^{\mu}k_{3}^{\nu}\epsilon_{3}^{\rho}\alpha_{\gamma\zeta}k_{3}^{\gamma}k_{3}^{\zeta}-k_{3}^{\mu}k_{3}^{\nu}\alpha^{\rho\zeta}k_{3\zeta}(\epsilon_{3}\cdot k_{2})\big]\nonumber \\
 & \qquad\qquad+(2\alpha')\big[2k_{3}^{\mu}\alpha^{\nu\rho}(\epsilon_{3}\cdot k_{2})-2k_{3}^{\mu}\epsilon_{3}^{\nu}\alpha^{\rho\zeta}k_{3\zeta}+k_{3}^{\mu}k_{3}^{\nu}\alpha^{\rho\zeta}\epsilon_{3\zeta}-4k_{3}^{\nu}\epsilon_{3}^{\rho}\alpha^{\mu\zeta}k_{3\zeta}\nonumber \\
 & \qquad\qquad+2k_{3}^{\nu}k_{3}^{\rho}\alpha^{\mu\zeta}\epsilon_{3\zeta}+2\epsilon_{3}^{\nu}k_{3}^{\rho}\alpha^{\mu\zeta}k_{3\zeta}\big]+2\epsilon_{3}^{\mu}\alpha^{\nu\rho}\Big\}\,,\\
\mathcal{A}(\pi_{1},\Phi_{2},\epsilon_{3}) & ={\rm Tr}(T^{1}\{T^{2},T^{3}\})\frac{g_3}{2\sqrt{3}}k_{1}^{\lambda}\varepsilon_{\lambda(\mu|\rho\gamma|}\pi_{\phantom{\gamma}\nu)}^{\gamma}\Big\{(2\alpha')\big[4\epsilon_{3}^{\mu}k_{3}^{\nu}k_{3}^{\rho}-6k_{3}^{\mu}k_{3}^{\nu}\epsilon_{3}^{\rho}+2k_{3}^{\mu}\eta^{\nu\rho}(\epsilon_{3}\cdot k_{2})\big]\nonumber \\
 & +2\epsilon_{3}^{\mu}\eta^{\nu\rho}\pm i\big(2\alpha'k_{3}^{\mu}k_{3}^{\nu}\eta^{\rho\gamma}\varepsilon_{\gamma\zeta\tau\lambda}\epsilon_{3}^{\zeta}k_{3}^{\tau}k_{2}^{\lambda}-2k_{3\mu}\varepsilon_{\nu\rho\tau\lambda}\epsilon_{3}^{\tau}k_{2}^{\lambda}+2\epsilon_{3\mu}\varepsilon_{\nu\rho\tau\lambda}k_{3}^{\tau}k_{2}^{\lambda}\big)\Big\}\,,\\
\mathcal{A}(\pi_{1},\bar{\chi}_{2},u_{3}) & =T_{\alpha_{2}\alpha_{3}}^{a}\frac{g_3\alpha'}{\sqrt{3}}k_{1}^{\lambda}\varepsilon_{\lambda(\mu|\rho\gamma|}\pi_{\phantom{\gamma}\nu)}^{\gamma}
\big[\alpha'k_{3}^{\mu}k_{3}^{\nu}(u_{3}^{a}\sigma_{a\dot{a}}^{\rho}\bar{\chi}_{2}^{\lambda\dot{a}}k_{1\lambda}-u_{3}^{a}\slashed k_{1a\dot{a}}\bar{\chi}_{2}^{\rho\dot{a}}) \nonumber \\
&\qquad\qquad\qquad\qquad \qquad
+k_{3}^{\mu}u_{3}^{a}\sigma_{a\dot{a}}^{\rho}\bar{\chi}_{2}^{\nu\dot{a}}
+\frac{1}{3} u_3^a (\sigma^{\nu} \sigma^{\rho} \slashed k_2)_{a \dot{a}} \bar{\chi}_{2}^{\mu \dot{a}} \big]\,,\\
\mathcal{A}(\pi_{1},\bar{a}_{2},u_{3}) & =T_{\alpha_{2}\alpha_{3}}^{a}\frac{g_3\alpha^{\prime\frac{3}{2}}}{\sqrt{6}}k_{1}^{\lambda}\varepsilon_{\lambda(\mu|\rho\gamma|}\pi_{\phantom{\gamma}\nu)}^{\gamma} ( k_{3}^{\mu}k_{3}^{\nu}u_{3}^{b}\sigma_{b\dot{b}}^{\rho}\bar{a}_{2}^{\dot{b}}-k_{3}^{\mu}u_{3}^{a}\sigma_{a\dot{a}}^{\rho}\bar{\sigma}^{\nu\dot{a}b}\slashed k_{2b\dot{b}}\bar{a}_{2}^{\dot{b}} ) \,.
\end{align}
Applying helicity techniques and using the reference momenta we
have chosen above, we obtain
\begin{align}
\mathcal{A}\big[\pi_{1}(-2),\alpha_{2}(+1),\epsilon_{3}^{+}\big] & =\frac{8g_3}{\sqrt{6\alpha'}}d_{a_{1}a_{2}a_{3}}\,,\\
\mathcal{A}\big[\pi_{1}(-1),\alpha_{2}(\ 0\ ),\epsilon_{3}^{+}\big] & =\frac{2\sqrt{2}g_3}{\sqrt{\alpha'}}d_{a_{1}a_{2}a_{3}}\,,\\
\mathcal{A}\big[\pi_{1}(\ 0\ ),\alpha_{2}(-1),\epsilon_{3}^{+}\big] & =\frac{2g_3}{\sqrt{\alpha'}}d_{a_{1}a_{2}a_{3}}\,,\\
\mathcal{A}\big[\pi_{1}(+1),\alpha_{2}(-2),\epsilon_{3}^{+}\big] & =\frac{2g_3}{\sqrt{3\alpha'}}d_{a_{1}a_{2}a_{3}}\,,
\end{align}
and
\begin{align}
\mathcal{A}\big[\pi_{1}(-1),\Phi_{2+},\epsilon_{3}^{+}\big] & =\frac{2g_3}{\sqrt{3\alpha'}}d_{a_{1}a_{2}a_{3}}\,,\\
\mathcal{A}\big[\pi_{1}(-1),\Phi_{2-},\epsilon_{3}^{+}\big] & =0\,,
\end{align}
which match exactly with Eqs.~\eqref{Spin3toalphag2}
--\eqref{Spin3toalphag5} and Eqs.~\eqref{Spin3toPhig}, respectively.
For the fermionic decay channels, we have
\begin{align}
\mathcal{A}\big[\pi_{1}(+2),\bar{\chi}_{2}(-\tfrac{3}{2}),u_{3}(-\tfrac{1}{2})\big] & =\frac{g_3}{\sqrt{6\alpha'}}T_{\alpha_{2}\alpha_{3}}^{a}\,,\\
\mathcal{A}\big[\pi_{1}(+1),\bar{\chi}_{2}(-\tfrac{1}{2}),u_{3}(-\tfrac{1}{2})\big] & =\frac{g_3}{2\sqrt{\alpha'}}T_{\alpha_{2}\alpha_{3}}^{a}\,,\\
\mathcal{A}\big[\pi_{1}(\ 0\ ),\bar{\chi}_{2}(+\tfrac{1}{2}),u_{3}(-\tfrac{1}{2})\big] & =\frac{\sqrt{3}g_3}{4\sqrt{\alpha'}}T_{\alpha_{2}\alpha_{3}}^{a}\,,\\
\mathcal{A}\big[\pi_{1}(-1),\bar{\chi}_{2}(+\tfrac{3}{2}),u_{3}(-\tfrac{1}{2})\big] & =\frac{g_3}{2\sqrt{3\alpha'}}T_{\alpha_{2}\alpha_{3}}^{a}\,,,
\end{align}
and
\begin{align}
\mathcal{A}\big[\pi_{1}(+1),\bar{a}_{2}(-\tfrac{1}{2}),u_{3}(-\tfrac{1}{2})\big] & =\frac{g_3}{2\sqrt{3\alpha'}}T_{\alpha_{2}\alpha_{3}}^{a}\,,\\
\mathcal{A}\big[\pi_{1}(\ 0\ ),\bar{a}_{2}(+\tfrac{1}{2}),u_{3}(-\tfrac{1}{2})\big] & =\frac{g_3}{4\sqrt{\alpha'}}T_{\alpha_{2}\alpha_{3}}^{a}\,,
\end{align}
which match the results of Eqs.~\eqref{Spin3tochiq1}-\eqref{Spin3tochiq4},
and Eqs.~\eqref{Spin3toaq1} and \eqref{Spin3toaq2} precisely.
Thus we also confirm the partial decay widths of these channels
obtained from factorization in the previous section.

In closing, it is important to stress that the bosonic states we
considered in Secs.~\ref{s4} and \ref{s5} are gluons, the color
singlet $C_\mu$, and their excitations. As a result, we have taken 
the QCD coupling $g_3$ in all the amplitudes. The derivation of the
amplitudes, however, is valid for any vector boson. To obtain the
amplitudes involving (excited) bosons on other stacks, one can just
simply replace $g_3$ by the corresponding coupling constant in 
all the formulae.

\section{Discovery reach at HL-LHC, HE-LHC, and VLHC } 
\label{discoveryreach}

\subsection{Bump hunting}
We have seen that particles created by vibrations of relativistic
strings populate Regge trajectories relating their spins and
masses. Most apparently, one would expect that lowest massive Regge
excitations would be visible in data binned according to the invariant
mass $M$ of dijets, after setting
cuts on the different jet rapidities, $|y_1|, \, |y_2| < y_{\rm max} =2.5$  and
both transverse momenta $p_T > 30~{\rm GeV}$~\cite{Chatrchyan:2013qha}. With the
definitions $Y\equiv \thalf (y_1 + y_2)$ and $y \equiv
\thalf(y_1-y_2)$, the cross section per interval of $M$ for $p
p\rightarrow {\rm dijet}$ is given by
\begin{eqnarray}
\frac{d\sigma}{dM} & = & M\tau\ \sum_{ijkl}\left[
\int_{-Y_{\rm max}}^{0} dY \ f_i (x_a,\, M)  \right. \ f_j (x_b, \,M ) \
\int_{-(y_{\rm max} + Y)}^{y_{\rm max} + Y} dy
\left. \frac{d\sigma}{d\hat t}\right|_{ij\rightarrow kl}\ \frac{1}{\cosh^2
y} \nonumber \\
& + &\int_{0}^{Y_{\rm max}} dY \ f_i (x_a, \, M) \
f_j (x_b, M) \ \int_{-(y_{\rm max} - Y)}^{y_{\rm max} - Y} dy
\left. \left. \frac{d\sigma}{d\hat t}\right|_{ij\rightarrow kl}\
\frac{1}{\cosh^2 y} \right]
\label{longBH}
\end{eqnarray}
where $\tau = M^2/s$, $x_a =
\sqrt{\tau} e^{Y}$,  $x_b = \sqrt{\tau} e^{-Y},$
and
\begin{equation}
  |{\cal M}(ij \to kl) |^2 = 16 \pi \hat s^2 \,
  \left. \frac{d\sigma}{d\hat t} \right|_{ij \to kl} \, .
\label{Lucero}
\end{equation}
In this section we reinstate the caret notation ($\hat s,\ \hat t,\
\hat u$) to specify partonic processes. The $Y$ integration range in
Eq.~(\ref{longBH}), $Y_{\rm max} = {\rm min} \{ \ln(1/\sqrt{\tau}),\ \
y_{\rm max}\}$, comes from requiring $x_a, \, x_b < 1$ together with
the rapidity cuts $y_{\rm min} <|y_1|, \, |y_2| < y_{\rm max}$. The
kinematics of the scattering also provides the relation $M = 2p_T
\cosh y$, which when combined with $p_T = M/2 \ \sin \theta^* = M/2
\sqrt{1-\cos^2 \theta^*}$ yields $\cosh y = (1 - \cos^2
\theta^*)^{-1/2}$, where $\theta^*$ is the center-of-mass scattering
angle.  Finally, the Mandelstam invariants occurring in the cross
section are given by $\hat s = M^2,$ $\hat t = -\thalf M^2\ e^{-y}/
\cosh y,$ and $\hat u = -\thalf M^2\ e^{+y}/ \cosh y.$ An equivalent
expression can be obtained for $pp \to \gamma$ +
jet~\cite{Anchordoqui:2008ac}. Following Ref.~\cite{Aad:2013cva}, we take
$p_T^\gamma, p_T^{\rm jet} > 125~{\rm GeV}$, $y_{\rm max}^\gamma =
1.37$, and $y_{\rm max}^{\rm jet} = 2.8$.

\begin{figure}[t]
\begin{center}
\includegraphics[width=0.49\linewidth]{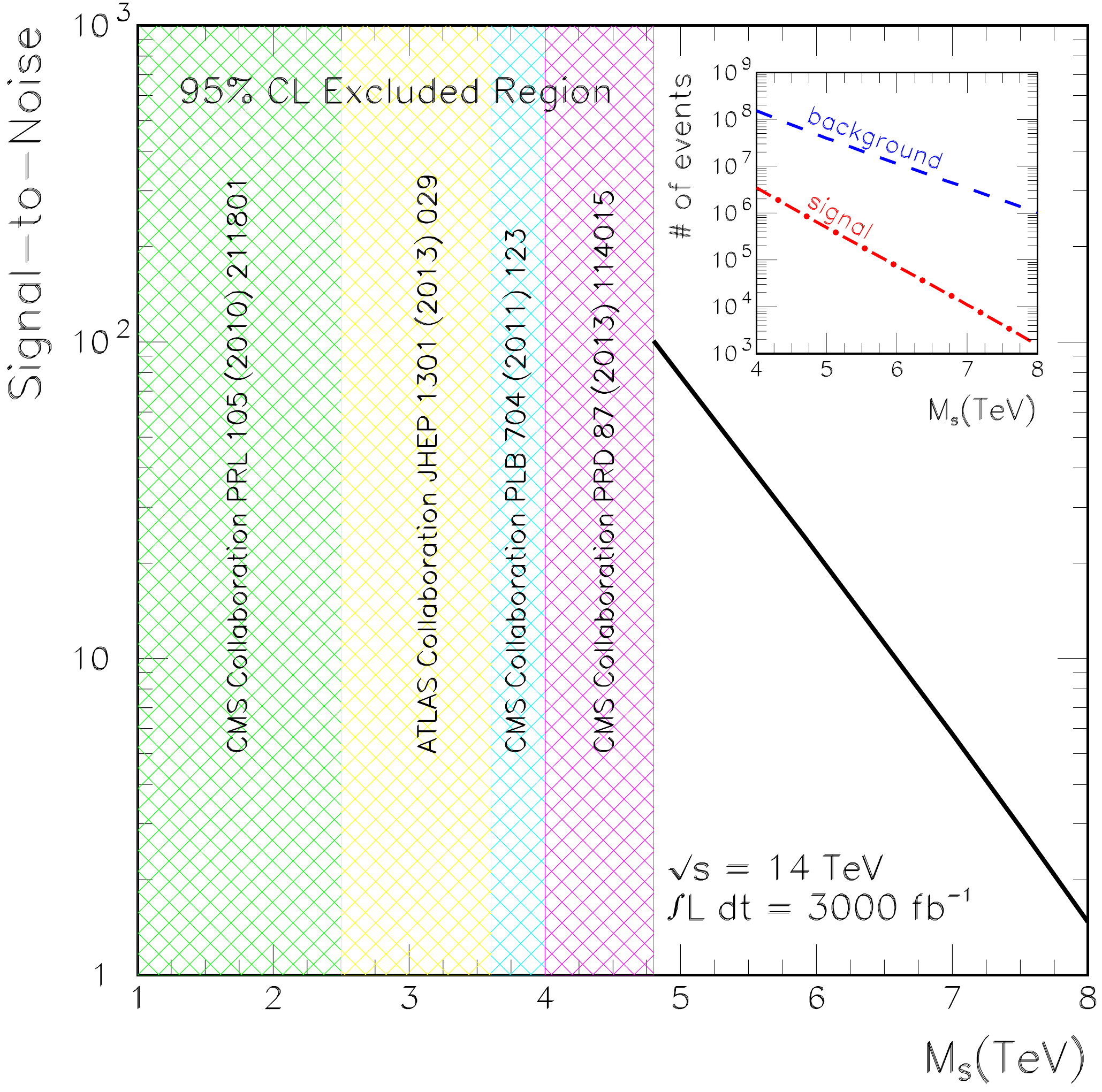}
\includegraphics[width=0.49\linewidth]{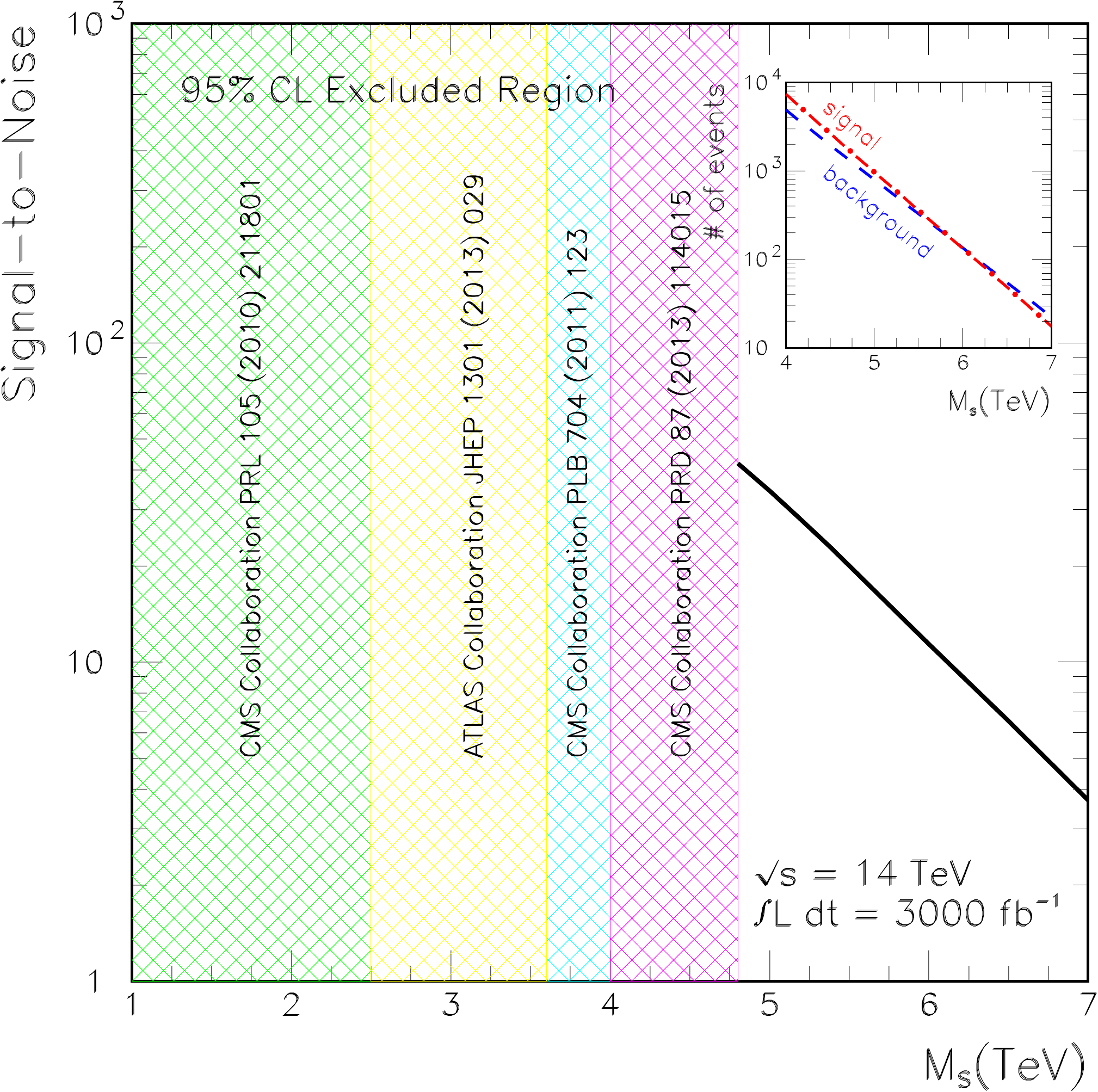}
\end{center}
\caption{Signal-to-noise ratio of the lowest massive Regge excitations
  for the HL-LHC in the dijet (left) and $\gamma$ + jet (right)
  topologies. For comparison,  we also show ATLAS and CMS
  upper limits on $M_s$ from unsuccessful searches of
  new particles decaying to pairs of partons (quarks, antiquarks, or
  gluons)~\cite{Khachatryan:2010jd,ATLAS:2012pu,Chatrchyan:2011ns,Chatrchyan:2013qha}. For
LHC phase I, the signal-to-noise ratio is suppressed by $\simeq 0.32$.}
\label{figure:uno}
\end{figure}

\begin{figure}[t]
\begin{center}
\includegraphics[width=0.49\linewidth]{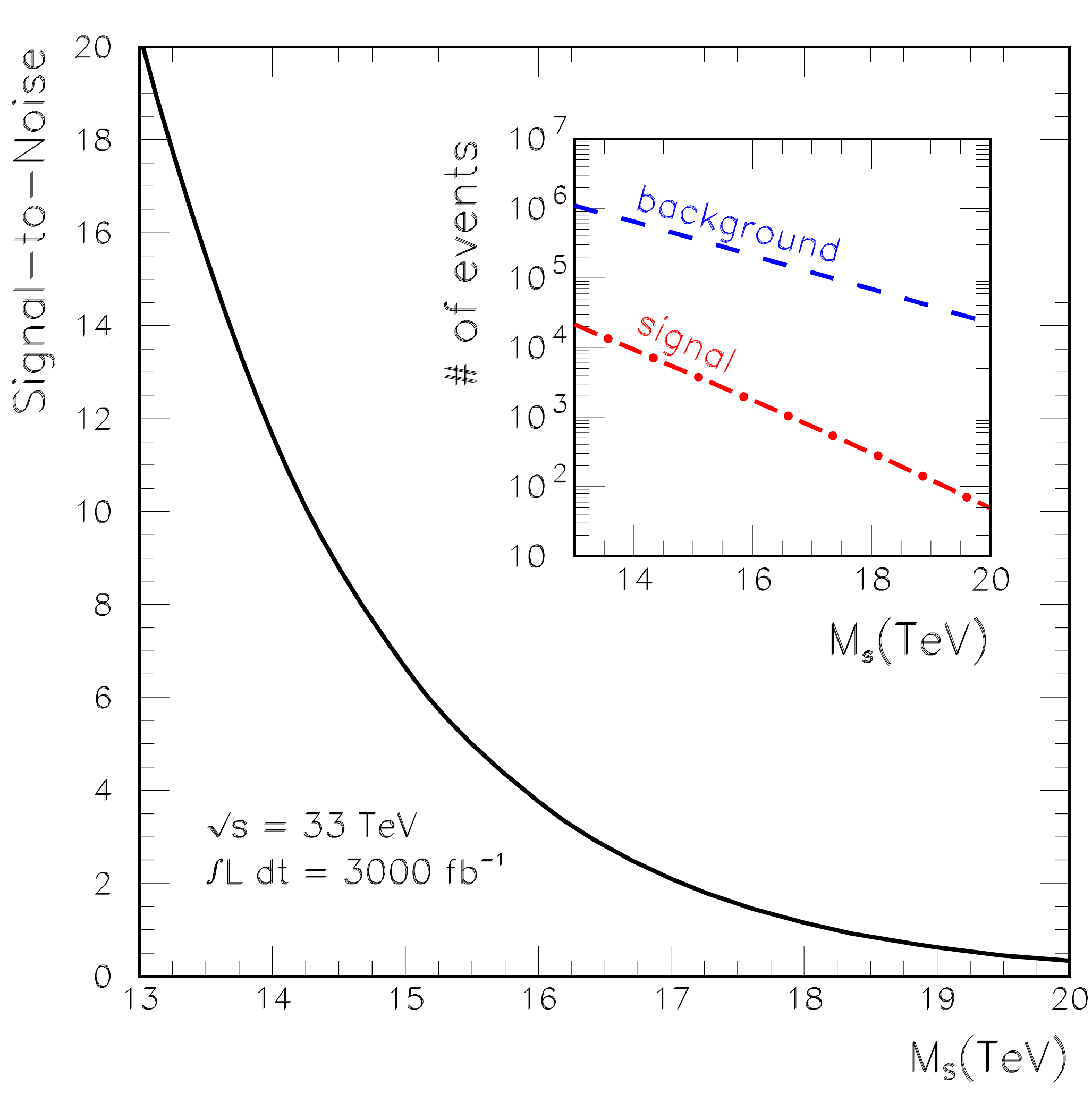}
\includegraphics[width=0.49\linewidth]{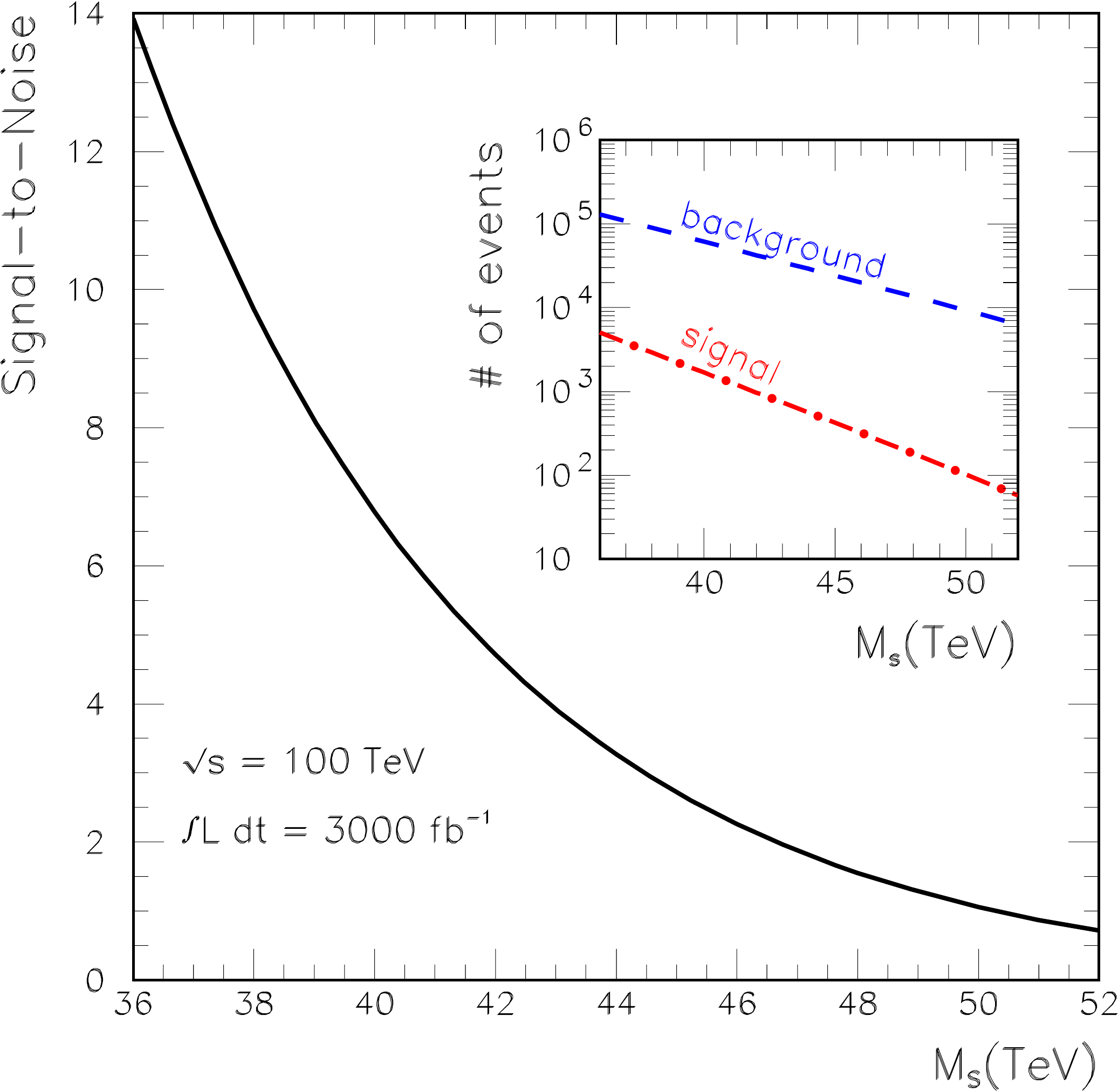}
\end{center}
\caption{Dijet signal-to-noise ratio of the lowest massive Regge excitations
  for HE-LHC (left) and VLHC (right). }
\label{figure:dos}
\end{figure}

The QCD background is calculated at the partonic level making use of
the CTEQ6l1 parton distribution functions (PDFs)~\cite{Pumplin:2002vw}. Standard bump-hunting
methods, such as obtaining cumulative cross sections, 
\begin{equation}
\sigma (M_0) = \int_{M_0}^\infty \frac{d\sigma}{dM} \, dM \,,
\end{equation}
from the data and searching for regions with significant deviations
from the QCD background, may reveal an interval of $M$ suspected of
containing a bump. With the establishment of such a region, one may
calculate a signal-to-noise ratio, with the signal rate estimated in
the invariant mass window $[M_s -2\Gamma,\, M_s + 2\Gamma]$. The noise is defined as
the square root of the number of background events in the same dijet
mass interval for the same integrated luminosity. The HL-LHC dijet
discovery reach 
of lowest massive Regee excitations (at the parton level) is
encapsulated in Fig.~\ref{figure:uno}. It is remarkable that string
scales as large as 7.1~TeV  are open to discovery at the $\geq
5\sigma$ level. Next, we duplicate the calculation for the HE-LHC and
VLCH. The results are shown in Fig.~\ref{figure:dos}. The $5\sigma$ discovery
reach exceedingly improves, reaching 15~TeV at the HE-LHC and 41~TeV
at the VLHC. Once more, we stress that all these results contain no
unknown parameters. They depend only on the D-brane construct for the
SM and {\it are independent of compactification details}. 

We now turn to the study of $pp \to \gamma$ + jet. Armed with
(\ref{mhvlow2}) and (\ref{qgqz}), we first compute the signal for an
integrated luminosity of $20~{\rm fb}^{-1}$ at $\sqrt{s} = 8~{\rm
  TeV}$. Using the 95\% C.L. upper limits on the production cross
section $\times$ branching of excited quarks (into $\gamma$ + jet), as
reported by the ATLAS and CMS
collaborations~\cite{Aad:2013cva,Khachatryan:2014aka}, we derived an
upper limit on the string scale for $\kappa = 0.14$, $M_s = 4~{\rm
  TeV}$ at 95\% C.L..  This limit, however, does not include detailed
detector modeling. It is worth noting that this number is not far
from the dijet limit reported by ATLAS and CMS collaboration using the
dijet channel. The signal-to-noise ratio for the HL-LHC is displayed in
Fig.~\ref{figure:uno}. For string scales as high as 6.5~TeV,
observations of resonant structures in $pp \to \gamma +$ jet can
provide interesting corroboration for stringy physics.

\begin{figure}[t]
\begin{center}
\includegraphics[width=0.49\linewidth]{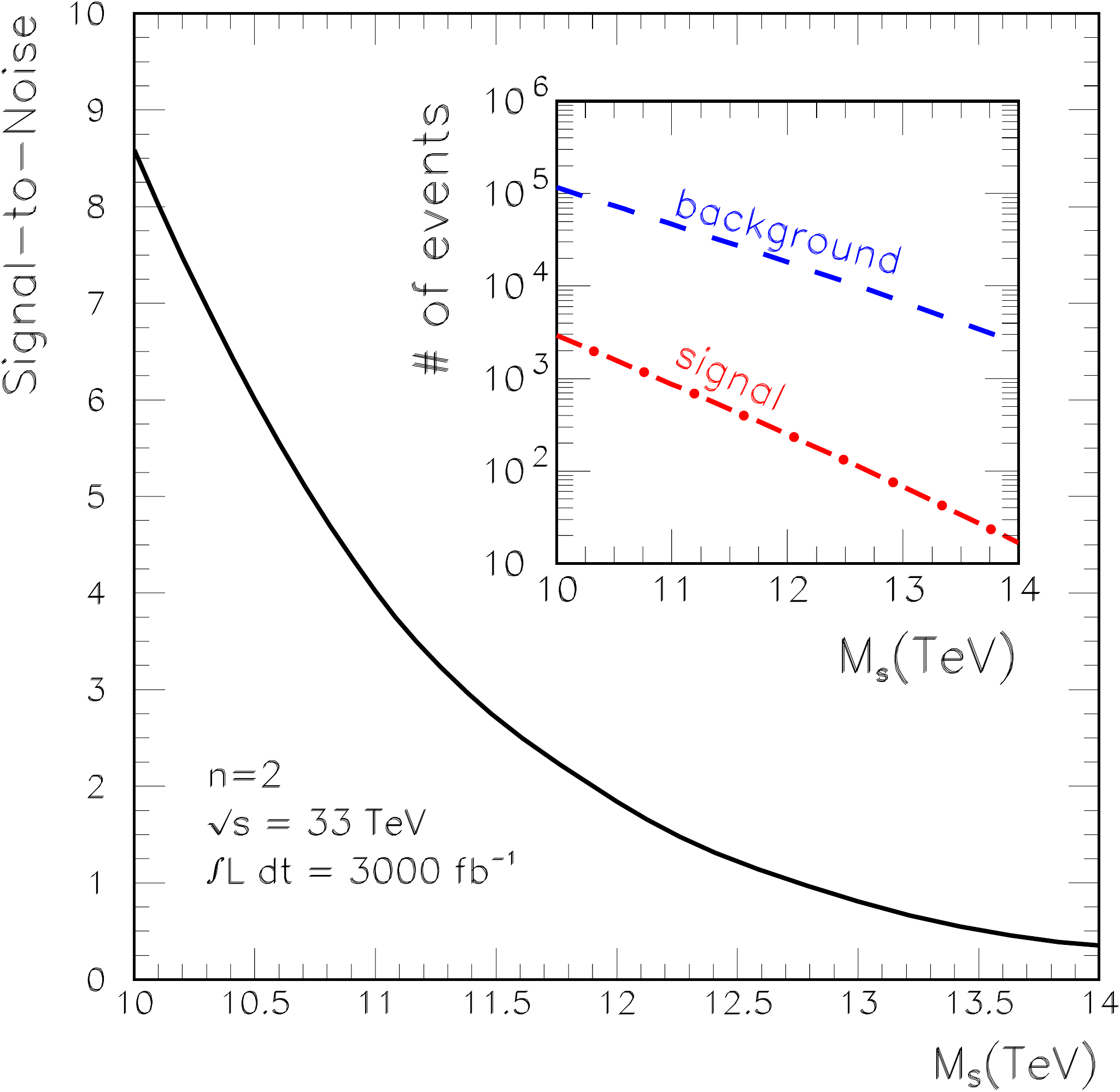}
\includegraphics[width=0.49\linewidth]{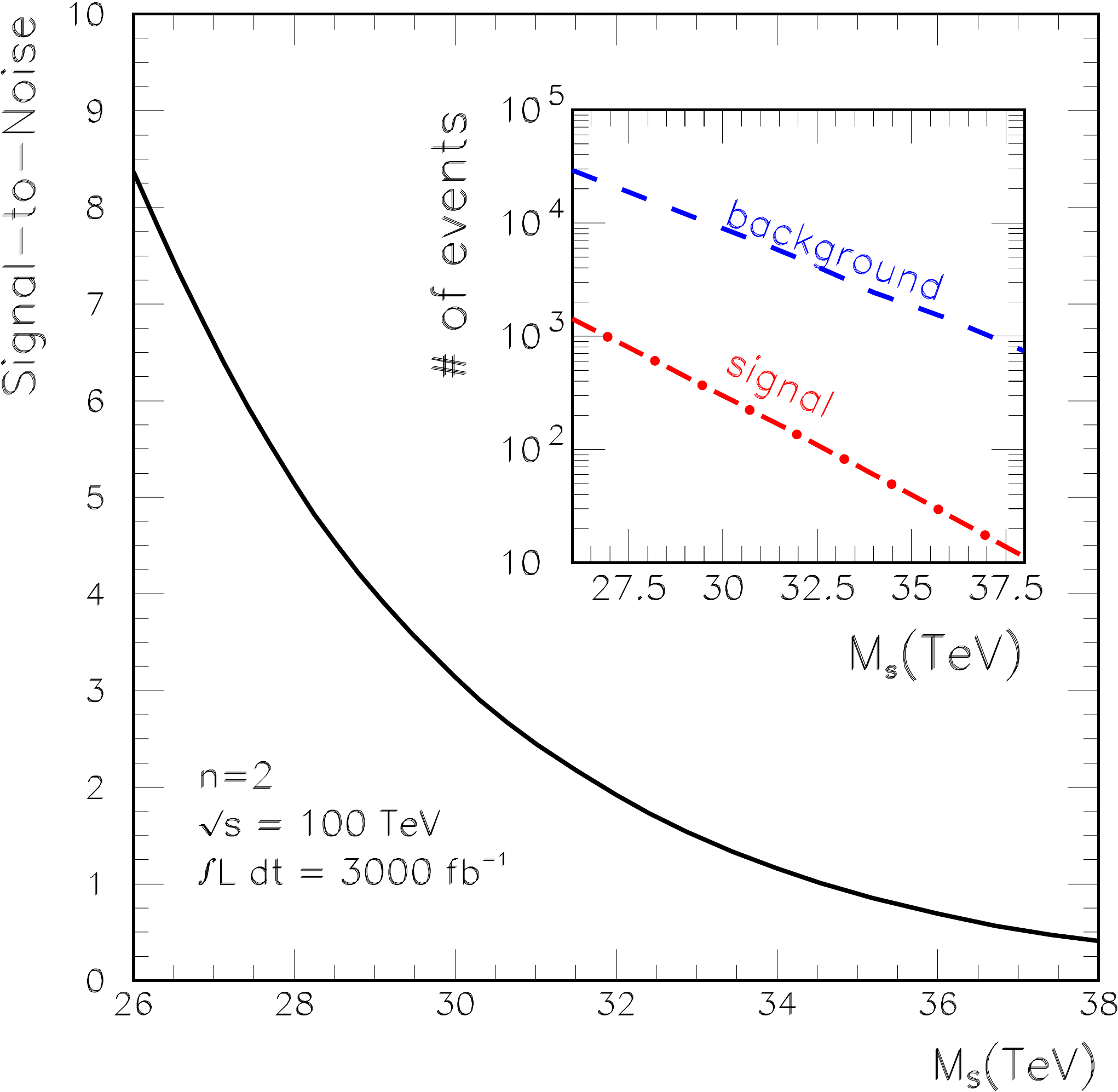}
\end{center}
\caption{Dijet signal-to-noise ratio of $n=2$ Regge excitations
  for the HE-LHC (left) and VLHC (right). }
\label{figure:tres}
\end{figure}

Excitations of the second massive string state may become visible at
the HE-LHC and VLHC. The relevant resonant amplitudes around $s=2M_s$
are as follows:
\begin{eqnarray} 
 \mathcal{M}(g_1^{-}, g_2^{-}, g_3^{+}, g_4^{+})
&=&  \frac{8g_3^2M_s^2 \cos(\theta)}{s-2M_s^2} \Tr( [T^{a_1},T^{a_2}] [T^{a_3},T^{a_4}])       \nonumber \\
&=& -\frac{8g_3^2 M_s^2}{s-2M_s^2} d^1_{0,0}(\theta) f^{a_1a_2a}f^{a_3a_4a}\,,
\end{eqnarray}

\begin{eqnarray}
\mathcal{M}(g_1^{-},g_2^+,g_3^{+},g_4^-)
&=& -\frac{8g_3^2M_s^2}{s-2M_s^2} \left( \frac{1+\cos \theta }{2} \right)^2\cos \theta ~ f^{a_1a_2a}f^{a_3a_4a} \nonumber \\
&=&  -\frac{8g_3^2M_s^2}{s-2M_s^2} \left( \frac{1}{3}d^3_{+2,+2}(\theta)+\frac{2}{3}d^2_{+2,+2}(\theta) \right)  f^{a_1a_2a}f^{a_3a_4a}\,,
\end{eqnarray}

\begin{eqnarray}
\mathcal{M}(g_1^{-},g_2^+,g_3^{-},g_4^+)
&=& -\frac{8g_3^2M_s^2}{s-2M_s^2} \left( \frac{1-\cos \theta }{2} \right)^2\cos \theta ~ f^{a_1a_2a}f^{a_3a_4a} \nonumber \\
&=&  -\frac{8g_3^2M_s^2}{s-2M_s^2} \left( \frac{1}{3}d^3_{+2,-2}(\theta)-\frac{2}{3}d^2_{+2,-2}(\theta) \right)  f^{a_1a_2a}f^{a_3a_4a}\,,
\end{eqnarray}

\begin{eqnarray}
\mathcal{M} (q_1^{-}, \bar{q}_2^+,g_3^{-},g_4^+) 
=   \frac{4g_3^2M_s^2}{s-2M_s^2} \left( \frac{1}{3}\sqrt{\frac{2}{5}} d^3_{+2,+1}(\theta) +\frac{1}{6} d^2_{+2,+1}(\theta) \right) [T^{a_3},T^{a_4}]_{\alpha_1\alpha_2},
\end{eqnarray}

\begin{eqnarray}
\mathcal{M} (q_1^{-}, \bar{q}_2^+,g_3^{+},g_4^-) 
=  \frac{4g_3^2M_s^2}{s-2M_s^2}\left( \frac{1}{3}\sqrt{\frac{2}{5}} d^3_{+2,-1}(\theta) -\frac{1}{6} d^2_{+2,-1}(\theta) \right) [T^{a_3},T^{a_4}]_{\alpha_1\alpha_2},
\end{eqnarray}

\begin{equation}
 \mathcal{M} ( q_1^\pm, g_2^\pm,\bar q_3^\mp, g_4^\mp)
 = - \frac{4g_3^2 M_s^2}{s-2M_s^2}
                 \biggl[
                        \frac{1}{3}
                        d^{J=1/2}_{\mp1/2,\mp1/2}(\theta)
                    + \frac{2}{3}
                        d^{J=3/2}_{\mp1/2,\mp1/2}(\theta)
                 \biggr]
                 \bigl(T^{a_4}T^{a_2}\bigr)_{\alpha_3\alpha_1} \,,
\end{equation}
\begin{equation}
 \mathcal{M} ( q_1^\pm, g_2^\mp, \bar q_3^\mp, g_4^\pm )
 = -\frac{4g_3^2 M_s^2}{s-2M_s^2}
                 \biggl[
                        \frac{3}{5}
                        d^{J=3/2}_{\pm3/2,\pm3/2}(\theta)
                    + \frac{2}{5}
                        d^{J=5/2}_{\pm3/2,\pm3/2}(\theta)
                 \biggr]
                 \bigl(T^{a_4}T^{a_2}\bigr)_{\alpha_3\alpha_1} \,.
\end{equation}
For phenomenological purposes, the poles need to be softened to a
Breit--Wigner form. We can tell what the intermediate states are from
the Wigner $d$ matrices and put in the corresponding total decay
widths. After this is done, the contributions of the various channels
to the dijet production are as follows:
\begin{eqnarray}
  &&|\mathcal{M}(gg\rightarrow gg)|^2  = \frac{9g_3^4}{16M_s^4}\lsb~\frac{4M_s^4(t-u)^2}{(s-2M_s^2)^2+2(\Gamma_{G^{(2)}}^{J=1}M_s)^2} +\frac{4}{9}~\frac{ t^4-6 t^3 u+6 t^2 u^2-6 t u^3+u^4}{(s-2M_s^2)^2+2\Gamma_{G^{(2)}}^{J=3}\Gamma_{G^{(2)}}^{J=2}M_s^2}\rc
                    \nonumber \\
&&\lc +\frac{4}{9}\frac{ t^4+u^4}{(s-2M_s^2)^2+2(\Gamma_{G^{(2)}}^{J=2}M_s)^2} +\frac{1}{36M_s^4}\frac{ t^6-10 t^5 u+25 t^4 u^2+25 t^2 u^4-10 t u^5+u^6}{(s-2M_s^2)^2+2(\Gamma_{G^{(2)}}^{J=3}M_s)^2}\rsb,
 \end{eqnarray}

\begin{eqnarray}
  &&|\mathcal{M}( q\bar{q}\rightarrow gg)|^2  = \frac {4 g_3^4 }{9 M_s^4}\lsb
                   \frac{1}{6 M_s^4} ~\frac{t u (t^4 - 4 t^3 u + 8 t^2 u^2 - 4 t u^3 + u^4)}{(s-2M_s^2)^2+2(\Gamma_{G^{(2)}}^{J=3}M_s)^2} \rc \nonumber \\
                  && \lc + \frac{1}{6 }~\frac{ t u \left(t^2+u^2\right)}{(s-2M_s^2)^2+2(\Gamma_{G^{(2)}}^{J=2}M_s)^2}
                   + \frac{2}{3}~\frac{t u \left(t^2-3 t u+u^2\right)}{(s-2M_s^2)^2+2\Gamma_{G^{(2)}}^{J=3}\Gamma_{G^{(2)}}^{J=2}M_s^2}\rsb\, \,,
 \end{eqnarray}

\begin{eqnarray}
  &&|\mathcal{M}(gg\rightarrow q\bar{q})|^2  = \frac {g_3^4 N_f}{16 M_s^4}\lsb
                   \frac{1}{6 M_s^4} ~\frac{t u (t^4 - 4 t^3 u + 8 t^2 u^2 - 4 t u^3 + u^4)}{(s-2M_s^2)^2+2(\Gamma_{G^{(2)}}^{J=3}M_s)^2} \rc \nonumber \\
                  &&\lc + \frac{1}{6 }~\frac{ t u \left(t^2+u^2\right)}{(s-2M_s^2)^2+2(\Gamma_{G^{(2)}}^{J=2}M_s)^2}
                   + \frac{2}{3}~\frac{t u \left(t^2-3 t u+u^2\right)}{(s-2M_s^2)^2+2\Gamma_{G^{(2)}}^{J=3}\Gamma_{G^{(2)}}^{J=2}M_s^2}\rsb \,,
 \end{eqnarray}

\begin{eqnarray}
 && \bigl| \mathcal{M} ( q_Lg \rightarrow q_Lg ) \bigr|^2 = \bigl| \mathcal{M} ( \bar{q}_Rg \rightarrow \bar{q}_Rg ) \bigr|^2 \nn
 & = & \frac{8g_3^4}{9M_s^2}\left\{\lsb\frac{1}{9}\frac{ -M_s^4 u }{ \lb {s} - 2M_s^2 \rb^2+ 2 \lb \Gamma^{J=1/2}_{Q^{(2)}_L} M_s \rb^2 } +\frac{1}{9}\frac{  -{u}  ( 2{t} - {u} )^2 }{\lb {s} - 2M_s^2 \rb^2+ 2 \lb \Gamma^{J=3/2}_{\tilde Q^{(2)}_L}M_s  \rb^2 } \rsb \rc \nn
& + &\frac{1}{4M_s^4}\lsb\frac{9}{25}\frac{ -M_s^4 u^3 }{ \lb {s} - 2M_s^2 \rb^2+ 2 \lb \Gamma^{J=3/2}_{Q^{(2)}_L}  M_s\rb^2 }+ \frac{1}{25} \frac{  -{u} ^3 ( 4{t} - {u} )^2 }{ \lb {s} - 2M_s^2 \rb^2+ 2 \lb \Gamma^{J=5/2}_{Q^{(2)}_L} M_s \rb^2 }\rsb \nn
&+&\lc \lsb\frac{2}{9}\frac{ M_s^2 (-2tu+u^2)  }{ \lb {s} - 2M_s^2 \rb^2 + 2 \Gamma^{J=1/2}_{Q^{(2)}_L} \Gamma^{J=3/2}_{\tilde Q^{(2)}_L} M_s  ^2 } +\frac{3}{50 }\frac{ M_s^{-2} ( -4{t}{u} ^3 + {u}^4) }{\lb {s} - 2M_s^2 \rb^2+ 2 \Gamma^{J=3/2}_{Q^{(2)}_L} \Gamma^{J=5/2}_{Q^{(2)}_L}M_s^2 } \rsb \right\} ,
\end{eqnarray}

\begin{eqnarray}
 && \bigl| \mathcal{M} ( q_Rg \rightarrow q_Rg ) \bigr|^2 = \bigl| \mathcal{M} ( \bar{q}_Lg \rightarrow \bar{q}_Lg ) \bigr|^2 \nn
 & = & \frac{8g_3^4}{9M_s^2}\left\{\lsb\frac{1}{9}\frac{ -M_s^4 u }{ \lb {s} - 2M_s^2 \rb^2+ 2 \lb \Gamma^{J=1/2}_{Q^{(2)}_R} M_s \rb^2 } +\frac{1}{9}\frac{  -{u}  ( 2{t} - {u} )^2 }{\lb {s} - 2M_s^2 \rb^2+ 2 \lb \Gamma^{J=3/2}_{\tilde Q^{(2)}_R}M_s  \rb^2 } \rsb \rc \nn
& + &\frac{1}{4M_s^4}\lsb\frac{9}{25}\frac{ -M_s^4 u^3 }{ \lb {s} - 2M_s^2 \rb^2+ 2 \lb \Gamma^{J=3/2}_{Q^{(2)}_R}  M_s\rb^2 }+ \frac{1}{25} \frac{  -{u} ^3 ( 4{t} - {u} )^2 }{ \lb {s} - 2M_s^2 \rb^2+ 2 \lb \Gamma^{J=5/2}_{Q^{(2)}_R} M_s \rb^2 }\rsb \nn
&+&\lc \lsb\frac{2}{9}\frac{ M_s^2 (-2tu+u^2)  }{ \lb {s} - 2M_s^2 \rb^2 + 2 \Gamma^{J=1/2}_{Q^{(2)}_R} \Gamma^{J=3/2}_{\tilde Q^{(2)}_R} M_s  ^2 } +\frac{3}{50 }\frac{ M_s^{-2} ( -4{t}{u} ^3 + {u}^4) }{\lb {s} - 2M_s^2 \rb^2+ 2 \Gamma^{J=3/2}_{Q^{(2)}_R} \Gamma^{J=5/2}_{Q^{(2)}_R}M_s^2 } \rsb \right\} ,
\end{eqnarray}
The total decay widths for $n=2$ string resonances can be computed using the formulas in
Table~\ref{decayratetable1}. We note that the widths of $Q^{(2)}$ are
model dependent since they can decay into the $U(1)$ gauge bosons. In
the $U(3) \times Sp(1) \times U(1)$ D-brane model, we have (at $M_s
\sim 15$~TeV)

\begin{align}
& \Gamma_{G^{(2)}}^{J=3} =58\, (M_s/{\rm TeV})~{\rm GeV}, \quad \Gamma_{G^{(2)}}^{J=2} = 53 \, (M_s/{\rm TeV})~{\rm GeV},\nn
& \Gamma_{G^{(2)}}^{J=1} = 67 \, (M_s/{\rm TeV})~{\rm GeV},\quad
\Gamma_{Q^{(2)}_L}^{J=5/2} = 30 \, (M_s/{\rm TeV})~{\rm GeV},\nn
& \Gamma_{Q^{(2)}_L}^{J=3/2} = 26 \, (M_s/{\rm TeV})~{\rm GeV},\quad \Gamma_{\tilde
Q^{(2)}_L}^{J=3/2} = 38 \, (M_s/{\rm TeV})~{\rm GeV}\nn
& \Gamma_{Q^{(2)}_L}^{J=1/2} = 37 \, (M_s/{\rm TeV})~{\rm GeV},\quad \Gamma_{Q^{(2)}_R}^{J=5/2} = 26 \, (M_s/{\rm TeV})~{\rm GeV}\nn
& \Gamma_{Q^{(2)}_L}^{J=3/2} = 22 \, (M_s/{\rm TeV})~{\rm GeV},\quad \Gamma_{\tilde
Q^{(2)}_L}^{J=3/2} = 32 \, (M_s/{\rm TeV})~{\rm GeV}\nn
& \Gamma_{Q^{(2)}_L}^{J=1/2} = 31 \, (M_s/{\rm TeV})~{\rm GeV}\,.
\end{align}
At higher string scales, the decay widths slightly
decrease because of the running of the couplings. For  $M_s \sim 40$~TeV,
we obtain
\begin{align}
& \Gamma_{G^{(2)}}^{J=3} =50\, (M_s/{\rm TeV})~{\rm GeV}, \quad \Gamma_{G^{(2)}}^{J=2} = 46 \, (M_s/{\rm TeV})~{\rm GeV},\nn
& \Gamma_{G^{(2)}}^{J=1} = 59 \, (M_s/{\rm TeV})~{\rm GeV},\quad
\Gamma_{Q^{(2)}_R}^{J=5/2} = 27 \, (M_s/{\rm TeV})~{\rm GeV},\nn
& \Gamma_{Q^{(2)}_R}^{J=3/2} = 23 \, (M_s/{\rm TeV})~{\rm GeV},\quad \Gamma_{\tilde
Q^{(2)}_R}^{J=3/2} = 34 \, (M_s/{\rm TeV})~{\rm GeV}\nn
& \Gamma_{Q^{(2)}_R}^{J=1/2} = 33 \, (M_s/{\rm TeV})~{\rm GeV},\quad \Gamma_{Q^{(2)}_R}^{J=5/2} = 23 \, (M_s/{\rm TeV})~{\rm GeV}\nn
& \Gamma_{Q^{(2)}_R}^{J=3/2} = 19 \, (M_s/{\rm TeV})~{\rm GeV},\quad \Gamma_{\tilde
Q^{(2)}_R}^{J=3/2} = 28 \, (M_s/{\rm TeV})~{\rm GeV}\nn
& \Gamma_{Q^{(2)}_R}^{J=1/2} = 27 \, (M_s/{\rm TeV})~{\rm GeV}\,.
\end{align}
The dijet signal-to-noise ratio for $n=2$ is shown in
Fig.~\ref{figure:tres}. For $M_s \lesssim 10.5~{\rm TeV}$ 
the second massive Regge excitations could also  be observed with a
statistical significance  $\geq
5\sigma$ at the HE-LHC and for $M_s \lesssim 28~{\rm TeV}$ at the
VLHC. Measurement of both resonant peaks would constitute 
definitive evidence for string physics.

\subsection{Angular distributions}

In what follows we briefly comment on the angular distributions.  QCD
parton-parton cross sections are dominated by $t$-channel exchanges
that produce dijet angular distributions which peak at small
center-of-mass scattering angles. In contrast, nonstandard contact
interactions or excitations of resonances result in a more isotropic
distribution. In terms of rapidity variable for standard transverse
momentum cuts, dijets resulting from QCD processes will preferentially
populate the large rapidity region, while the new processes generate
events more uniformly distributed in the entire rapidity region. To
analyze the details of the rapidity space the D\O\ Collaboration
introduced a new parameter~\cite{Abbott:1998wh},
\begin{equation}
R = \frac{d\sigma/dM|_ {(|y_1|,|y_2|< 0.5)}}{d\sigma/dM|_{(0.5 < |y_1|,|y_2| < 1.0)}} \, ,
\end{equation}
the ratio of the number of events, in a given dijet mass bin, for both
rapidities $|y_1|, |y_2| < 0.5$ and both rapidities $0.5 < |y_1|,
|y_2| < 1.0$.  The ratio $R$ is
a genuine measure of the most sensitive part of the angular
distribution, providing a single number that can be measured as a
function of the dijet invariant mass. An illustration of the use of
this parameter in a heuristic model where standard model amplitudes
are modified by a Veneziano form factor has been
presented in Ref.~\cite{Meade:2007sz}. 

It is important to note that although there are no $s$-channel resonances
in $qq\rightarrow qq$ and $qq'\rightarrow qq'$ scattering,
KK modes in the $t$ and $u$ channels generate calculable
effective four-fermion contact terms. These in turn are manifest in  a
small departure from the QCD value of $R$ outside the resonant
region~\cite{Anchordoqui:2009mm}. In an optimistic scenario,
measurements of this modification could shed light on the D-brane
structure of the compact space. It could also serve to differentiate
between a stringy origin for the resonance as opposed to an isolated
structure such as a $Z'$, which would not modify $R$ outside the
resonant region. While the signal of quark scattering is suggestive,
the analysis in Ref.~\cite{Anchordoqui:2009mm} did not take into account
all of the potential detector effects, which is necessary to be
confident that the effect is real. In the next  section we describe
the first steps toward a more realistic description of the string physics
processes.

\section{SEGI}
\label{s7}

SEGI is a modification of the original BlackMax event
generator~\cite{Dai:2007ki,Dai:2009by}, which is extensively used by
ATLAS and CMS collaborations in search for exotic physics. At its
inception, BlackMax could simulate only black hole production in
particle collisions (including all the greybody factors known to
date)~\cite{Banks:1999gd,Dimopoulos:2001hw,Giddings:2001bu,Kanti:2002ge,Anchordoqui:2003ug,Stojkovic:2004hp,Dai:2006dz,Creek:2007tw}. Then it gradually grew into a very
comprehensive generator that can accommodate different signatures of
quantum gravity, {\it e.g.}, stringball evaporation in a two-body final
state~\cite{Dimopoulos:2001qe}.  With the current
modification, BlackMax will be able to simulate production and decay
of lowest massive Regge excitations yielding $\gamma$ + jet, $Z$ +
jet, and dijet events.

\begin{figure}[t]
\begin{center}
\includegraphics[width=0.8\linewidth]{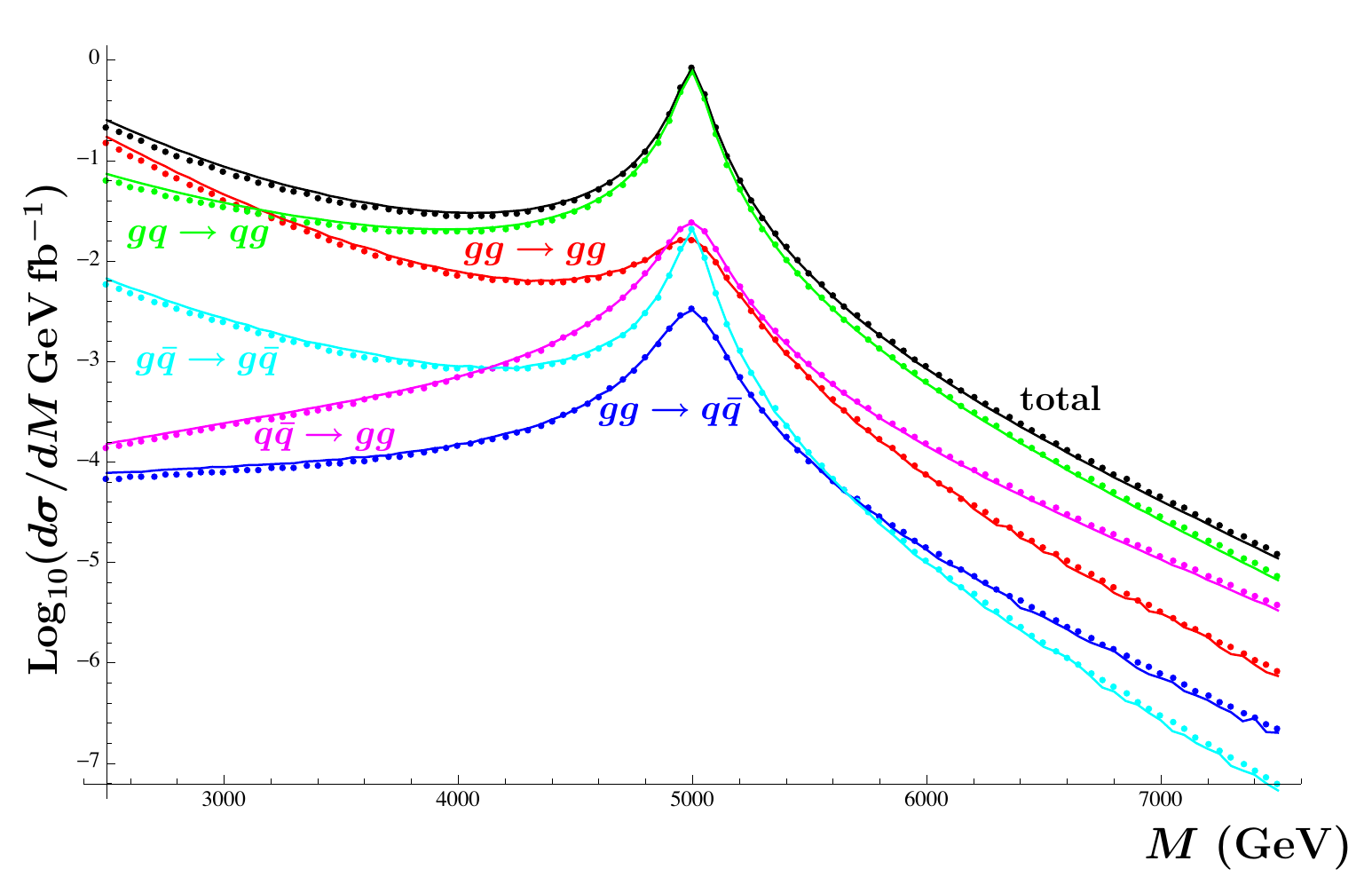}
\end{center}
\caption{$d\sigma/dM$ vs. $M$ of first resonance string signal as
  obtained through the semianalytic parton model calculation (dots)
  and with SEGI (solid). We have taken $M_s = 5~{\rm TeV}$.}
\label{segi-uno}
\end{figure}

A necessary input for the event generator is the amplitudes for
perturbative string mediated processes.  The parton-parton subprocesses
of lowest massive Regge excitations decaying to dijets are given in
Eqs.~(\ref{gggg2}), (\ref{LCdos}), (\ref{LCtres}), and (\ref{qgqg2}),
whereas those decaying into $\gamma$ + jet are giving in
Eqs.~(\ref{mhvlow2}) and (\ref{qgqz}).\footnote{Ignoring the $Z$ mass and assuming that cross
  sections $\times$ branching into lepton pairs are large enough for complete
  reconstruction of $pp \to Z +$ jet, the contribution to the
  signal is suppressed relative to the photon signal by a factor of
  $\tan^2 \theta_W = 0.29$.}  The cross section can be written as a
convolution of (\ref{Lucero}) with PDFs
{\it e.g.}, for dijets,
\begin{equation}
\sigma_{pp \to {\rm dijet}} = \sum_{ij} \int^{\hat{s}_{{\rm
      max}}/s}_{\hat{s}_{{\rm min}}/s}
d\tau \int^1_\tau \frac{dx_a}{x_a}  \ \sigma_{ij \to kl} \ f_i
(x_a,\hat{s}) \  f_j (\tau/x_a,\hat{s}) \,,
\end{equation}
where $\hat{s}_{\rm max}$ and $\hat{s}_{\rm min}$ are the maximum and
minimum square center-of-mass energy of the colliding partons.  The
code iterates $10^6$ times to calculate the Monte Carlo integral. As
an illustration, in Fig.~\ref{segi-uno} we show a comparison of the
invariant mass distribution, setting $M_s = 5~{\rm TeV}$, as obtained
by SEGI and with the semianalytic (parton model) approach adopted in
the preceding section.

\begin{figure}[t]
\begin{center}
\includegraphics[width=0.49\linewidth]{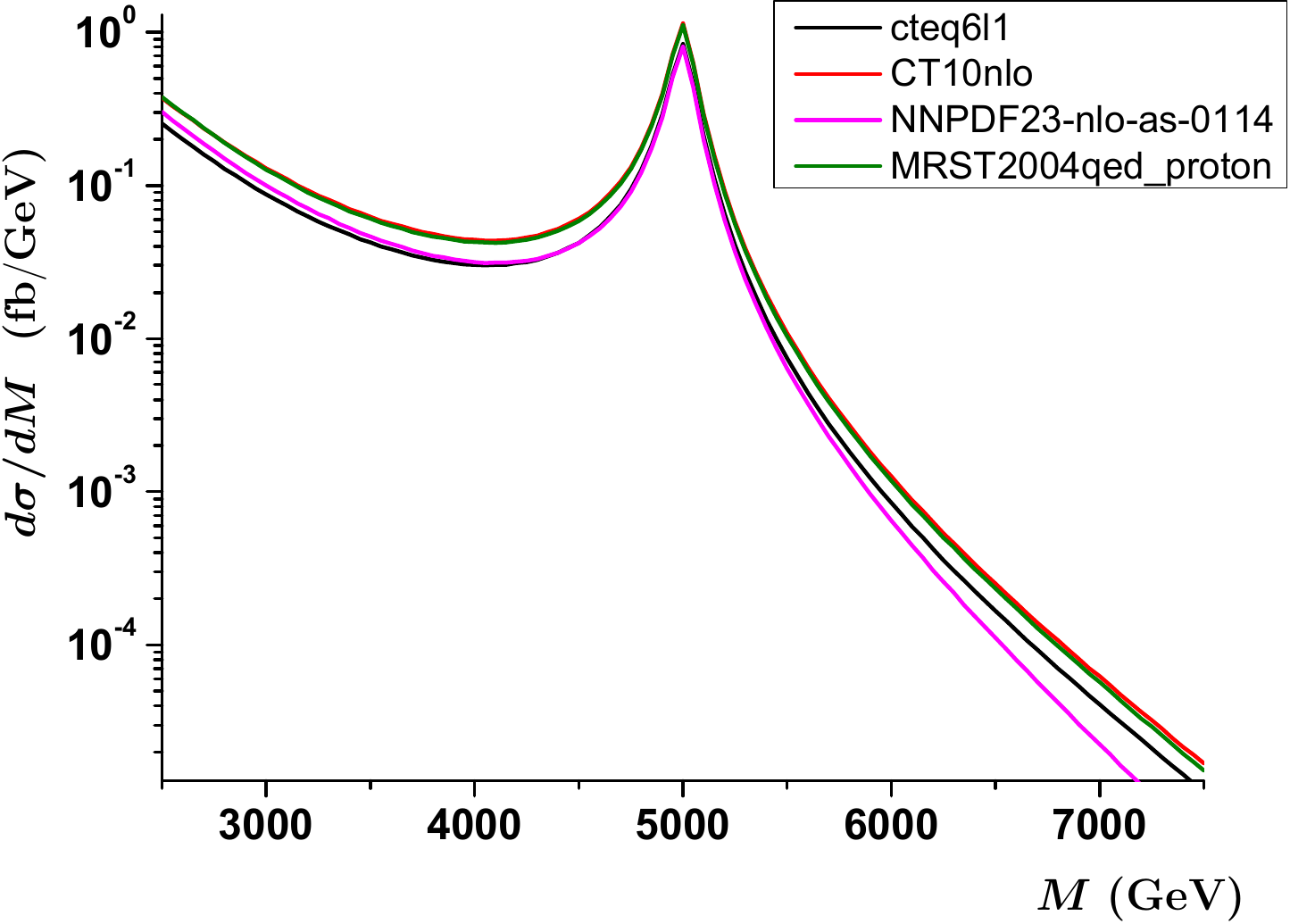}
\includegraphics[width=0.49\linewidth]{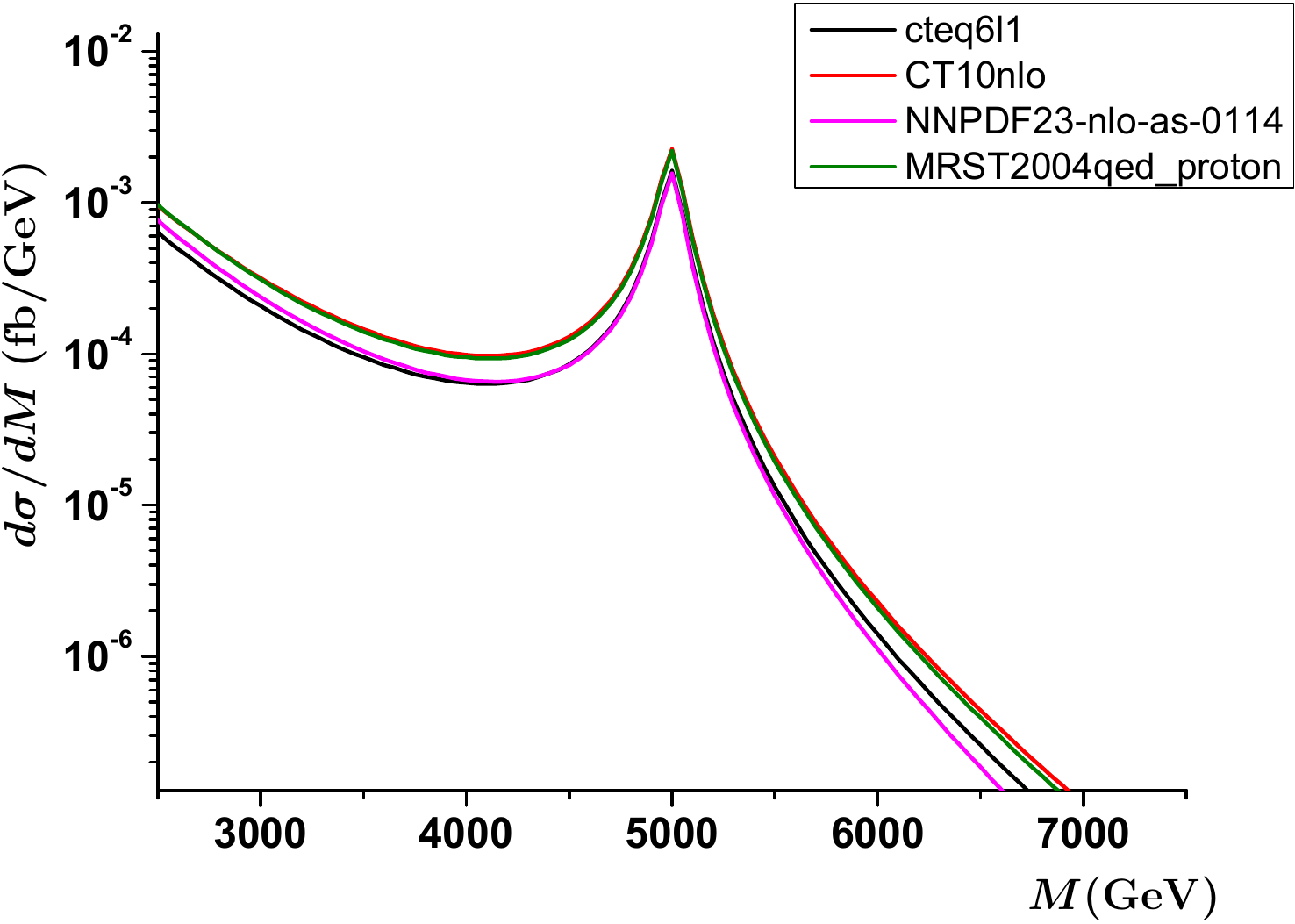}
\end{center}
\vspace{-0.3cm}
\caption{Systematic uncertainty
of the dijet (left) and $\gamma$ + jet (right) string signal due to
PDFs 
as obtained with SEGI.}
\label{segi-dos}
\end{figure}

The input parameters for the generator are read from the file
parameter.txt (see Appendix D for how to access the file). In the
following list we provide an explanation for the relevant
input parameters:
\begin{itemize}
\item{\tt Number\_of\_simulations} This parameter is the number of events to be generated.
\item{\tt Type\_of\_incoming\_particles} This parameter determines the type of 
incoming particles: 
\begin{enumerate}
\item {\tt pp} 
\item ${\tt p}\bar{\tt p}$
\item {\tt e}$^+${\tt e}$^-$
\end{enumerate}
\item{\tt Center\_of\_mass\_energy\_of\_incoming\_particles} This is the 
center-of-mass energy of the two incoming particles in units of GeV.
\item{\tt Choose\_a\_case} This parameter determines which type of
  events are simulated:
\begin{enumerate}
\item {\tt nonrotating\_black\_hole\_on\_a\_tensionless\_brane}
\item {\tt nonrotating\_black\_hole\_on\_a\_nonzero\_tension\_brane}
\item {\tt rotating\_black\_hole\_on\_a\_tensionless\_brane}
\item {\tt nonrotating\_black\_hole\_with\_fermion\_tensionlees\_brane\_splitting}
\item {\tt stringballs\_two\_particle\_final\_states}
\item {\tt lowest\_massive\_Regge\_excitations\_decaying\_to\_dijets}
\item {\tt lowest\_massive\_Regge\_excitations\_decaying\_to\_gamma+jet}
\item {\tt lowest\_massive\_Regge\_excitations\_decaying\_to\_Z+jet}
\end{enumerate}
\item{\tt Choose\_a\_pdf\_file
    (200\_to\_240\_CETQ6\_or\_$>$10000\_for\_LHAPDF)} This parameter
  determines which PDF is used in the
  simulation. The code includes CETQ6 PDFs by
  default. In that case this parameter should be set from 200 to
  240. For different PDFs one must install
  LHAPDF. The impact of the different PDFs and induced systematics in
  the production and decay of Regge recurrences is shown in
  Fig.~\ref{segi-dos}.
\item{\tt Minimum\_mass} This is the minimum mass that one wants to include in the simulation in units of GeV.
\item{\tt Maximum\_mass} This is the maximum mass that one wants to include in the simulation in units of GeV.
\item{\tt String\_scale} This parameter is the string scale $M_s$ in units of GeV.
\item{\tt string\_coupling} This parameter is the string coupling; the default is set to  $g_s = 0.1$.
\item{\tt kappa} This is the $C-Y$mixing parameter; the default is set to $\kappa = 0.14$.
\end{itemize}
All the other BlackMax parameters are irrelavant for simulation of
Regge recurrences.

\begin{figure}[t]
\begin{center}
\includegraphics[width=0.49\linewidth]{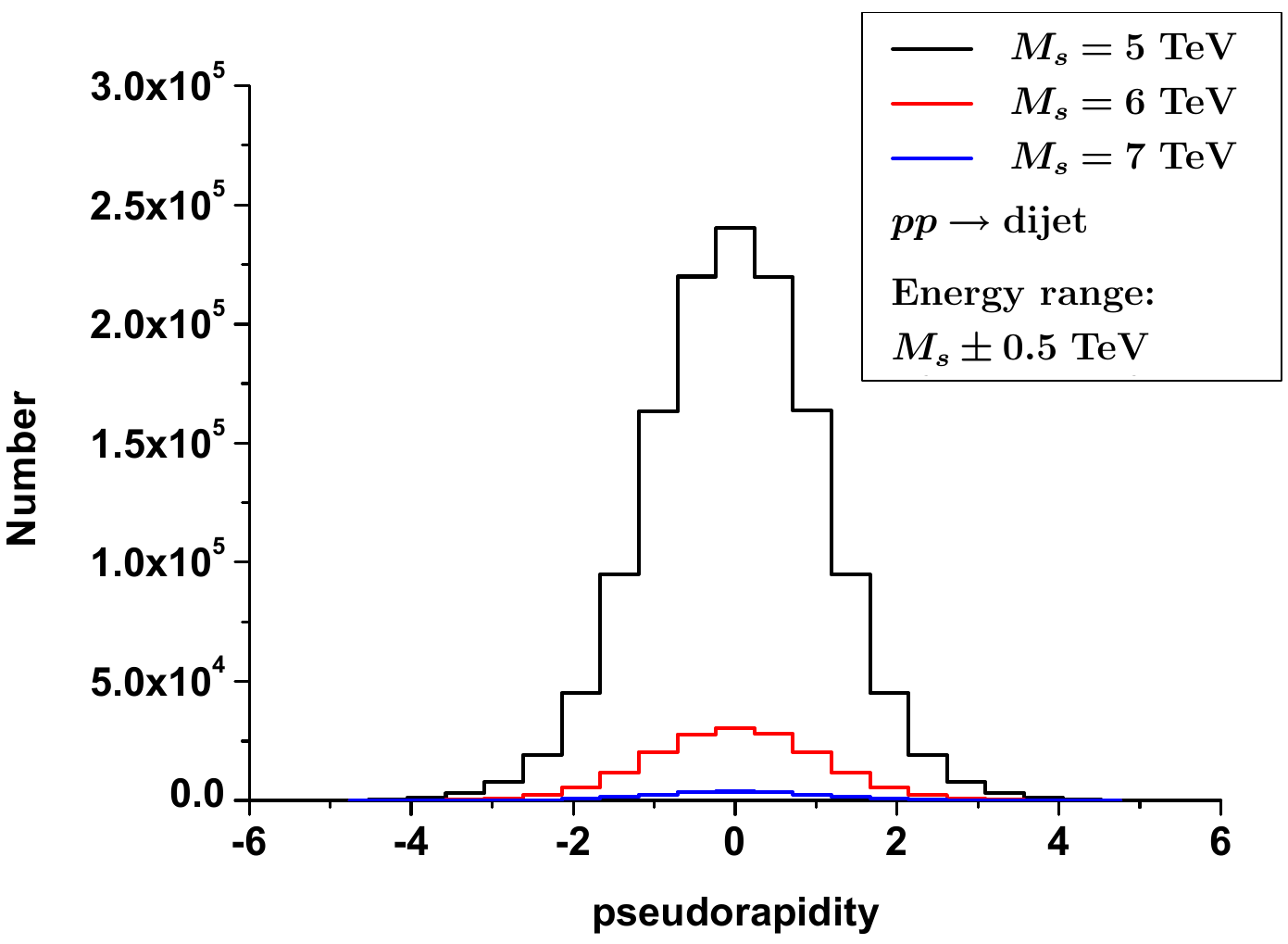}
\includegraphics[width=0.49\linewidth]{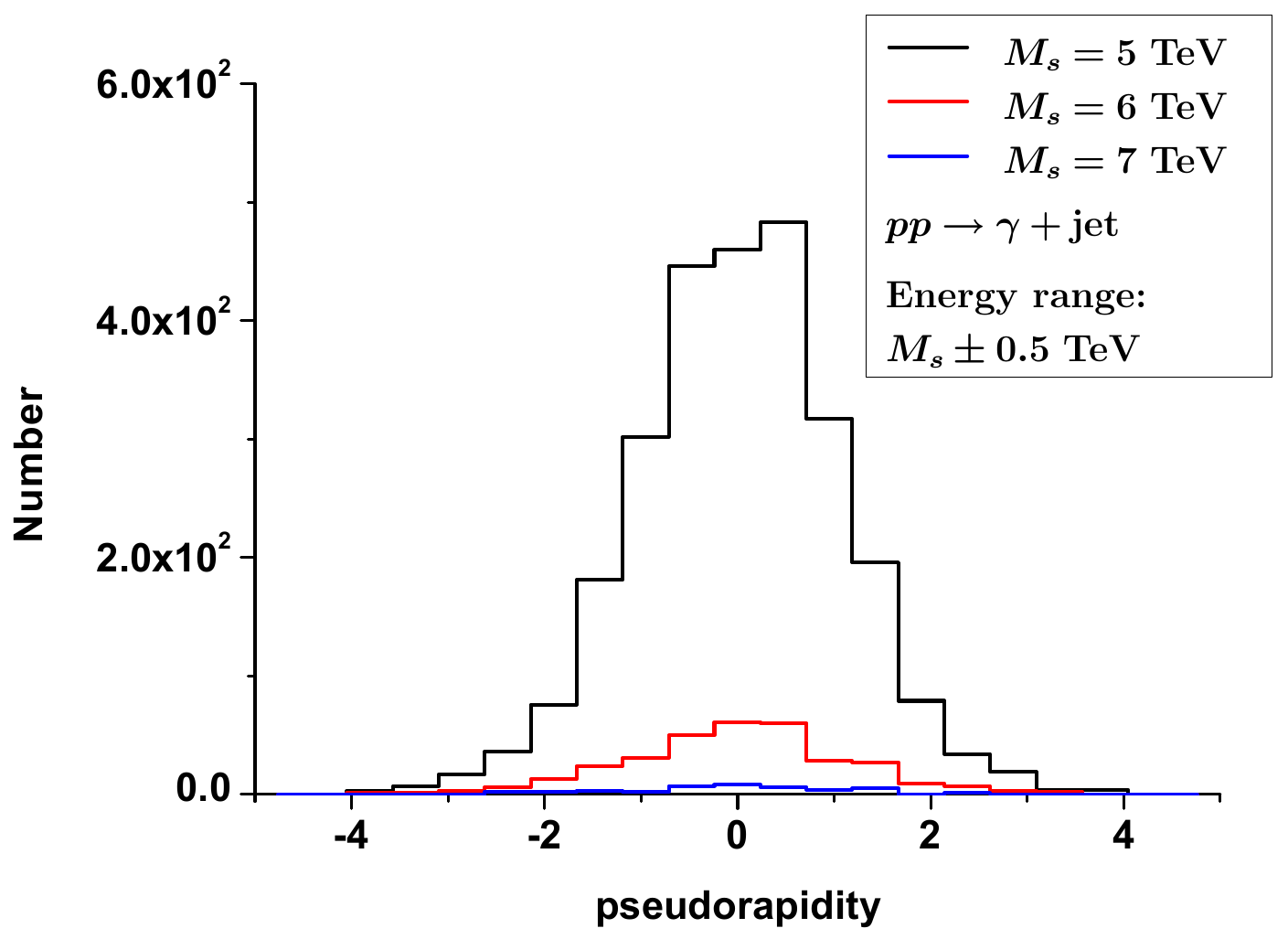}
\end{center}
\caption{Dijet (left) and $\gamma$ + jet (right) pseudorapidity distributions.}
\label{figure-pseudo}
\end{figure}

The generator gives the output.txt file. This file contains the cross
sections and the energy momentum distributions of the incoming and
outgoing particles (pseudorapidity distributions are displayed in
Fig.~\ref{figure-pseudo} for illustrative purposes only). The incoming
particles are marked as {\tt parent}. The outgoing particles are
marked as {\tt elast}. The meaning of each column is the same as in
the original BlackMax event generator~\cite{Dai:2007ki,Dai:2009by}.
The most up-to-date source code and TarBall can be downloaded from:\\
\\
\url{http://projects.hepforge.org/blackmax/}\\
\\
The details for SEGI installation can be found in the BlackMax
manual~\cite{Dai:2009by}. For completness, a brief summary of the
installation process is provided in Appendix~\ref{installation}.

Thus far we have included in SEGI string excitations only up to $n=1$. In
future versions we plan to extend the code to account for higher order
excitations of the string, as well as $qq \to qq$ and $q q' \to qq'$ interactions.

\section{Conclusions}
\label{s8}

We have explored the discovery potential of existing and proposed
hadron colliders to unmask excitations of the string.  We have studied
the direct production of Regge recurrences, focusing on the first and
second excited levels of open strings localized on the world volume of
D-branes.  In this framework, $U(1)_B$ and $SU(3)_C$ appear as subgroups
of $U(3)$ associated with open strings ending on a stack of three 
D-branes. In addition, the minimal models contain two other stacks to
accommodate the electroweak $SU(2)_L \subset U(2)$ and the
hypercharge $U(1)_Y$. For such D-brane models, the resonant parts of
the relevant string theory amplitudes are {\it universal} to leading
order in the gauge coupling. As a consequence, it is feasible to
extract genuine string effects which are independent of the
compactification scheme.  In this paper we have made use of the
amplitudes evaluated near the first and second resonant poles to
report on the discovery potential for Regge excitations of the quark,
the gluon, and the color singlet living on the QCD stack of D-branes.

To calculate the string signal for $n=1$ resonances, we used the
partial decay widths obtained elsewhere~\cite{Anchordoqui:2008hi}.  To
compute the signal for $n=2$ resonances, we have presented here a
complete calculation of all relevant decay widths of the second
massive-level string states, including decays into massless particles
and a massive $n=1$ and a massless particle. The latter were obtained
from factorizing four-point amplitudes with one first massive-level
string state computed in Ref.~\cite{Feng:2010yx}. The partial decay widths
of the spin-3 and spin-2 bosons from the second massive level
were also obtained from direct string amplitude computations and match
exactly with the results obtained from factorization. We also
constructed the helicity wave functions of arbitrary higher spin
massive boson.

Our phenomenological study among the various processes indicates that:
\begin{itemize}
\item For $M_s \lesssim 7.1~{\rm TeV}$, the HL-LHC will be able to
  discover (with statistical \mbox{significance $> 5\sigma$}) the lowest massive
  Regge excitations in dijet events. For string scales as high as 6.1~TeV, observations of resonant structures in $pp \to \gamma$ + jet can provide
  interesting corroboration (with statistical \mbox{significance $> 5\sigma$}) of low-mass-scale string physics.
\item The dijet discovery potential exceedingly improves at the HE-LHC and
  VLHC. For $n=1$, the HE-LHC will be able to discover string
  excitations up to $M_s \approx 15~{\rm TeV}$, whereas the VLHC will attain
  $5\sigma$ discovery  up to $M_s \approx 41~{\rm TeV}$.  Moreover,  for $n=2$,
 the HE-LHC will reach $5\sigma$ discovery for $M_s \lesssim 10.5~{\rm
    TeV}$, while the VLHC will be able to discover Regge
  excitations for $M_s \lesssim 28~{\rm TeV}$.
\item Keeping only transverse $Z$'s and assuming that cross sections
  $\times$ branching into lepton pairs are large enough for complete
  reconstruction of $pp \to Z$ + jet, the D-brane contribution to the
  signal is suppressed relative to $pp \to \gamma$ + jet  by a factor of
  $\tan^2 \theta_W = 0.29$. This differs radically from stringball
  evaporation in two-body final state. In such a case, emissions of $\gamma$ +
  jet and $Z$ + jet are comparable. The suppression of  $Z$ +jet 
  production, the origin of which  lies in the particular structure of the
  D-brane model, will hold true for all the low-lying levels of the
  string.
\end{itemize}

Our calculations have been performed using a semianalytic parton
model approach which is cross checked against an original software
package.  The string event generator interfaces with HERWIG and Pythia
through BlackMax. The source code is publically available in the
hepforge repository.

In summary, in this paper we have provided a concrete starting point
for understanding the string physics potential of proposed machines
that would collide protons at energies approaching the boundary of
what (wo)mankind can daydream to achieve. The results presented herein will
help to lay out opportunities, connections, and challenges for future
LHC upgrades.

\section*{Acknowledgements}
We thank Oliver Schlotterer for valuable discussions.  L.A.A. is
supported in part by the U.S. National Science Foundation (NSF) CAREER
Award PHY-1053663 and by the National Aeronautics and Space
Administration (NASA) Grant No. NNX13AH52G.  I.A. is supported in part
by the European Commission under the ERC Advanced Grant 226371.
D.C.D. is supported by Shanghai Institutions of Higher Learning, the
Science and Technology Commission of Shanghai Municipality, Grant
No. 11DZ2260700. W.Z.F. is supported by the Alexander von Humboldt
Foundation. H.G. and T.R.T. are supported by NSF Grant
No. PHY-1314774. X.H. is supported by the MOST Grant
103-2811-M-003-024. D.L. is partially supported by the ERC Advanced
Grant ``Strings and Gravity'' (Grant.No. 32004) and by the DFG cluster
of excellence ``Origin and Structure of the Universe.'' D.S. is
supported by NSF Grant No. PHY-1066278. Any opinions, findings, and
conclusions or recommendations expressed in this material are those of
the authors and do not necessarily reflect the views of the National
Science Foundation.

\appendix

\section{Notation of group factors}\label{App1}

We define the structure constant $f^{abc}$ and the total symmetric
group factor $d^{abc}$ as
\begin{align}
[T^{a},T^{b}] & =i\sum_{c}f^{abc}T^{c}\,,\\
\{T^{a},T^{b}\} & =4\sum_{c}d^{abc}T^{c}\,.
\end{align}
With the notation ${\rm Tr}(T^{a}T^{b})=\frac{1}{2}\delta^{ab}$,
we could obtain
\begin{align}
{\rm Tr}([T^{a},T^{b}]T^{c}) & =\tfrac{i}{2}f^{abc}\,,\label{fabc}\\
{\rm Tr}(\{T^{a},T^{b}\}T^{c}) & =2d^{abc}\,.\label{dabc}
\end{align}
We could also obtain
\begin{equation}
{\rm Tr}(T^{ab}T^{cd})=2\sum_{e}{\rm Tr}(T^{ab}T^{e}){\rm Tr}(T^{cd}T^{e})\,,
\end{equation}
where $T^{ab}$ or $T^{cd}$ presents either $[T^{a},T^{b}]$ or $\{T^{a},T^{b}\}$.

We thus arrive at
\begin{align}
{\rm Tr}([T^{a},T^{b}][T^{c},T^{d}]) & =-\tfrac{1}{2}\sum_{e}f^{abe}f^{cde}\,,\\
{\rm Tr}(\{T^{a},T^{b}\}\{T^{c},T^{d}\}) & =8\sum_{e}d^{abe}d^{cde}\,,\\
{\rm Tr}([T^{a},T^{b}]\{T^{c},T^{d}\}) & =2i\sum_{e}f^{abe}d^{cde}\,.
\end{align}

\section{Spinor helicity formalism for massless fields}\label{AppB}

\subsection{Helicity wave functions for massless spin-$\frac{1}{2}$ fermions}

For massless spin-$\frac{1}{2}$ spinors, we use the notation following
Ref.~\cite{Feng:2010yx},
\begin{flalign}
|i\rangle & =|k_{i}\rangle=u_{+}(k_{i})=v_{-}(k_{i})=\binom{0}{k_{i}^{*\dot{a}}}\,,\\
|i] & =|k_{i}]=u_{-}(k_{i})=v_{+}(k_{i})=\binom{k_{i,a}}{0}\,,\\
[i| & =[k_{i}|=\bar{u}_{+}(k_{i})=\bar{v}_{-}(k_{i})=(k_{i}^{a},0)\,,\\
\langle i| & =\langle k_{i}|=\bar{u}_{-}(k_{i})=\bar{v}_{+}(k_{i})=(0,k_{i,\dot{a}}^{*})\,,
\end{flalign}
where the momenta with spinor indices are two-component commutative spinors,
which are defined by
\begin{gather}
P^{\dot{a}a}=p_{\mu}\bar{\sigma}^{\mu\dot{a}a}=-p^{*\dot{a}}p^{a}\,,\\
P_{a\dot{a}}=p_{\mu}\sigma_{a\dot{a}}^{\mu}=-p_{a}p_{\dot{a}}^{*}\,,
\end{gather}
where $p^{*\dot{a}}=(p^{a})^{*}$ and $p_{\dot{a}}^{*}=(p_{a})^{*}$.
Spinor indices could be raised (lowered) by $\varepsilon^{ab}$
($\varepsilon_{ab}$) or $a,b$ with dots,
\begin{equation}
p^{a}=\varepsilon^{ab}p_{b}\,,\qquad p^{*\dot{a}}=\varepsilon^{\dot{a}\dot{b}}p_{\dot{b}}^{*}\,.
\end{equation}
The spinor products are defined by
\begin{gather}
\langle pq\rangle=\langle p|q\rangle=\bar{u}_{-}(p)u_{+}(q)=p_{\dot{a}}^{*}q^{*\dot{a}}\,,\\
{}[pq]=[p|q]=\bar{u}_{+}(p)u_{-}(q)=p^{a}q_{a}\,,
\end{gather}
and we have the following relations:
\begin{gather}
[pq]=-[qp]\,,\qquad
\langle pq\rangle=-\langle qp\rangle\,, \qquad
\langle pp\rangle=[pp]=0\,,\\
\langle pq\rangle^{*}=[qp]\,, \qquad \langle pq\rangle[qp]=-2(p\cdot q)\,.
\end{gather}

\subsection{Helicity wave functions for massless spin-1 gauge boson}

The gauge transformation for a spin-1 gauge boson reads
$\epsilon^{\mu} \rightarrow\epsilon^{\mu}+\Lambda k^{\mu}$.  The
massless spin-1 gauge boson only has 2 degrees of freedom, which
are helicity up ($+$) and down ($-$).  The helicity wave functions
(polarization vectors) of a massless spin-1 gauge boson can be
written as
\begin{align}
\epsilon_{\mu}^{+}(k,r)&=\frac{\langle r|\gamma_{\mu}|k]}{\sqrt{2}\langle rk\rangle}=\frac{r_{\dot{a}}^{*}\bar{\sigma}_{\mu}^{\dot{a}a}k_{a}}{\sqrt{2}\langle rk\rangle}\,,\\
\epsilon_{\mu}^{-}(k,r)&=-\frac{[r|\gamma_{\mu}|k\rangle}{\sqrt{2}[rk]}=-\frac{r^{a}\sigma_{\mu a\dot{a}}k^{*\dot{a}}}{\sqrt{2}[rk]}\,,
\end{align}
where $k$ is the momentum of the gauge boson and $r$ is the reference
momentum which can be chosen to be any lightlike momentum except $k$.
The final results of the helicity amplitudes are independent
of the choice of reference momentum $r$.

\section{Helicity wave functions for massive spin-$\frac{1}{2}$ and -$\frac{3}{2}$ fermions}\label{AppC}

The wave functions of massive spin-$\frac{1}{2}$ and
spin-$\frac{3}{2}$ fermions were constructed in Ref.~\cite{Novaes:1991ft}.

\subsection{Helicity wave functions for massive spin-$\frac{1}{2}$ fermions}

Massive spin-$\frac{1}{2}$ fermions wave functions satisfy the Dirac equation
\begin{align}
(\slashed k+m)u(k) & =0\,,\\
(\slashed k-m)v(k) & =0\,,
\end{align}
where $u(k)$ and $v(k)$ are positive and negative energy solutions
with the momentum $k^{\mu}$, which correspond to fermion and antifermion
wave functions, respectively.
After decomposing $k$ into two lightlike momenta $p,q$, up to a phase factor,
the helicity wave function of the massive spin-$\frac{1}{2}$ fermions can be written as
\begin{gather}
u_{+}(k)  =\binom{\frac{\langle qp\rangle}{m}q_{a}}{p^{*\dot{a}}}\,,\qquad
u_{-}(k)  = \binom{p_{a}}{\frac{[qp]}{m}q^{*\dot{a}}}\,, \\
v_{+}(k)  = \binom{p_{a}}{\frac{[pq]}{m}q^{*\dot{a}}}\,,\qquad
v_{-}(k)  = \binom{\frac{\langle pq\rangle}{m}q_{a}}{p^{*\dot{a}}}\,.
\end{gather}

\subsection{Massive spin-$\frac{3}{2}$ fermions wave functions}

A massive spin-$\frac{3}{2}$ fermion could be described by Rarita--Schwinger
spinor-vector $\Psi^{A,\mu}$ which satisfies equations
\begin{align}
(i\slashed\partial-m)_{\phantom{A}B}^{A}\Psi^{B,\mu} & =0\,,\label{eq:32M-C1}\\
(\gamma_{\mu})_{\phantom{A}B}^{A}\Psi^{B,\mu} & =0\,,\label{eq:32M-C2}\\
\partial_{\mu}\Psi^{B,\mu} & =0\,,\label{eq:32M-C3}
\end{align}
where $A$ and $B$ are spinor indices which run from 1 to 4. We can
rewrite the first equation in terms of positive and negative solutions
of Dirac equation, i.e., $U$ and $V$, which read
\begin{align}
(\slashed k+m)_{\phantom{A}B}^{A}U(k)^{B,\mu} & =0\,,\\
(\slashed k-m)_{\phantom{A}B}^{A}V(k)^{B,\mu} & =0\,.
\end{align}
Using the same decomposition $k=p+q$, where $p,q$ are lightlike
reference momenta, we have, up to a phase factor,
\begin{align}
U^{A,\mu}(+\tfrac{3}{2}) & =\frac{1}{\sqrt{2}m}\binom{\frac{\langle qp\rangle}{m}q_{a}}{p^{*\dot{a}}}(p_{\dot{b}}^{*}\bar{\sigma}^{\mu\dot{b}b}q_{b})\,,\\
U^{A,\mu}(+\tfrac{1}{2}) & =\frac{\bar{\sigma}^{\mu\dot{b}b}}{\sqrt{6}m}\left[\binom{\frac{\langle qp\rangle}{m}q_{a}}{p^{*\dot{a}}}(p_{\dot{b}}^{*}p_{b}-q_{\dot{b}}^{*}q_{b})+\binom{\frac{\langle qp\rangle}{m}p_{a}}{-q^{*\dot{a}}}(p_{\dot{b}}^{*}q_{b})\right]\,,\\
U^{A,\mu}(-\tfrac{1}{2}) & =\frac{\bar{\sigma}^{\mu\dot{b}b}}{\sqrt{6}m}\left[\binom{p_{a}}{\frac{[qp]}{m}q^{*\dot{a}}}(p_{\dot{b}}^{*}p_{b}-q_{\dot{b}}^{*}q_{b})+\binom{-q_{a}}{\frac{[qp]}{m}p^{*\dot{a}}}(q_{\dot{b}}^{*}p_{b})\right]\,,\\
U^{A,\mu}(-\tfrac{3}{2}) & =\frac{1}{\sqrt{2}m}\binom{p_{a}}{\frac{[qp]}{m}q^{*\dot{a}}}(q_{\dot{b}}^{*}\bar{\sigma}^{\mu\dot{b}b}p_{b})\,,
\end{align}
and
\begin{align}
V^{A,\mu}(+\tfrac{3}{2}) & =\frac{1}{\sqrt{2}m}\binom{p_{a}}{\frac{[pq]}{m}q^{*\dot{a}}}(q_{\dot{b}}^{*}\bar{\sigma}^{\mu\dot{b}b}p_{b})\,,\\
V^{A,\mu}(+\tfrac{1}{2}) & =\frac{\bar{\sigma}^{\mu\dot{b}b}}{\sqrt{6}m}\left[\binom{p_{a}}{\frac{[pq]}{m}q^{*\dot{a}}}(p_{\dot{b}}^{*}p_{b}-q_{\dot{b}}^{*}q_{b})+\binom{-q_{a}}{\frac{[pq]}{m}p^{*\dot{a}}}(q_{\dot{b}}^{*}p_{b})\right]\,,\\
V^{A,\mu}(-\tfrac{1}{2}) & =\frac{\bar{\sigma}^{\mu\dot{b}b}}{\sqrt{6}m}\left[\binom{\frac{\langle pq\rangle}{m}q_{a}}{p^{*\dot{a}}}(p_{\dot{b}}^{*}p_{b}-q_{\dot{b}}^{*}q_{b})+\binom{\frac{\langle pq\rangle}{m}p_{a}}{-q^{*\dot{a}}}(p_{\dot{b}}^{*}q_{b})\right]\,,\\
V^{A,\mu}(-\tfrac{3}{2}) & =\frac{1}{\sqrt{2}m}\binom{\frac{\langle pq\rangle}{m}q_{a}}{p^{*\dot{a}}}(p_{\dot{b}}^{*}\bar{\sigma}^{\mu\dot{b}b}q_{b})\,.
\end{align}

\section{SEGI installation}
\label{installation}

The first step is to download the zipped tar file which has to be
unzipped to
extract the files and make the program executable:\\
\\
{\tt
  gunzip 	 BlackMax-2.00.tar.gz\\
  tar -xvf BlackMax-2.00.tar\\
}
\\
Before compilation one has to check the compiler version of gcc by executing the  command\\
\\
{\tt gcc --version} \\
\\
which generates the output\\
\\
{\tt gcc (GCC) 3.4.6 20060404 (Red Hat 3.4.6-10)\\
  Copyright (C) 2006 Free Software Foundation, Inc.\\
  ...}\\
\\
This second step is required because the latest gcc compiler version (4.1.2)
has changed the names of some system libraries needed to compile
Fortran with C code. The download is configured to use gcc version
4. If an older gcc version ({\it e.g.} 3.4.6) is in operation, then
one needs to modify the BlackMax Makefile. This can be accomplished by
uncommenting the following lines in the Makefile\\
\\
{\tt
  F77LIB =g2c\\
  F77COMP=g77}\\
\\
After that SEGI is ready for compilation. There are three different ways to run SEGI:
{\it (i)}  standalone mode for which no additional libraries are required,
{\it (ii)} accessing PDFs from LHAPDF, or {\it (iii)} accessing PDFs from
LHAPDF and simultaneous hadronization from Pythia. In each case a different
compilation/linking step is required  to produce the executable. For all three
options, the default format of the event output is the Les Houches
Accord format~\cite{Boos:2001cv}. This text file can be used as input
into HERWIG and Pythia to hadronize the SEGI events.

\subsection{ Standalone mode}

In this version the proton parton densities are taken from CTEQ6m
which
are packaged with BlackMax. After unpacking, the command:\\
\\
{\tt gmake BlackMaxOnly}\\
\\
has to be executed and the file parameter.txt has to be modified to
select one of the 41 CTEQ6m PDF sets that has been bundled
with BlackMax, {\it e.g.},\\
\\
{\tt choose$\_$a$\_$pdf$\_$file(200$\_$to$\_$240$\_$cteq6)Or$\_>$10000$\_$for$\_$LHAPDF}\\
200\\
\\
After that, the executable can be run\\
\\
{\tt BlackMax $>\&!$ out}
\\

\subsection{LHAPDF} 
\label{subsec:lhapdf}
This version uses the proton parton densities from the LHAPDF library,
which must be
downloaded from\\
\\
\url{http://projects.hepforge.org/lhapdf/}\\
\\
Of course, one has to install the package in a directory with write
permission. One can do this by specifying an installation directory (for additional information, the reader is referred to the LHAPDF manual). Then the BlackMax Makefile must
be edited to insert the library locations. One has to verify that the
${\tt LD\_LIBRARY\_PATH}$ environment variable includes the location
of the newly built LHAPDF library:\\
\\
export ${\tt LD\_LIBRARY\_PATH=\$LD\_LIBRARY\_PATH:/data/rizvi/atlas/lhapdf-5.3.0/lhapdf/lib}$\\
export ${\tt LHAPATH=/data/rizvi/atlas/lhapdf-5.3.0/lhapdf/share/lhapdf/PDFsets}$\\
\\
The next step is to select a valid PDF set in parameter.txt, {\it
  e.g.}, the LHAPDF
partons from the H1 PDF2000 fit of HERA data:\\
\\
{\tt choose$\_$a$\_$pdf$\_$file(200$\_$to$\_$240$\_$cteq6\_or$\_>$10000$\_$for$\_$LHAPDF)}\\
70050\\

After unpacking the source files one can compile the program\\
\\
{\tt gmake BlackMax}\\
\\
After that, the executable can be run\\
\\
{\tt BlackMax $>\&!$ out}

\subsection{LHAPDF with simultaneous Pythia hadronization}

To hadronize the events BlackMax comes with
an interface to Pythia. To generate fully hadronized events one needs
to download and install the latest versions of LHAPDF and PYTHIA. They
are available at\\
\\
\url{http://www.hepforge.org/downloads/pythia6}\\
and
\url{http://www.hepforge.org/downloads/lhapdf}\\
\\
BlackMax has been tested wth Pythia 6.4.10 and LHAPDF 5.3.0.  After
that, one has to create
the Pythia libraries and remove both the following four dummy routines\\
\\
upinit.f\\
upevnt.f\\
pdfset.f\\
structm.f\\
\\
and  the pdfset.f routine from the Pythia Makefile.  The
four routines above are all dummy routines which actually exist in
LHAPDF.  Next, one must edit the BlackMax Makefile to insert the
library locations, while checking that the ${\tt
  LD\_LIBRARY\_PATH}$ environment variable includes the location
of the newly built Pythia and LHAPDF libraries:\\
\\
export ${\tt LD\_LIBRARY\_PATH=\$LD\_LIBRARY\_PATH:/data/rizvi/atlas/lhapdf-5.3.0/lhapdf/lib}$\\
export ${\tt LHAPATH=/data/rizvi/atlas/lhapdf-5.3.0/lhapdf/share/lhapdf/PDFsets}$\\
\\
Finally, one has to create the BlackMax executable using the target ``all'' which will
link to the Pythia and LHAPDF libraries,\\
\\
{\tt gmake all}\\
\\
and select  a valid PDF set in parameter.txt, {\it e.g.},\\
\\
${\tt choose\_a\_pdf\_file(200\_to\_240\_cteq6)Or\_>10000\_for\_LHAPDF}$\\
$10050$\\
\\
After that, the exectuable can be run\\
\\
{\tt BlackMax $>\&!$ out}
\\

\end{document}